%% file: main.tex
\documentclass[journal]{IEEEtran}

\usepackage{calc,amsfonts,amssymb,amsmath,bm,url,color,graphicx,cite,epstopdf,nicefrac,bbold, amsthm}
\usepackage{psfrag,subfigure,float}
\usepackage[algoruled,linesnumbered]{algorithm2e}
\usepackage{multirow}
\usepackage{algorithmic}
\usepackage[normalem]{ulem}
\usepackage{stmaryrd}
\usepackage{xcolor}
\usepackage{psfrag,framed,mdframed}
\usepackage{comment}
\usepackage{mdframed}
\usepackage{cleveref}
\usepackage{bibunits}
\usepackage[utf8]{inputenc}

\DeclareMathOperator*{\argmin}{arg\,min}
\usepackage{mathtools}

\definecolor{orange}{RGB}{255,107,0}
\definecolor{green}{RGB}{0,100,0}

\makeatletter
\def\endthebibliography{%
	\def\@noitemerr{\@latex@warning{Empty `thebibliography' environment}}%
	\endlist
}
\makeatother

\makeatletter
\newcommand{\nosemic}{\renewcommand{\@endalgocfline}{\relax}}
\newcommand{\dosemic}{\renewcommand{\@endalgocfline}{\algocf@endline}}
\let\oldnl\nl
\newcommand{\nonl}{\renewcommand{\nl}{\let\nl\oldnl}}
\makeatother

\newlength\myindent
\input{def}
\begin{document}

	\title{Deep Spectrum Cartography: Completing Radio Map Tensors Using Learned Neural Models}

	\author{Sagar Shrestha, Xiao Fu, and Mingyi Hong
	\thanks{
	S. Shrestha and X. Fu are with the School of Electrical Engineering and Computer Science, Oregon State University, Corvallis, OR 97331, USA. 
	
	M. Hong is with the Department of Electrical and Computer Engineering, University of Minnesota, Minneapolis, MN 55455, USA.

	The work of S. Shrestha and X. Fu is supported in part by the NSF MLWiNS Program NSF CNS-2003082, and in part by NSF ECCS-2024058. The work is also supported by a gift from Intel through the MLWiNS Program.
	
	The work of M. Hong is supported in part by the NSF MLWiNS Program NSF CNS-2003033, and in part by ARO W911NF-19-1-0247. The work is also supported by a gift from Intel through the MLWiNS Program.}

	}
	
	\maketitle
	
	\begin{abstract} 
		The \textit{spectrum cartography} (SC) technique constructs multi-domain (e.g., frequency, space, and time) radio frequency (RF) maps from limited measurements, which can be viewed as an ill-posed tensor completion problem. Model-based cartography techniques often rely on handcrafted priors (e.g., sparsity, smoothness and low-rank structures) for the completion task. Such priors may be inadequate to capture the essence of complex wireless environments---especially when severe shadowing happens. To circumvent such challenges, offline-trained deep neural models of radio maps were considered for SC, as deep neural networks (DNNs) are able to ``learn'' intricate underlying structures from data. 
		However, such deep learning (DL)-based SC approaches encounter serious challenges in both off-line model learning (training) and completion (generalization), possibly because the latent state space for generating the radio maps is prohibitively large. In this work, an emitter radio map disaggregation-based approach is proposed, under which only individual emitters' radio maps are modeled by DNNs. This way, the learning and generalization challenges can both be substantially alleviated. Using the learned DNNs, a fast nonnegative matrix factorization-based two-stage SC method and a performance-enhanced iterative optimization algorithm are proposed. Theoretical aspects---such as recoverability of the radio tensor, sample complexity, and noise robustness---under the proposed framework are characterized, and such theoretical properties  have been elusive in the context of DL-based radio tensor completion. Experiments using synthetic and real-data from indoor and heavily shadowed environments are employed to showcase the effectiveness of the proposed methods.
	\end{abstract}
	

	\IEEEpeerreviewmaketitle

	\section{Introduction}\label{sec:intro}
	    Radio frequency (RF) awareness is the stepping stone towards smart, intelligent, and high-efficiency wireless communication systems.
	    Sensing the RF environment has been at the heart of the physical layer technologies, especially since the vision of cognitive radio was advocated by the {\it Federal Communications Commission} (FCC) \cite{fcc2004notice}. Starting from the early 2000s, a plethora of \textit{spectrum sensing} techniques have been proposed to estimate the spectral usage; see, e.g.,  \cite{tian2007compressed, fu2015factor, bazerque2010distributed} and a recent survey \cite{arjoune2019comprehensive}.
	    A step forward is the \textit{spectrum cartography} (SC) technique, which aims at crafting a radio power propagation map across a geographical region from (often sparsely acquired and limited) sensor measurements \cite{jayawickrama2013improved, boccolini2012wireless, bazerque2011group}. 
	    SC produces radio maps that reveal critical information of the RF environment across multiple domains, e.g., time, frequency, and space. Hence, SC is considered useful for a wide range of tasks in wireless communications---e.g., spectrum surveillance, opportunistic access, interference-free routing and networking, and mobile relay placement; see \cite{bi2019engineering}. 
	    
	    On the other hand, SC is a highly nontrivial task. Sensors are often sparsely deployed in the geographical region of interest, and can only acquire power measurements {\it locally} in space. Reconstructing the full radio map across multiple domains poses a multi-aspect high-dimensional data inverse problem using heavily down-sampled measurements---which is known to be challenging in both theory and methods \cite{kanatsoulis2019tensor, yuan2016tensor}.

	    Many SC approaches have emerged since the late 2000s. 
	    For example,  the work in \cite{boccolini2012wireless} employed a widely used geostatistical tool, namely, Kriging interpolation \cite{boccolini2012wireless}, which treats the SC problem as a 2D field interpolation problem (implicitly) under statistical priors (e.g., the Gaussian prior).
	    The work in \cite{jayawickrama2013improved, jayawickrama2014iteratively} proposed a compressive sensing-based SC framework that exploits the spatial sparsity of emitter locations---and the work in \cite{bazerque2010distributed} additionally exploits the sparsity of spectral usage by the emitters. The methods in \cite{kim2013cognitive} took a dictionary learning/sparse coding perspective for SC, which exploit spatio-temporal correlations and the associated sparse representations. 
	    To further exploit the spatial smoothness/correlation of the radio maps, the approaches in \cite{bazerque2011group,hamid2017non} used representation tools such as {\it thin plate splines} (TPS) and {\it radial basis functions} (RBF) to enable interpolation; also see a sensor signal-quantized version in \cite{romero2017learning}. 
	    Recently, the work in \cite{zhang2020spectrum} models the radio map as a low-rank block-term tensor and offered a tensor recovery-based SC approach, which is again a means to advantage of the spatial smoothness of radio maps.

	    The existing SC approaches are effective to a certain extent, but a number of challenges remain. 
	    Some approaches may need a number of environmental parameters, e.g., the point-to-point propagation model \cite{jayawickrama2013improved}, basis function for representing the emitters' power spectral density (PSD) \cite{bazerque2011group}, or even the exact PSDs of the emitters \cite{romero2017learning}---which are not always easy to acquire, especially in fast changing environments.
	    More importantly, handcrafted model priors (e.g., sparsity, smoothness, low rank) are often inadequate to capture the complex underlying structure of the radio maps. Such modeling mismatches could lead to substantial performance degradation. For example, the tensor-based method in \cite{zhang2020spectrum} is built upon the premise that the spatial power propagation pattern of each emitter gives rise to a low-rank matrix. However, this assumption is often violated in geometrically complicated environments, e.g., urban and indoor areas, where many blockages and barriers exist and the shadowing effect is severe (cf. Fig. \ref{fig:low-rank}). Many temporal/spatial smoothness based approaches, e.g., \cite{bazerque2011group,hamid2017non}, share the same challenges.
	    
	    To circumvent such modeling uncertainties, data-driven methods were recently proposed in \cite{han2020power, teganya2020data} for SC.
	    Instead of hinging on pre-specified and handcrafted model priors, 
	    these methods ``learn'' from training samples and leverage the expressiveness of deep neural networks (DNNs) to represent radio data with complex underlying structures.  The methods in \cite{han2020power, teganya2020data} learn ``completion networks'' offline, and then feed the sensor-acquired measurements through the network to produce an estimated radio map. 
	    
	    On one hand, DL-based SC has potentials in dealing with intricate RF environments. On the other hand, purely data-driven methods have their own challenges in both the offline network learning (training) and online radio map completion (generalization) stages.
	    Training a DNN to represent a multi-domain high-dimensional radio tensor is computationally challenging.
	    In particular, the state space of the radio maps is often prohibitively large. The size of the state space is affected by the number of emitters, the frequency usage of them, the locations of them---and all possible combinations. Consequently, one may need an unexpectedly large number of training samples to learn the network. 
	    Hence, the training stage may involve a large number of samples to `cover' representative states---which induce heavy computational loads and overall harder optimization problems.
	    In addition, the generalization stage also has a number of challenges.
	    Specifically, even if very large training sets have been used,
	    many ``out-of-training-distribution'' cases can still easily happen in the completion stage---e.g., when $R$ emitters are used for generating the training samples but some extra emitters move into the scene, as a consequence, the networks trained with a smaller number of emitters often produce radio maps that miss some emitters.

	    \noindent
	    {\bf Contributions.} In this work, we propose a {\it hybrid} model and data-driven SC framework.
	    Specifically, our method combines a radio map disaggregation model with DNN-based spatial power propagation pattern learning.
	    The idea is only using DNNs to represent the most complex part in radio map models, instead of the entire radio map---in order to alleviate the training and generalization challenges. Our detailed contributions are as follows:
	    
	    \noindent
	    $\bullet$ {\bf Emitter Disaggregated Neural Modeling.}   Motivated by the fact that the overall interference radio map is an aggregation of a number of emitters' individual radio maps, we propose to learn and represent each emitter's spatio-spectral radio map (as opposed to the ambient radio map itself) using deep networks. Intuitively, {\it single} emitter's radio maps reside in a much smaller state space compared to that of the aggregated maps. Using our disaggregated modeling also effectively overcomes the out-of-distribution generalization problem in the completion (test) stage.
	    
	     \noindent
	    $\bullet$ {\bf Recoverability-Guaranteed Realizations.}
	    After  the disaggregated deep neural model is trained,
	    we offer two algorithms to realize SC. (i) We propose a two-stage method. The first stage leverages the sparsity of spectral occupancy of the emitters and a greedy {\it nonnegative matrix factorization} (NMF) approach \cite{gillis2013fast,fu2014self} to decompose the sensor measurements to emitter components. Then, the second stage uses an off-line trained deep completion network to estimate the individual emitters' radio maps. (ii) When the spectral usage of the emitters is not sparse, we propose a deep generative model regularized radio map completion criterion for our task---with the price paid on a heavier computational burden. We provide performance characterizations for our methods---including recoverability, noise robustness, and sample complexity---which have been elusive in the context of DL-based SC.
	    
	      \noindent
	    $\bullet$ {\bf Synthetic and Real Data Validation.} We validate the proposed methods over a variety of synthetic data experiments, as well as a real data experiment with radio map measurements from an indoor area at University of Mannheim, Germany \cite{king2008crawdad}. Both the simulations and the real experiments corroborate our design, and show that our framework offers promising SC performance in challenging scenarios.
	    
	    \smallskip
	    
	    A conference version of this work  has appeared in ICASSP 2021 \cite{icassp2021submission}, which includes the NMF-assisted SC approach under sparse emitter spectral occupancy. In this journal version, we additionally include a new iterative algorithm that deals with more challenging dense spectral occupancy cases, its radio map recoverablity and stability analyses, more extensive simulations, and a real data experiment.

	    \noindent
		\textbf{Notation.} $x \in \bbR, \x \in \bbR^K, \X \in \bbR^{I \times K}, \tX \in \bbR^{I \times J \times K}$ represent a scalar, a vector, a matrix and a tensor, respectively. 
		$\x_i$ represents the $i$th column of matrix $\X$. The \texttt{Matlab} notations $\X(:,j)$ and $\X(i,:)$ represent the $j$th column and $i$th row of matrix $\X$, respectively; similar \texttt{Matlab} notations are used for vectors and tensors as well. $\circ$ and $\circledast$ represent the outer product and Hadamard product, respectively; in particular, if $\X\in\mathbb{R}^{I\times J}$ and $\bm y\in\mathbb{R}^K$, then $\tZ=\X\circ \y$ is an $I\times J\times K$ tensor such that $\tZ(i,j,k)=\X(i,j)\bm y(k)$. $\|\x\|_2$ and $\|\X\|_2$ denote the vector 2-norm and the matrix spectral norm, respectively. $\|.\|_{\rm F}$ represents the matrix and tensor Euclidean norm. $\X^\T$ denotes the transpose of $\X$. $\geq$, $\leq$, $>$ and $<$ represent element-wise inequality. $\zero$ and $\one$ denote all-zero and all-one matrices/vectors with proper size, respectively. $|{\cal X}|$ denotes the cardinality of set ${\cal X}$. $[I] = \{1,2, \dots, I\}$ where $I$ is an integer. 
		A list of the abbreviations used in this paper can be found in Table~\ref{tab:abbrev}.
		
		\begin{table}[t!]
            \centering
            \caption{List of Abbreviations.}
            \begin{tabular}{|c|c|}
                \hline
                \textbf{Abbreviations} & \textbf{Definitions} \\ \hline 
                SLF & Spatial Loss Function \\ \hline
                PSD & Power Spectral Density \\ \hline
                SC & Spectrum Cartography \\ \hline
                DNN & Deep Neural Network \\ \hline
                DL & Deep Learning \\ \hline
                NNLS & Nonnegativity constrained Least Squares \\ \hline
                SPA & Successive Projection Algorithm \\ \hline
                AO & Alternating optimization \\ \hline
                NMF & Nonnegative Matrix Factorization \\ \hline
                NAE & Normalized Absolute Error \\ \hline
                SRE & Squared Reconstruction Error  \\ \hline
            \end{tabular}
            \label{tab:abbrev}
        \end{table}

		\begin{figure}[t!]
			\centering
			\includegraphics[width=0.3\linewidth]{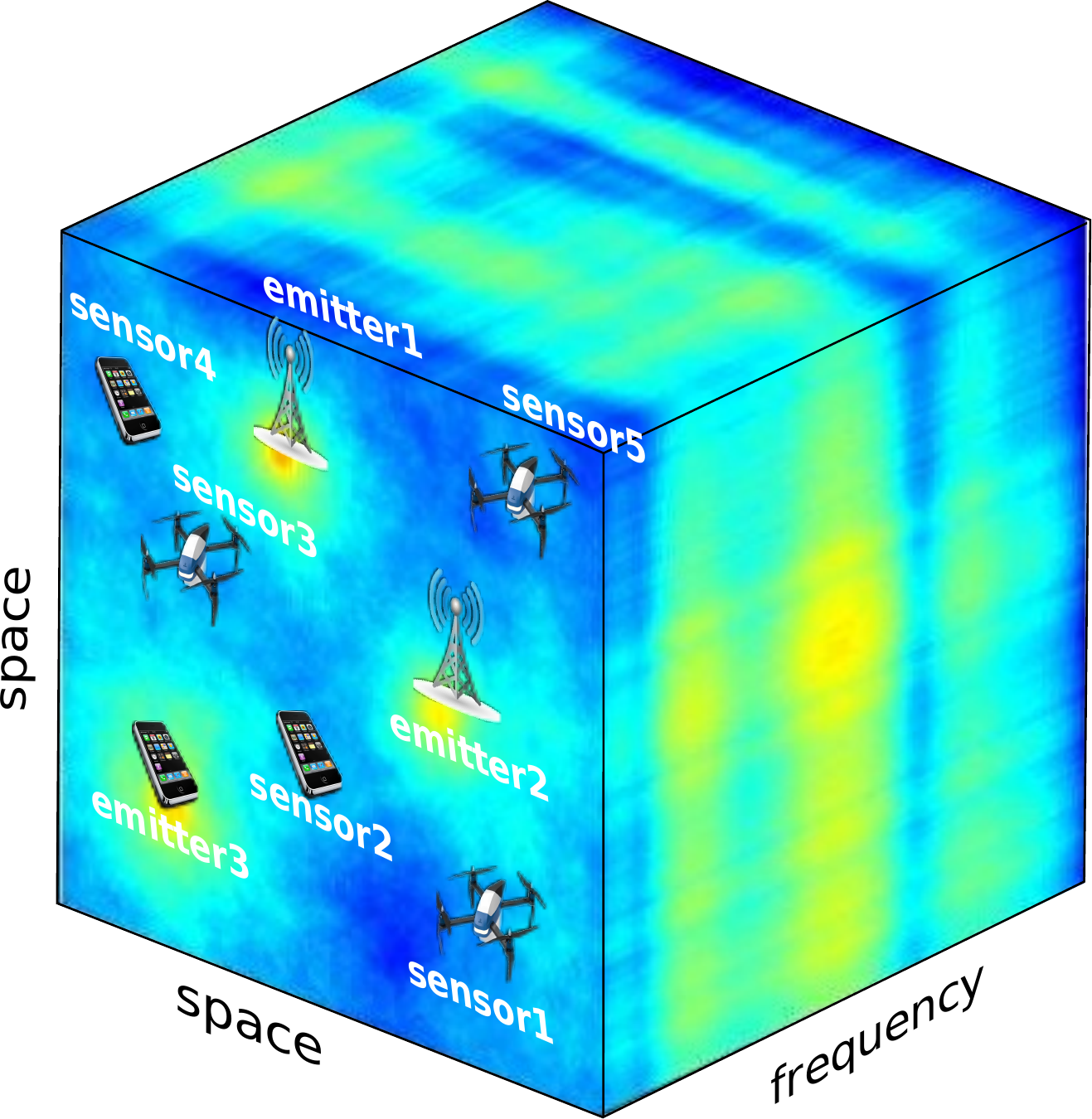}
			\quad
		   	\includegraphics[width=.55\linewidth]{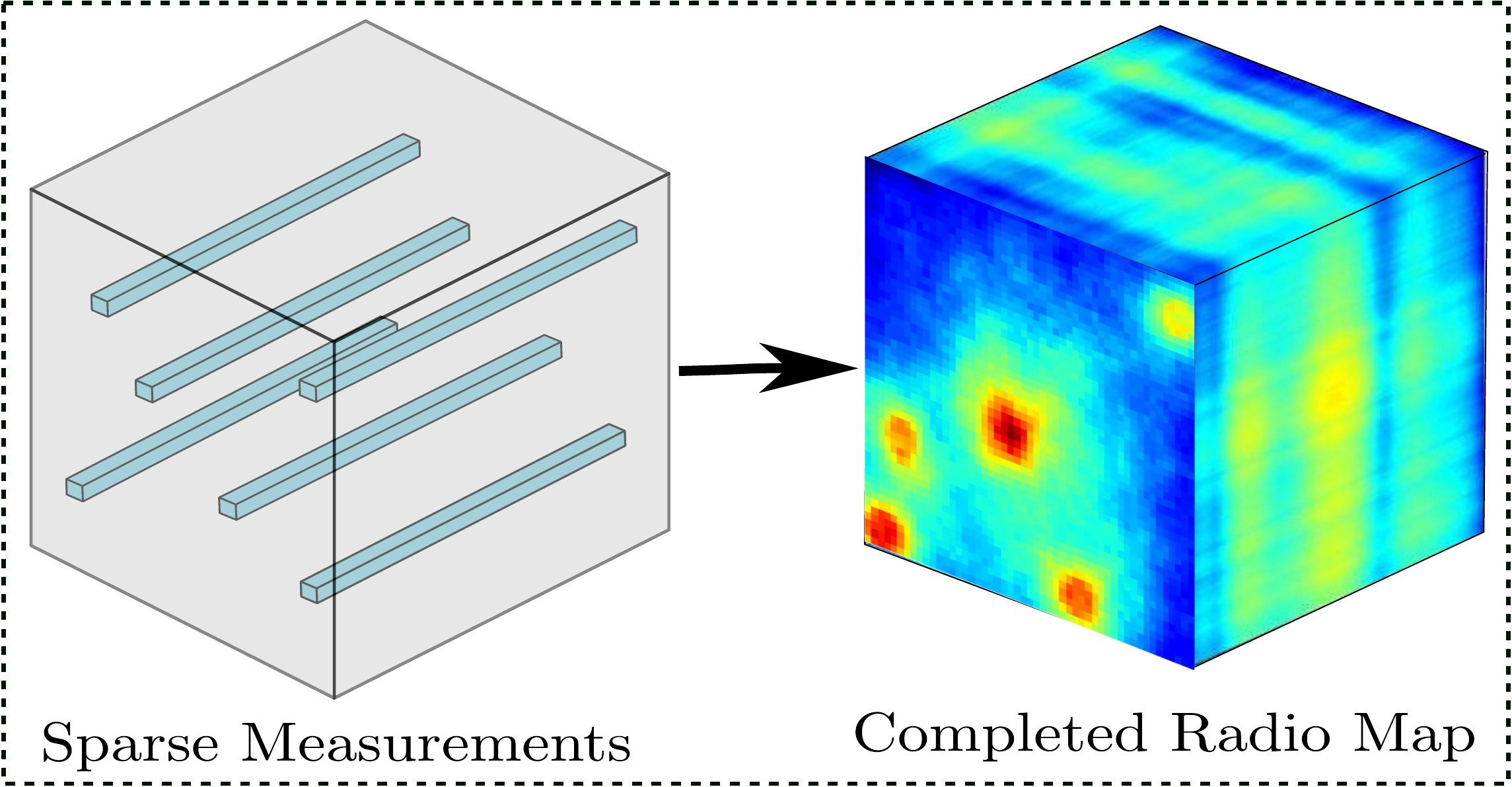}	
			\caption{(Left) Illustration of a spatio-spectral radio map tensor. (Right) The SC scenario of interest: completing the radio map from sensed tensor ``fibers''.}
			\label{fig:spectrum3d}
		\end{figure}

	\section{Problem Statement and Background}
	In this section, we present the problem setup and briefly introduce existing approaches.
	
	\subsection{Problem Setup and Signal Model}
	{
		We are interested in estimating a spatio-spectral radio map of a certain geographical region across a number of frequency bins. 
		The region has
		$R$ emitters that transmit their signals across certain (possibly overlapping) frequency bands, which together create a power interference field across space and frequency. We consider a rectangular region discretized into $I \times J$ grids and $K$ frequency bins. Hence, the (discretized) {\it power spectral density} (PSD) measured over space and frequency gives rise to a 3D
		radio map that can be expressed as an $I\times J\times K$ tensor $\tX \in \mathbb{R}^{I\times J\times K}$, where $\tX(i,j,k)$ denotes the PSD of the signal received at location $(i,j)$ and frequency $k$; see  Fig.~\ref{fig:spectrum3d} (left). We should mention that although the instantaneous radio map may change quickly, the PSD captures the average spectrum usage over a certain period of time, if the signals are stationary random processes. Hence, estimating PSD-based radio maps is often considered more realistic than estimating the instantaneous situation; see, e.g., \cite{mehanna2013frugal,fu2016power,fu2015factor,bazerque2010distributed,bazerque2011group}. 
		
		Suppose that there are a number of sensors deployed within this region.
		We assume that a sensor located at position $(i,j)$ where $i\in[I]$ and $j\in[J]$ can measure the power spectral density (PSD) of its received signal locally over all $K$ frequency bins; i.e., the sensor located at $(i,j)$ can acquire the PSD $\bm p_{(i,j)}\in\mathbb{R}^K$ of the received signal such that
		$ {\bm p}_{(i,j)}=\tX(i,j,:) \in\mathbb{R}^K,$
		where $K$ is the number of frequency bins.
		Note that a vector $\tX(i,j,:)$ is also sometimes called a ``{\it tensor fiber}'' \cite{fu2020block,fu2020computing}.
		If every location $(i,j)$ has a sensor, a radio map tensor $\tX\in\mathbb{R}^{I\times J\times K}$ would be constructed. However, in practice, placing $IJ$ sensors to ``cover'' the geographical region may not be an viable option. 
		Instead, only a small number of sensors are sparsely placed within the region of interest. In other words, letting	$$\bOmega = \{(i,j)|i \in [I],j\in [J]\}\subseteq [I]\times [J]$$ denote the set of sensor locations, we often have $|\bm \varOmega| \ll IJ$. Hence, the task of spatio-spectral SC is to estimate the full radio map tensor  $\tX \in \bbR^{I \times J \times K}$ using the sensor-acquired measurements (i.e., tensor fibers) $\tX(i,j,:) \in \bbR^K, \forall (i,j) \in \bOmega$---see Fig. \ref{fig:spectrum3d} (right) for an illustration. 
		}
			
\begin{figure}[t]
			    \centering
			    \includegraphics[width=0.7\linewidth]{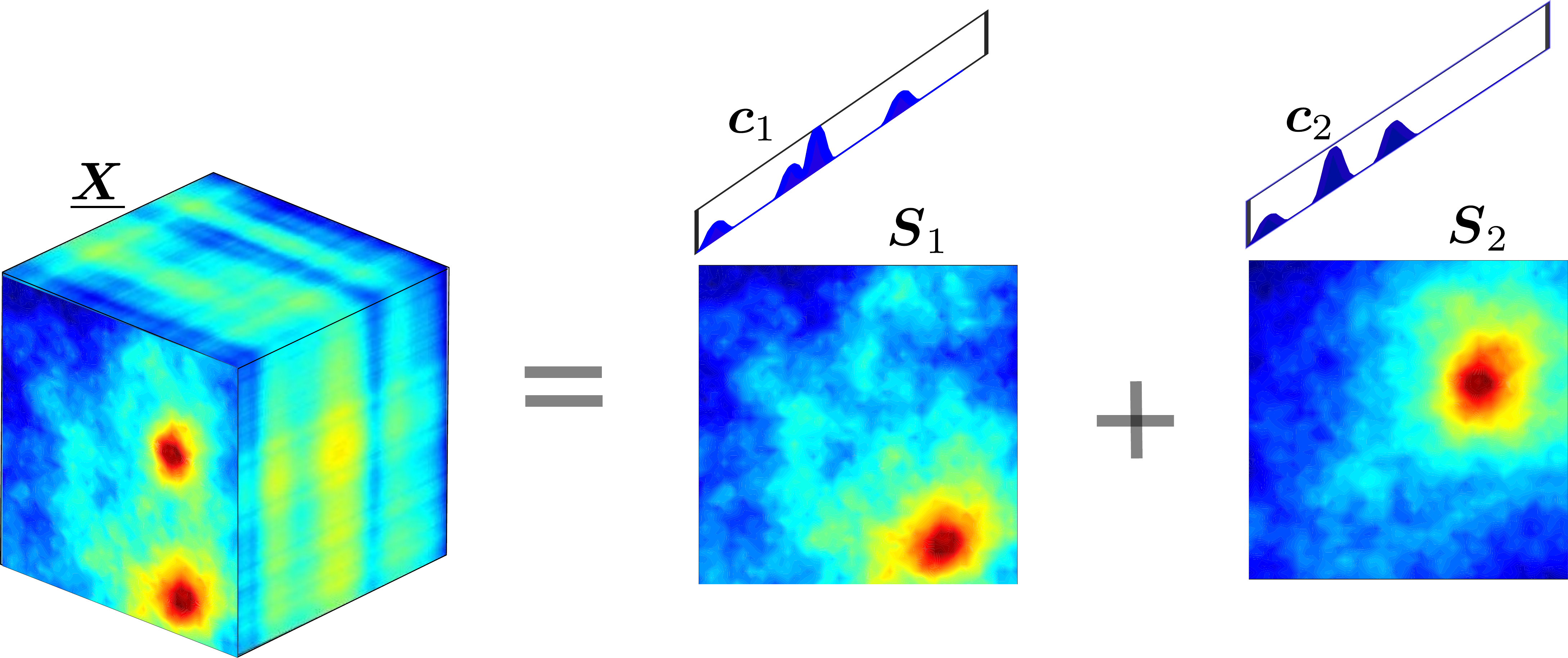}
			    \caption{Illustration of \eqref{eq:sigmodel} for $R=2$ case.}
			    \label{fig:disaggregationmodel}
			\end{figure}

			 If the spectral band of interest is not large relative to the central carrier frequency (e.g., 20MHz bandwidth at a central carrier frequency within 2-5 GHz.), the spatial propagation pattern of an emitter across different frequencies are identical up to scaling differences caused by the emitter's PSD. Consequently, if noise is absent, the full radio map can be expressed as follows \cite{zhang2020spectrum,bazerque2011group}:
			\begin{equation}
			\tX(i,j,k) =\sum_{r=1}^R\S_r(i,j)\c_r(k) \Longleftrightarrow	\tX =  \sum_{r=1}^R \S_r \circ \c_r,
				\label{eq:sigmodel}
			\end{equation}
		where $\S_r\in\mathbb{R}^{I\times J}$ denotes the {\it spatial loss function} (SLF) of the power of emitter $r$, $\c_r\in\mathbb{R}^K$ represents the PSD of emitter $r$, $\circ$ is the outer product, and $R$ is the number of emitters in the region of interest. An illustration is shown in Fig.~\ref{fig:disaggregationmodel}. A remark is that the factorization based expression of a radio map also makes storing $\tX$ easier. The memory cost of $\tX\in\mathbb{R}^{I\times J\times K}$ under \eqref{eq:sigmodel} is only $O((IJ+K)R)$ other than $O(IJK)$. The memory is even lower when $\S_r$ can be parameterized, e.g., using a low-rank matrix as in \cite{zhang2020spectrum} or a deep generative model as will be seen in this work.

	   \subsection{Prior Art}	
	   {   
	   In essence, the SC problem is an ill-posed inverse problem. Hence, the same insights employed in classic inverse problems, e.g., compressive sensing and low-rank matrix/tensor completion may be used for SC. Indeed, early works exploited the spatial sparsity of emitter locations \cite{jayawickrama2013improved}, sparse representation of emitter PSDs in a certain learned domain \cite{kim2013cognitive, bazerque2010distributed}, spatial smoothness of the SLFs \cite{bazerque2011group}, and a low-rank tensor structure \cite{zhang2020spectrum} to come up with radio map recovery formulations. Notably, the most recent work in \cite{zhang2020spectrum} assumed that ${\rm rank}(\S_r)\leq L$ where $L\ll \min\{I,J\}$ and recast the SC problem as a low-rank {\it block-term tensor} \cite{de2008decompositions2} completion problem. This way, recoverability guarantees of the radio map from sparse sensor measurements can be established.
	   
	   Structural priors/constraints-assisted inverse problem formulations are useful, but they are often inadequate in dealing with complex scenarios. For example, the low-rank assumption on $\S_r$ is plausible in free space or in rural areas where electromagnetic waveforms could propagate without encountering many barriers. However, the assumption becomes unrealistic when urban or indoor regions are considered. The same challenge is also shared by methods that use spatial smoothness; see, e.g., \cite{bazerque2011group, yucek2009survey}. Fig. \ref{fig:low-rank} shows how heavy shadowing (caused by barriers like buildings and walls) makes the low-rank and spatial smoothness assumptions on $\S_r$ less accurate. In this example, the first 10 principal components of a $50\times 50$ SLF with little shadowing contain more than 90\% energy of the SLF. However, the same number of principal components only contain less than 70\% energy of the SLF when the shadowing effect is severe.
	   
	   	\begin{figure}[t!]
				\centering
				\includegraphics[width=
				\linewidth]{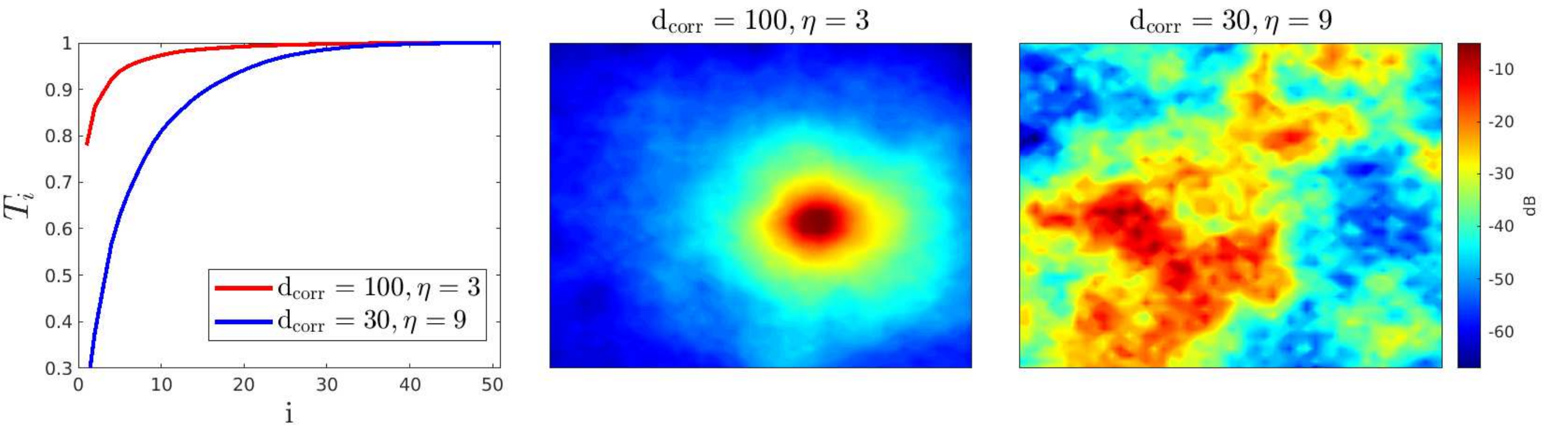}
				\caption{Shadowing effect affects the soundness of the low-rank assumption on the $50\times 50$ SLFs. 
				(Left) $T_i$ for all $i\in[50]$ where $T_i=\nicefrac{\sum_{k=1}^i \sigma_k}{ \sum_{k=1}^I \sigma_k  }$ and $\sigma_i$ is the $i$th singular value of the SLF. 
				(Middle) An SLF with little shadowing effect. (Right) An SLF with heavy shadowing effect.
				The SLFs are generated following the model in \cite{goldsmith2005wireless}; ${\rm d}_{\rm corr}$ is the decorrelation distance   and $\eta$ is the shadowing variance (see Sec.~\ref{sec:exp})}.
				\label{fig:low-rank}
			\end{figure}
	   
	   Very recently, a couple of data-driven methods were proposed to circumvent the model mismatch challenge. The idea is to learn the complex data generative process (from partial observations) using DNNs through a large number of realistic simulated data samples, and then use the learned neural model to assist radio map completion \cite{teganya2020data, han2020power}. For example, the work in \cite{teganya2020data} generates a training set, denoted by $\{\tX_n\}_{n=1}^N$, where each $\tX_n\in\mathbb{R}^{I\times J\times K}$ is a radio map tensor following a spatial propagation model \cite{goldsmith2005wireless}. Then, a deep completion network is trained using the following:
	   \begin{equation}\label{eq:deepform}
	     \widehat{\btheta}\leftarrow\arg \min_{\bm \theta}~\sum_{n=1}^N\left\|  \bm f_{\bm \theta}\left(\underline{\bm M}_n \circledast \tX_n  \right) -\tX_n \right\|_{\rm F}^2,
	   \end{equation}
	   
	   where $\bm f_{\bm \theta}(\cdot):\mathbb{R}^{I\times J \times K}\rightarrow \mathbb{R}^{I\times J \times K}$ is a DNN whose network weights are collected in $\bm \theta$; $\underline{\bm M}_n\in\mathbb{R}^{I\times J\times K}$ is a mask tensor (or, a sampler) with $\underline{\bm M}_n(i,j,k)=1$ if the entry in the radio map is sensed, and $\underline{\bm M}_n(i,j,k)=0$ otherwise\footnote{In practice, the idea in \eqref{eq:deepform} is often carried out with different variations, e.g., slab by slab realization in \cite{teganya2020data} and \cite{han2020power}. The mask information is also fed to the DNNs together with the sensed data; see details in \cite{teganya2020data}.}.
	   After a network $\bm f_{\widehat{\bm \theta}}$ is learned (via any off-the-shelf DNN training algorithms, e.g., \cite{kingma2015adam}), a radio map can be simply estimated via $\widehat{\tX} = \bm f_{\widehat{\bm \theta}}\left( \underline{\bm M}_{\rm sens.}\circledast \underline{\bm X}_\natural \right)$, where  $\tX_\natural$ is the ground-truth radio map to estimate, $\underline{\bm M}_{\rm sens.}$ is the mask tensor that reflects the sensor locations, and $ \underline{\bm M}_{\rm sens.}\circledast \underline{\bm X}_\natural $ represents the sensed measurements (i.e., tensor fibers).
	   Specifically, we have $\tM_{\rm sens.}(i,j,k)=1,~\forall k,~\text{if}~(i,j)\in \bm \varOmega$, and $\tM_{\rm sens.}(i,j,k)=0$ otherwise.
	   The major idea of these methods is summarized in Fig.~\ref{fig:deepcomp_schematic}.

        \begin{figure}[t]
            \centering
            \includegraphics[width=0.8\linewidth]{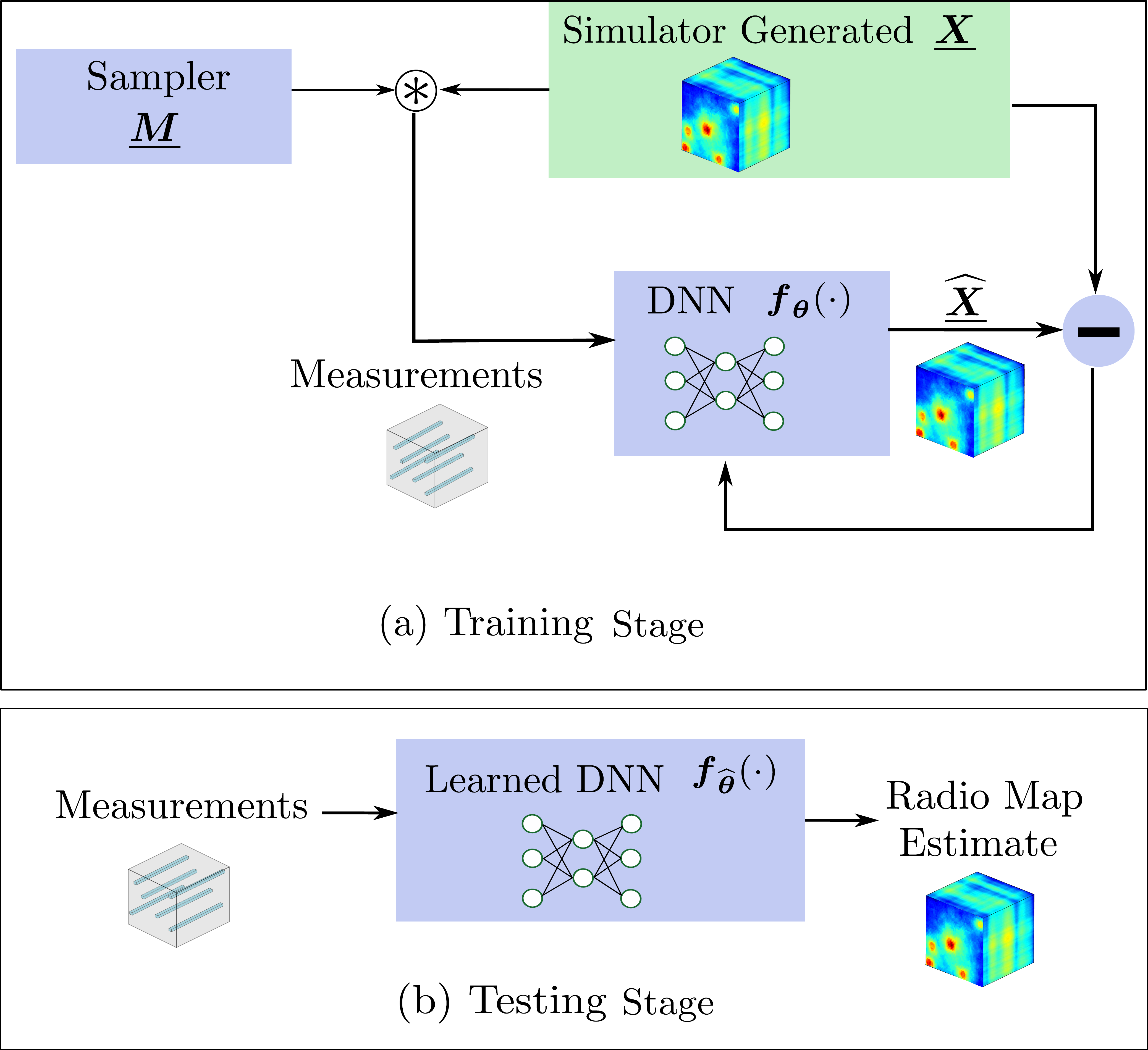}
            \caption{Schematic of deep learning based SC methods in existing works \cite{teganya2020data,han2020power}.}
            \label{fig:deepcomp_schematic}
        \end{figure}

	   \subsection{Challenges of Deep Learning-Based SC}
	   The most appealing feature of the DL-based SC approaches is that the network $\bm f_{\widehat{\bm \theta}}$ could potentially complete complex radio maps (e.g., those with heavy shadowing) that handcrafted priors may struggle to model. However, a couple of challenges remain.
	   
	   \noindent $\bullet$ {\bf Training Challenge.}
	   One premise for learning a high-quality  $\bm f_{\widehat{\bm \theta}}$ is that the training samples $\{\tX_n\}_{n=1}^N$ should ``cover'' most representative radio maps under the scenario of interest; otherwise, the network could not make accurate inferences, if $ \underline{\bm X}_\natural$ is not close to any sample in the training set. In SC, the state space  contains} all possible $\tX_n$'s, which may be too large. The reason is that there are too many combinations of key components, e.g., the number of emitters, their locations, transmission power levels, just to name a few. 
	   For example, if $R\leq 5$ and $I=J=100$, the number of different emitter locations exceeds $8\times 10^{17}$---without considering any other key parameters of the radio map.
	   As a consequence, one may need a prohibitively large $N$ to attain reasonable SC performance---making the training stage costly. 
	   
	     \noindent $\bullet$ {\bf Generalization Challenge.}
	   In addition, ``out-of-training-distribution'' test data can appear quite often.
	   For instance, when $R'$ emitters emerge in the region of area, but the network is trained with at most $R$ emitters, where $R<R'$. This situation is particularly difficult for the off-line trained system to adapt because re-training can take a long time. 
	   However, the system designers have no control of the number of emitters, which means $R<R'$ may always happen. One may use a very large $R$ in the training stage to circumvent this situation, but this may make the already hard and costly training problem even more challenging.

	\section{Proposed Approach}
	In this work, we propose an alternative paradigm for deep learning-assisted SC. In a nutshell, instead of using a purely data-driven method,
	we utilize the aggregation part in \eqref{eq:sigmodel}, as this part is intuitive and relatively reliable. In addition,	we use deep networks to help learn the hard to model part in \eqref{eq:sigmodel}, i.e., the SLFs of individual emitters. As one will see, this model-assisted data-driven approach can effectively circumvent the training and generalization challenges encountered in pure DL methods, while keeping the expressive power of deep networks for accurately representing radio maps.

		\subsection{Training DNN for Completing Individual SLFs}
		Recall that the radio map can be modeled as $\tX=\sum_{r=1}^R\S_r\circ \c_r$, i.e., aggregated power intensity over space and frequency from $R$ emitters. Due to the difficulty of modeling the $\S_r$'s using handcrafted priors, the performance of model-based methods often degrades in challenging cases, e.g., when heavy shadowing happens. 
		Nonetheless, the model that a radio map is a superposition of individual radio maps associated with the emitters is still accurately reflecting the underlying physics.
		Hence, instead of learning a DNN to complete the entire aggregated radio map tensor $\tX$, our idea is to learn a network for completing the individual $\S_r$'s. If the $\c_r$'s can be somehow estimated, then one can assemble the estimated SLFs and PSDs to recover the entire radio map using \eqref{eq:sigmodel}.
		
		Based on the discussion above, we propose the following process to learn a completion network for individual SLFs:
			\begin{equation}\label{eq:slf_training}
			     \widehat{\btheta}\leftarrow\arg \min_{\bm \theta}~\frac{1}{N}\sum_{n=1}^N\left\| \bm f_{\bm \theta}\left( \bm M_n\circledast \Q_n   \right) - \Q_n \right\|_{\rm F}^2,
			\end{equation}
			where $n$ is the training sample index, $\M_n\in\mathbb{R}^{I\times J}$ is the $n$th randomly generated 2D mask,  $\Q_n\in\mathbb{R}^{I\times J}$ is a complete SLF of a {\it single} emitter, and $\bm f_{\bm \theta}(\cdot):\mathbb{R}^{I\times J}\rightarrow \mathbb{R}^{I\times J}$ is the SLF completion network to be learned.

            After $\bm f_{\widehat{\bm \theta}}$ is learned, we complete the SLFs by letting 
            \begin{equation}\label{eq:individualpredict}
                        \widehat{\S}_r  \leftarrow \f_{\widehat{\btheta}} \left(\M_{\rm sens.}\circledast {\S}_r \right),
            \end{equation}
            where $\M_{\rm sens.}$ is the 2D version of sensing mask, in which $\M_{\rm sens.}(i,j)=1$ if $(i,j)\in \bm \varOmega$ and $\M_{\rm sens.}(i,j)=0$ otherwise. The notation $\M_{\rm sens.}\circledast {\S}_r$ represents the incomplete individual SLF of emitter $r$---and how to obtain it will be discussed in the next subsection.

        \begin{remark}\label{rmk:statespace}
          Intuitively, the state space where the individual SLF resides is much smaller than that of the aggregated radio map $\tX$. Using the same example where $100\times 100$ grids are considered, one can see that there are only $10^4$ emitter locations if we only model a single SLF---as opposed to more than $8\times 10^{17}$ possible emitter location combinations if we consider up to 5 emitters in the same region. Such a substantial reduction of space size may make the training process more efficient and effective. In addition, if $R$ can be properly estimated (using any model-order selection methods in signal processing, e.g., those in \cite{stoica2005spectral}), then the ``out-of-distribution'' and emitter misdetection problem encountered in existing DL-based SC approaches can be easily circumvented.
        \end{remark}

			 In this work, we employ an autoencoder structure for $\bm f_{\bm \theta}$. Nonetheless, in principle, any generative model such as \textit{generative adversarial network} (GAN) \cite{goodfellow2014generative} and \textit{variational autoencoder} (VAE) \cite{kingma2013auto} can be readily employed into our framework. The autoencoder consists of two major parts, i.e., the encoder and the decoder, respectively. To be more precise, 
			 for an ideal anutoencoder that is applied to high-dimensional data $\Q\in\mathbb{R}^{I\times J}$, we wish the following holds:
			 \begin{equation}\label{eq:autoencodermodel}
			     \bm f_{\bm \theta}( \bm Q ) =  \bm g_{\bm \theta_{\rm d}} \left( \bm p_{\bm \theta_{\rm e}}(\bm Q)\right),
			 \end{equation}
			 in which $\bm \theta_{\rm e}$ and $\bm \theta_{\bm d}$ represents the network parameters of the encoder and decoder, respectively, $\bm \theta =[ \bm \theta_{\rm e}^\T, \bm \theta_{\rm d}^\T]^\T$, and 
			 \begin{equation}\label{eq:pandg}
			    \bm p_{\bm \theta_{\rm e}}(\cdot):\mathbb{R}^{I\times J}\rightarrow \mathbb{R}^D,~ 	    \bm g_{\bm \theta_{\rm d}}(\cdot):\mathbb{R}^{D}\rightarrow \mathbb{R}^{I\times J}
			 \end{equation}
			 denote the encoder and decoder networks, respectively. Notably, if $D\ll IJ$, the model in \eqref{eq:autoencodermodel} means that the data matrix  $\bm Q \in\mathbb{R}^{I\times J}$ can be represented in a low-dimensional latent domain, using its ``embedding'' $\bm z= \bm p_{\bm \theta_{\rm e}}(\bm Q) $. The decoder can be regarded as a {\it generator} that maps a latent embedding $\bm z$ to the observation domain, i.e., to ``generate'' the ambient data.
			 Hence, the deep autoencoders are also considered as deep generative model learners.

        \begin{figure}[t]
        \centering
           \includegraphics[width=8.5cm]{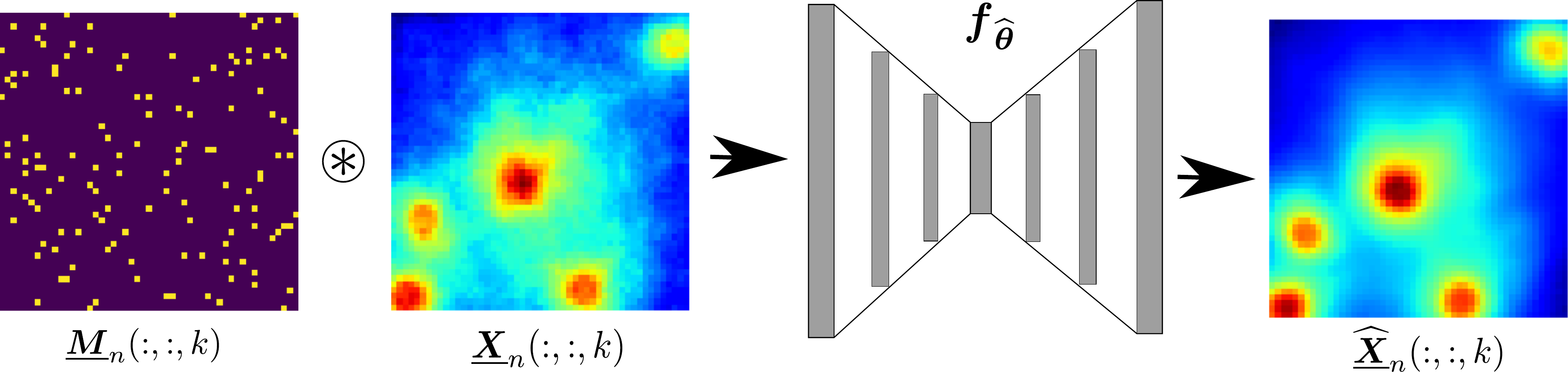}
        \includegraphics[width=8.5cm]{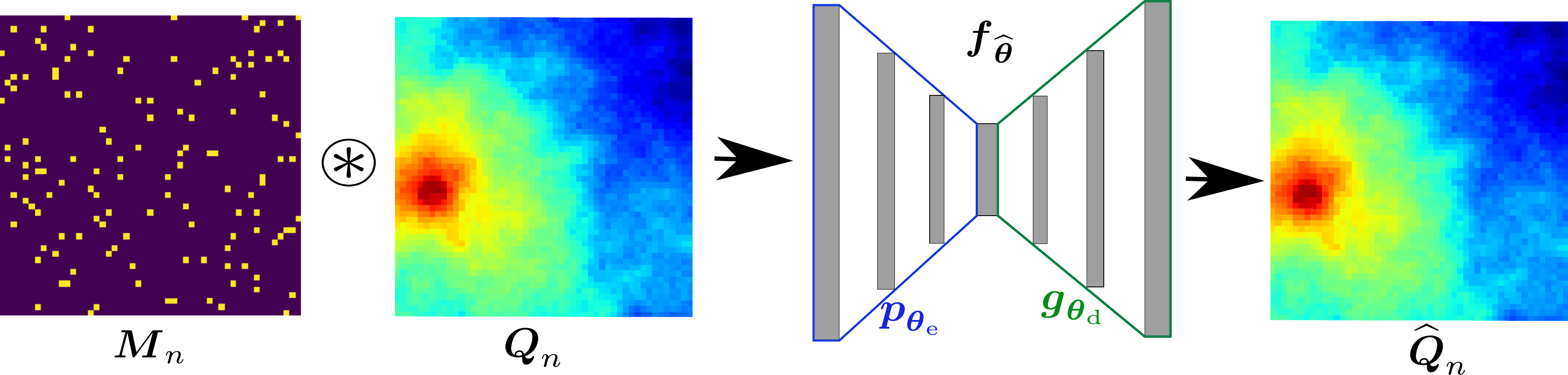}
        \caption{Training procedures of existing DL-based SC methods (e.g., \cite{teganya2020data}) that learn DNNs to complete the aggregated $\tX$ (top) and the proposed method (bottom) that uses DL to complete individual SLFs.} 
        \label{fig:compare}
    \end{figure}
	
		\subsection{ Emitter Disaggregation-Assisted Deep Completion}
		{ 
		The proposed individual SLF completion method could potentially reduce training costs and improve generalization. However, the inputs to such networks, i.e., $\M_{\rm sens.}\circledast{\S}_r$ for all $r$, are not observed.
		Instead, if noise is absent, we observe the following incomplete tensor:
	\begin{equation}\label{eq:sensing_noiseless}
	    \tY =  \underline{\bm M}_{\rm sens.} \circledast \tX_\natural,  
	\end{equation}  
		that is acquired by the sensors, where $\tX_\natural=\sum_{r=1}^R \S_r \circ \c_r$ is the target radio map that we aim to recover. 
		If the $\bm c_r$'s can be somehow estimated,
		our objective boils down to estimating the incomplete individual SLFs, i.e., $\M_{\rm sens.}\circledast{\S}_r$ from the observed measurements $\tY$.

		To this end, notice that the following relationship holds:
		\begin{equation}
		    \Y = {\rm unfold}(\tY) = {\rm unfold}\left(  \underline{\bm M}_{\rm sens.} \circledast \tX_\natural  \right); 
		\end{equation}
		where $\X_\natural = {\rm unfold}(\tX_\natural)\in\mathbb{R}^{K\times IJ}$ and the tensor unfolding operator is defined as $\X_\natural(:,q) = \tX_\natural(i,j,:),\quad q=I(i-1) +j$. 
		
		Clearly, if we define $\G := \Y(:,\bm \varOmega_{\rm col})$, where $\bm \varOmega_{\rm col}\subseteq [IJ]$ such that $q = I(i-1)+j \in \bm \varOmega_{\rm col},\quad \forall (i,j)\in \bm \varOmega$, then, we have the following expression for $\bm G$: 
		\begin{equation}\label{eq:factorization}
		    \G = \C\underbrace{\S(:,\bm \varOmega_{\rm col})}_{\bm H} = \bm C\bm H,
		\end{equation}
		where $\C = [\c_1, \ldots, \c_R] \in \bbR^{K \times R}$ and $\S = [{\rm vec}(\S_1), \ldots, {\rm vec}(\S_R)]^\T \in \bbR^{R \times IJ}$.

		If one could estimate $\H$, then define a matrix ${\S}_{\rm miss}\in\mathbb{R}^{R\times IJ}$ such that ${\S}_{\rm miss}(r,\bm \varOmega_{\rm col}):=\bm H(r,:)$. Further, let ${\S}_{\rm miss}(:,\bm \varOmega_{\rm col}^{\rm c})=\bm 0$, where $\bm \varOmega_{\rm col}^{\rm c}$ is the complement of $\bm \varOmega_{\rm col}$. As a consequence, we have
		\begin{equation}\label{eq:Pr}
		     \bm P_r = {\rm mat}({\S_{\rm miss}}(r,:)) = \M_{\rm sens.}\circledast \S_r,    
		\end{equation} 
		where the matricization operator ${\rm mat}(\cdot):\mathbb{R}^{IJ} \rightarrow \mathbb{R}^{I\times J}$ is the inverse operation of vectorization. Note that $\bm P_r$ is the incomplete individual radio map that we wish to feed to $\bm f_{\widehat{\bm \theta}}(\cdot)$ in \eqref{eq:individualpredict}.
		The procedure for creating the matrix $\bm G$ is illustrated in Fig. \ref{fig:factorization} (a).

		Note that $\C$ and $\H$ are are both nonnegative matrices, per their physical meanings. Hence, our task amounts to estimating $\bm C$ and $\bm H$ from $\bm G$---which is an NMF problem \cite{fun2019nonnegative}. Then, we take the estimated $\H$'s rows and apply deep network-based completion to estimate $\S_r$ for all $r$. With $\C$ and $\S_r$'s estimated, we can recover $\tX_\natural$; see the overall procedure in Fig.~\ref{fig:factorization}.
		
		\begin{figure*}[t!]
			\centering
			\includegraphics[width=\linewidth]{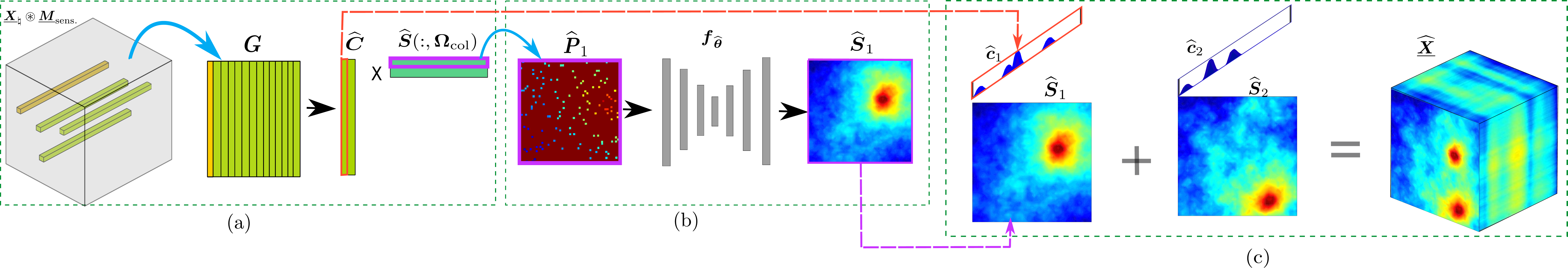}
			\caption{ Illustration for \texttt{Nasdac} algorithm. (a) Emitter Disaggregation model in \eqref{eq:factorization}. $\widehat{\C}$ and $\widehat{\S}(:,\bOmega_{\rm col})$ are estimated by sparsity based NMF algorithm. (b) Deep Completion of emitter SLFs. $r$-th row of $\widehat{\S}(:, \bOmega_{\rm col})$ is matricised and completed with completion network $\f_{\widehat{\btheta}}$ to obtain the estimate of completed $r$-th SLF, $\widehat{\S}_r$. (c) Estimated $\widehat{\C}$ and $\widehat{\S}$ are aggregated to recover the estimate $\widehat{\tX}$.}
			\label{fig:factorization}
		\end{figure*}

        Note that NMF is not always unique---i.e., estimating $\bm C$ and $\bm H$ from $\bm G$ may not be always possible; see discussions about NMF identifiability in the literature, e.g., \cite{fu2018identifiability,fun2019nonnegative,gillis2020nonnegative}. Nonetheless, if the emitters transmit sparsely across the spectral band, estimating $\bm C$ and $\bm H$ via NMF is viable. To see this, we make the following assumption:
		\begin{assumption}\label{ass:dominance}
		There exists a frequency $f_r$ for each emitter $r\in[R]$ such that $\C(f_r,r)>0$ and $\C(f_r,k)=0$ for $k\neq r$. 
		\end{assumption}
		Assumption~\ref{ass:dominance} is considered reasonable since emitters often take different carrier frequencies (or subbands in OFDM systems) \cite{fu2016power}. In addition, if the emitters use the $K$ frequencies in a sparse manner, such an assumption is often not hard to meet.
		Under Assumption~\ref{ass:dominance}, we show that:
		\begin{theorem}\label{thm:nmf}
				Under \eqref{eq:sigmodel} and the construction of $\G$  in \eqref{eq:factorization}, assume that the entries of $\S\in\mathbb{R}^{R\times IJ}$'s are drawn from any joint absolutely continuous distribution. Then, with probability one, there exists a polynomial time algorithm (e.g., the {\it successive projection algorithm} (SPA) \cite{gillis2013fast,fu2014self}) that estimates 
				$  \widehat{\bm C}=\bm C\bm \varPi\bm \varLambda,~\widehat{\bm H}= \bm \varLambda^{-1} \bm \varPi^\T \bm H$($\bm \varPi$ and $\bm \varLambda$ being permutation and full-rank diagonal scaling matrices, respectively),
				given that the number of sensors satisfies $|\bm \varOmega|\geq R$. 
			\end{theorem}
		}
			{
			The proof is relegated to Appendix~\ref{app:thmnmf}.
			Note that Assumption~\ref{ass:dominance} translates to the {\it separability} condition in the context of NMF, and the proof of Theorem~\ref{thm:nmf} is a direct application of identifiability of separable NMF  \cite{fun2019nonnegative, gillis2020nonnegative}. Note that under separability condition, NMF is solvable using either greedy algorithms \cite{gillis2013fast,fu2014self} or convex optimization \cite{gillis2014robust}, both of which are  polynomial-time algorithms. In this work, we employ the Gram-Schmidt-like {\it successive projection algorithm} (SPA) algorithm in \cite{gillis2013fast,fu2014self} due to its efficiency; see \cite{fun2019nonnegative,gillis2013fast,fu2014self} and Appendix~\ref{app:spa} for details of the SPA algorithm.

            With the learned $\bm f_{\widehat{\bm \theta}}$ and the estimated $\widehat{\bm P}_r$ [cf. Eq.~\eqref{eq:Pr}] from the NMF stage, we complete the SLFs by letting $\widehat{\S}_r  \leftarrow \f_{\widehat{\btheta}} \left(\widehat{\bm P}_r \right)$. The two-stage process is summarized in Algorithm~\ref{algo:nmf}, which we will refer to as the {\it \uline{n}onnegative matrix factorization \uline{as}sisted \uline{d}eep emitter sp\uline{a}tial loss \uline{c}ompletion} (\texttt{Nasdac}). One remark is that although all NMF algorithms admit inevitable scaling and permutation ambiguities (i.e., the $\bm \varLambda$ and $\bm \varPi$ matrices in Theorem~\ref{thm:nmf} cannot be removed), these ambiguities do not affect the reconstruction, if $ \widehat{\bm C}=\bm C\bm \varPi\bm \varLambda,~\widehat{\bm H}=\bm \varLambda^{-1}\bm \varPi^\T\bm H$} hold---since the ambiguity matrices cancel each other in the final assembling stage (i.e., line \ref{line:assemble} in Algorithm~\ref{algo:nmf}).

			\begin{algorithm}[t]\label{algo:nmf}
			\footnotesize
				\SetAlgoLined
				\KwData{${\tX}, \bOmega, R$, $\bm f_{\bm \theta}(\cdot)$.} 
				\KwResult{$\widehat{\tX}$, $\widehat{\S}$ and $\widehat{\C}$.}
				    \tcp{Stage 1: Emitter Disaggregation}
        				
					$\widehat{\C}, \widehat{\H}  \leftarrow \texttt{NMF\_SPA}(\G, R)$ (see Appendix~\ref{app:spa});\\
					\tcp{Stage 2: SLF completion}
					\For{$r=1:R$}{
					    $\widetilde{\S}(r, \bOmega_{\rm col}) \leftarrow \widehat{\H}(r,:)$; \\
					    $\widetilde{\S}(r, \bOmega_{\rm col}^{\rm c}) \leftarrow \zero$; \\
					    $\widehat{\S}_r \leftarrow \f_{\btheta} ({\rm mat}(\widetilde{\S}(r,:))$;
					    }
					 $\widehat{\tX} \leftarrow \sum_{r=1}^R\widehat{\S}_r\circ \widehat{\c}_r$;\label{line:assemble} \\
					return $\widehat{\tX}$, $\widehat{\S}$ and $\widehat{\C}$.
				\caption{\texttt{Nasdac}}
			\end{algorithm}

       		 Note that in \texttt{Nasdac}, the number of emitters $R$ is assumed to be previously estimated or known. Under the considered scenario, estimating $R$ corresponds to estimating the rank of the NMF model $\bm G\approx \C\H$ [cf. Eq.~\eqref{eq:nmf_factor_model}]. This can be accomplished by existing model-order selection methods (see \cite{stoica2004model} for an overview). Many methods for rank estimation under noisy NMF models can also be used, e.g., \cite{fu2014self,fu2015robust,bioucas2008hyperspectral,ambikapathi2012hyperspectral}.

		\section{Performance-Enhanced Approach}\label{sec:performance_enhancement}
		{
		Once the network $\bm f_{\widehat{\bm \theta}}$ is learned off-line,
		the \texttt{Nasdac} method offers a computationally simple way to complete the radio map tensors. Nonetheless, the method was proposed under the premise that $\C$ satisfies Assumption~\ref{ass:dominance}, or roughly speaking, the emitters use the frequency bins sparsely. 
		In this subsection, we take a step forward and address more challenging scenarios where Assumption~\ref{ass:dominance} is violated. We also consider the case where the sensor measurements are noisy, and we will offer stability guarantees.
		
		To proceed, we denote
		\begin{equation}\label{eq:noisy}
		    \tY(i,j,:) =\tX_\natural(i,j,:) + \tN(i,j,:),\quad \forall (i,j),
		\end{equation}
		where $\tN$ is a noise tensor and $\tX_\natural$ is the ground-truth tensor that we aim to recover.
		With $\tY(i,j,:)$'s for $(i,j)\in\bm \varOmega$ collected by the sensors, our formulation to estimate $\tX_\natural$ is as follows:
		\begin{equation}\label{eq:joint_opt}
		\begin{aligned}
		    \min_{\{\c_r, \bm z_r\}_{r\in [R]}}  \left\| \tM_{\rm sens.} \circledast \left(\tY - \sum_{r=1}^{R} \bm g_{\bm \theta_{\rm d}}(\bm z_r)\circ \c_r \right)\right\|_{\rm F}^2.
		\end{aligned}
		\end{equation}
		In the above, the neural network $\bm g_{\bm \theta_{\rm d}}({\bm z}_r)$ is a generative network that links a latent vector  $\bm z_r\in\mathbb{R}^D$} with the SLF $\S_r\in\mathbb{R}^{I\times J}$---see Eq.~\eqref{eq:pandg} and the discussions there. { Fig.~\ref{fig:dowjons_illus} illustrates the idea of modeling the radio map used in \eqref{eq:joint_opt}.

		Denote $	\{\widehat{\c}_r,\widehat{\z}_r\}_{r=1}^R $ as the solution obtained via solving \eqref{eq:joint_opt}. We reconstruct the radio map tensor by $\widehat{\tX}\leftarrow \sum_{r=1}^{R} \bm g_{\bm \theta_{\rm d}}(\widehat{\bm z}_r)\circ \widehat{\c}_r$.
		The idea is similar to low-rank matrix factorization based data completion, and in particular the learned generator $\bm g_{\bm \theta_{\rm d}}(\bm z_r)$ is used to represent the SLF of emitter $r$. This representation can be understood as imposing realistic data-driven constraints to the learned SLFs. Using the above, one does not need to resort to the two-stage approach as in \texttt{Nasdac}, which is prone to error propagation. More importantly, we will show that the criterion in \eqref{eq:joint_opt} does not need Assumption~\ref{ass:dominance} to guarantee recoverability of $\tX_\natural$, and that the criterion is robust to noise.

		\begin{figure}[t]
            \centering
            \includegraphics[width=\linewidth]{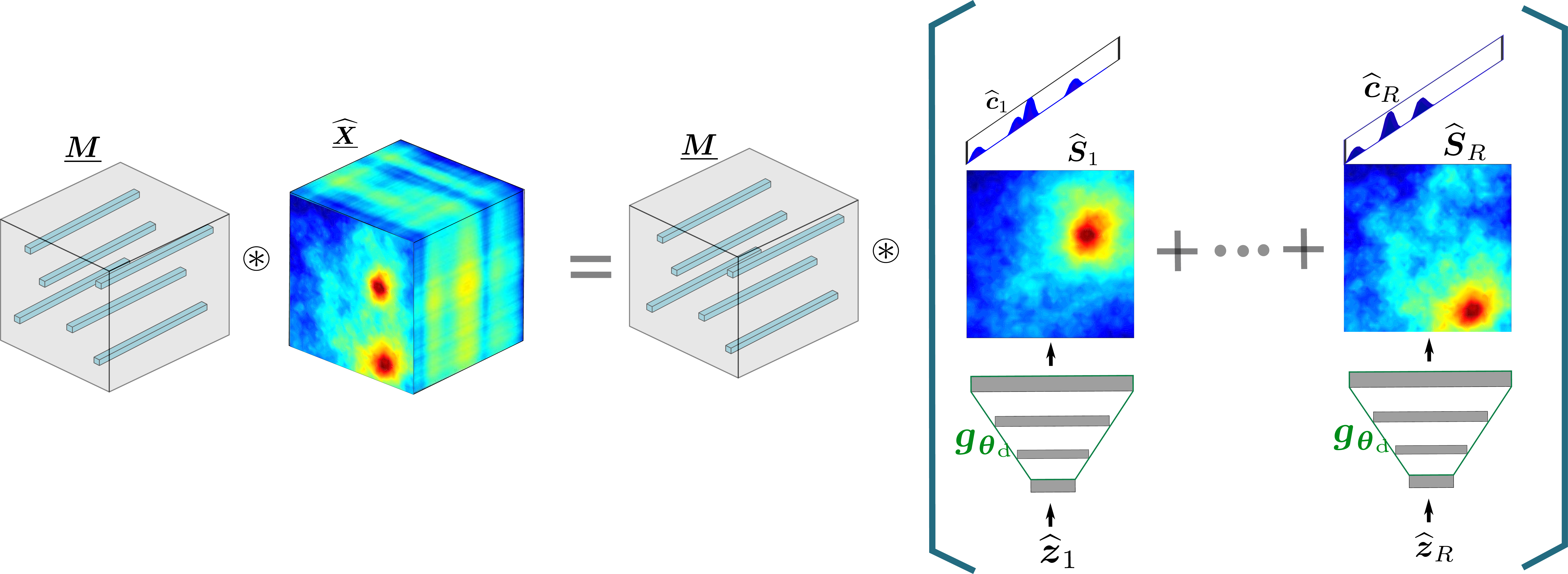}
            \caption{ Illustration of the modelling idea used for the performance enhanced approach (\texttt{DowJons}).} 
            \label{fig:dowjons_illus}
        \end{figure}

		\subsection{Recoverability Analysis}
	    In this subsection, we analyze the recoverability of the ground-truth radio map tensor $\tX_\natural$ under the criterion in \eqref{eq:joint_opt}. In our analysis, we treat $\tX_\natural$ as a deterministic term, while the set of sampling locations specified by $\bOmega$ are random. To proceed, first, consider the following assumption on the employed neural network in our framework: 
		\begin{assumption}[Deep Generator Structure] \label{ass:existence}
		Let ${\cal Z} = \{\z \in \bbR^D: \|\z\|_2 \leq q\}$ and  $\g_{\bm \theta_{d}}(\cdot): {\cal Z} \to \bbR^{I\times J}$ be a deep generative model with the network structure $\g_{\bm \theta_{d}}(\z) = {\rm mat} \left( \bm \zeta_L(\A_L (\ldots \bm \zeta_1(\A_1\z))\right)$, where $\A_\ell\in\mathbb{R}^{D_\ell\times D_{\ell-1}}$ is the network weight in the $\ell$th layer, $
		\ell=1,\ldots,L$, in which $D_0=D$ and $D_L= IJ$,  $\bm \zeta_i(\cdot)=[\zeta(\cdot),\ldots,\zeta(\cdot)]^\T:\mathbb{R}^{D_{\ell}} \to \mathbb{R}^{D_{\ell}}$ is a $\phi_i$-Lipschitz function, $\zeta(\cdot):\mathbb{R}\rightarrow \mathbb{R}$ is an entry-wise activation function (e.g., \texttt{ReLU} or \texttt{Sigmoid}) \cite{goodfellow2016deep}
		and $P= \prod_{i=1}^{L} \phi_i \|\A_i\|_2<\infty$.
		\end{assumption}
		Note that many neural network structures (e.g., the fully connected network (FCN) and the convolutional neural network (CNN)) satisfy the structure specified in Assumption~\ref{ass:existence}. Popular activation functions such as \texttt{Sigmoid} and \texttt{ReLU} are also Lipstchitz continuous.
		We will use the following definitions:
		\begin{definition}[Covering Number \cite{zhou2002covering}] 
				The covering number of a set ${\cal W}$ with parameter $\varepsilon>0$, denoted as ${\sf N}({\cal W,\varepsilon})$, is defined as the smallest number of $\varepsilon$-radius norm balls required to completely cover the set ${\cal W}$ (with possible overlaps). In other words, ${\sf N}({\cal W,\varepsilon}) = |\overline{\cal W}|$,	where $\overline{\cal W}$ is discrete, and for any $\bm w \in {\cal W}$, there exists a $ \overline{\bm w} \in \overline{\cal W}\subseteq {\cal W}$ such that $\|\overline{\bm w} - \bm w\| \leq \varepsilon$. The discrete set $\overline{\cal W}\subseteq{\cal W}$ is called an $\varepsilon$-net of ${\cal W}$ .
		\end{definition}
		In principle, any norm can be used to define an $\varepsilon$-net. We use the Euclidean norm in this work.
		The covering number is a way of measuring the ``complexity'' of a continuous set using its discretized version.
		
		\begin{definition}[Solution Set]\label{def:slnset}
			We define ${\cal X}_{R, \g_{\bm \theta_{d}}}\subset \mathbb{R}^{I\times J\times K}$ as the set that contains all possible $\widetilde{\tX}$ that could be expressed as $\widetilde{\tX} = \sum_{r=1}^R\widetilde{\S}_r\circ \widetilde{\c}_r$ with $\widetilde{\S}_r = \g_{\bm \theta_{d}}(\z_r)$,   $\z_r \in {\cal Z}$ (where ${\cal Z}$ and $ \g_{\bm \theta_{d}}$ are specified in Assumption~\ref{ass:existence}), $\|\bm g_{\bm \theta_{\rm d}}(\bm z_r)\|_{\rm F}\leq \beta$, and $\| \widetilde{\bm c}_r\|_2\leq \alpha$ for $r=1,\ldots,R$.
		\end{definition}
		
		\begin{definition}
				We define ${\sf Gap}(\widetilde{\tX},\bm \varOmega) =  \sqrt{	\widehat{\sf Loss}(\widetilde{\tX}) } - \sqrt{{\sf Loss}(\widetilde{\tX})}$, where $\widetilde{\tX}\in {\cal X}_{R, \g_{\bm \theta_{d}}}$ and 
			\begin{align*}
					\widehat{\sf Loss}(\widetilde{\tX}) = & \frac{1}{|\bOmega|K} \sum_{i,j \in \bOmega} \left\|\tY(i,j,:) - \widetilde{\tX}(i,j,:) \right\|_2^2, \\
					{\sf Loss}(\widetilde{\tX}) = & \frac{1}{IJK} \sum_{i,j \in [I] \times [J]} \left\|\tY(i,j,:) - \widetilde{\tX}(i,j,:) \right\|_2^2.
				\end{align*}
		\end{definition}
		Note that $\widehat{\sf Loss}$ can be understood as the ``empirical loss''  measured at a solution $\widetilde{\tX}\in {\cal X}_{R, \g_{\bm \theta_{d}}}$, and ${\sf Loss}$ is the ``true'' loss measured over the complete data (or, the generalization error). Under the above definitions, we first show the following Lemma:

			\begin{lemma}\label{lem:uc}
				Define ${\sf Gap}^\star(\bm \varOmega) = \sup_{\widetilde{\tX}\in {\cal X}_{R,\bm g_{\bm \theta_{\rm d}}}} | {\sf Gap}(\widetilde{\tX},\bm \varOmega) | $.
				Assume that Assumption~\ref{ass:existence} holds.
				For any $\widetilde{\tX}\in {\cal X}_{\bm g_{R,\bm \theta_{\rm d}}}$, under the sensing model in \eqref{eq:noisy},
				it holds with probability of at least $1 - \delta$ that
				\begin{align}
						   {\sf Gap}^\star(\bm \varOmega) \leq \frac{2 cR}{\sqrt{|\bOmega|}} + \label{eq:Gap}
					 \left( \frac{\xi^2 \omega}{2} \log\left(\frac{2{\sf N} \left(  {\cal X}_{\bm g_{R,\bm \theta_{\rm d}}},cR \right)}{\delta} \right)\right)^{\frac{1}{4}}, 
				\end{align}
				where $c>0$ is a positive scalar, 
				$\omega = (\frac{1}{|\bOmega|} - \frac{1}{IJ} + \frac{1}{IJ |\bOmega|})$,
				and $\xi = \frac{1}{K}(\sqrt{K}(\upsilon + \nu) + R\alpha\beta)^2$ in which $\upsilon = \max_{i,j,k} |\tX_\natural(i,j,k)|,~\nu = \max_{i,j,k} |\tN(i,j,k)|.$
			\end{lemma}
			
			We also show the following:
			\begin{lemma}	\label{lem:cover}
			For the set in Definition~\ref{def:slnset}, the covering number of its $\varepsilon$-net is
				\begin{equation}\label{eq:Nx}
					{\sf N} \left(  {\cal X}_{\bm g_{R,\bm \theta_{\rm d}}},\varepsilon \right) \leq \left(\frac{3R(\alpha + \beta)}{\varepsilon}\right)^{R(K+D)} \alpha^{RK} (Pq)^{RD}.
				\end{equation}
			\end{lemma}
		 Using the above lemmas and definitions, we present our main recoverability theorem:
			\begin{theorem}[Recoverability]
				\label{thm:rmse}
				Under the observation model in \eqref{eq:noisy},
				assume that $\tX^{\star}=\sum_{r=1}^R \bm g_{\bm \theta_{\rm d}}(\z_r^\star) \circ \bm c_r^\star$, where $\z_r^\star$ and $\c^\star$ are from any optimal solution of \eqref{eq:joint_opt}. Also assume that the network structure and $\bm z_r^\star$ and $\bm c_r^\star$ satisfy the specifications in Assumption~\ref{ass:existence} and the solution set in Definition~\ref{def:slnset}. Then, the following holds with probability of at least $1-\delta$:
				\begin{align*}
					& \frac{1}{\sqrt{IJK}} \|\tX^{\star} - \tX_{\natural}\|_{\rm F} \leq \frac{1}{\sqrt{IJK}} \|\tN\|_{\rm F} + {\sf Gap}^\star({\bm \varOmega}) \\
					& + \frac{1}{\sqrt{|\bOmega|K}} \left( \|\tM_{\rm sens.} \circledast \tN\|_{\rm F}+ {\sf Err}_{\rm rep}({\cal X}_{{R,\bm g_{\bm \theta_{\rm d}}}}) \right) ,
				\end{align*}
				where $
				    {\sf Err}_{\rm rep}({{\cal X}_{R,\bm g_{\bm \theta_{\rm d}}}}) = \| \widetilde{\tX}^\ast - \tX_{\natural}  \|_{\rm F}$, 
				    in which 
				    $
				        	    \widetilde{\tX}^\ast  = \argmin_{\widetilde{\tX} \in {{\cal X}_{R,\bm g_{\bm \theta_{\rm d}}}}} \|\widetilde{\tX} - \tX_\natural\|_{\rm F}^2, 
				    $
			and ${\sf Gap}^\star({\bm \varOmega})$ is upper bounded by \eqref{eq:Gap} with the covering number in \eqref{eq:Nx}.
			\end{theorem}

            Theorem~\ref{thm:rmse} asserts that the criterion in \eqref{eq:joint_opt} admits recoverability of the ground-truth radio map $\tX_\natural$---even if there exists sensing noise and the learned generative model is not perfect. The recovery error consists of three main sources, namely, the sensing noise $\tN$, sampling-induced noise (or, generalization error) ${\sf Gap}^\star({\bm \varOmega})$ and the generative model's representation error, i.e., $ {\sf Err}_{\rm rep}({{\cal X}_{R,\bm g_{\bm \theta_{\rm d}}}})$. If a more complex (i.e., deeper and/or wider) neural network is used---which means that the generative model can approximate more complex SLFs---then the representation error can be reduced. However, using a more complex neural network will make the covering number of the solution set increase, and thus ${\sf Gap}^\star({\bm \varOmega})$ increases---which presents a trade-off between network expressiveness and recovery accuracy.
            In addition, when $|\bm \varOmega|$ approaches $IJ$ and $I,J$ grow large, the sampling-induced error ${\sf Gap}^\star({\bm \varOmega})$ approaches zero [cf. Eq.~\eqref{eq:Gap}].
            
            A remark is that although the theorem statement assumes a known $R$, it can also be used to understand the cases where $R$ is not exactly known (or wrongly estimated). When $\widehat{R}=R'$ is underestimated, the energy of the missed components $\S_r\circ \c_r$ for $R'<r\leq R$ can be absorbed into the noise term $\tN$. When $\widehat{R}$ is overestimated, $\tX_{\natural}$ can still be expressed by using \eqref{eq:sigmodel}, with some redundant/spurious components. The redundant components do not create any new noise terms, but will increase the covering number of the model. This means that more samples will be required for attaining the same bound of recoverability accuracy---which is the price to pay for overestimating $R$.

		\subsection{Algorithm Design}
		We propose an alternating optimization algorithm to handle the formulated problem in \eqref{eq:joint_opt}; i.e., we tackle subproblems w.r.t. $\bm C$ and $\Z=[\z_1,\ldots,  \z_R]$ alternately, until a certain convergence criterion is met.
		
		Note that the $\C$-subproblem can be re-expressed as follows:
		\begin{equation}\label{eq:Cupdate}
		    \C^{(k+1)} \leftarrow \arg \min_{\C\geq \bm 0}~\| \G - \C\H^{(k)} \|_{\rm F}^2,
		\end{equation}
		where $\G$ is defined as before, $\H^{(k)}(r,:) = \S^{(k)}(r,\bm \Omega_{\rm col})$, and $\S^{(k)}(r,:)={\rm vec}(\g_{\bm \theta_{\rm d}}(\z_r))^\T$ is the (vectorized) current estimate for the $r$th SLF at iteration $k$. The problem in \eqref{eq:Cupdate} is a {\it nonnegativity-constrained least squares} (\texttt{NNLS}) problem---which admits a plethora of off-the-shelf solvers. In this work, we employ the \texttt{NNLS} algorithm proposed in \cite{lawson1995solving} that strikes a good balance between speed and accuracy.

		As for the $\bm z_r$-subproblem, since it involves deep neural networks, any back-propagation-based gradient descent-related algorithms can be used \cite{goodfellow2016deep}. To be specific, we use the update $\Z^{(k_{t+1})} \leftarrow \Z^{(k_t)} - \gamma^t \overline{\nabla}_{\Z} f(\Z^{(k_t)} ; \C^{(k+1)})$,
		and we let $\Z^{(k_0)}\leftarrow \Z^{(k)}$, $ \Z^{(k+1)} \leftarrow \Z^{(k_T)}$, where $T$ is the number of gradient iterations (inner iterations) used for handling the $\Z$-subproblem in the $k$th outer iteration (that alternates between $\Z$ and $\C$), $\gamma^t$ is the step size in the $t$th inner iteration, and $\Z=[\z_1,\ldots,\z_R]$. We have used the notation
		\[   f(\Z ; \C^{(k+1)}) =  \left\| \tM_{\rm sens.} \circledast \left(\tY - \sum_{r=1}^{R} \bm g_{\bm \theta_{\rm d}}(\bm z_r)\circ \c_r^{(k+1)} \right)\right\|_{\rm F}^2,  \]
		and $\overline{\nabla}_{\bm Z} f$ is a gradient-based direction, e.g., momentum-based gradient or scaled gradient (such as those used in \texttt{Adagrad} \cite{autograd} and \texttt{Adam} \cite{kingma2015adam}), which can often be constructed from the gradient ${\nabla}_{\bm Z} f$.
		The gradient of $f$ w.r.t. $\Z$ can be computed via the chain rule and back-propagation. In this work, we employ the popularized gradient evaluation tool, namely, \texttt{autograd} in \texttt{Pytorch} \cite{autograd}, to carry out the computation. The step size $\gamma^t$ and $\overline{\nabla}_{\bm Z} f$ follow the \texttt{Adam} strategy \cite{kingma2015adam}, and $\gamma^t$ has an initial value of $0.01$.
		The algorithm is summarized in Algorithm~\ref{algo:dow}, which is referred to as the  {\it Deep generative PriOr With Joint OptimizatioN for Spectrum cartography} (\texttt{DowJons}) algorithm.}

		The \texttt{DowJons} algorithm is an alternating optimization (AO) algorithm where the $\bm Z$ block is nonconex.
		This means that one often does not optimally solve the $\bm Z$-subproblem in practice. Nonetheless, we observe that \texttt{DowJons} converges well even if the $\bm Z$-subproblem is executed with only a few iterations. Indeed, under some conditions, one can show that such an ``inexact'' AO procedure produces a solution sequence that converges to the vicinity of a stationary point of \eqref{eq:joint_opt}.
		Discussions on \texttt{DowJons}'s convergence properties can be found in Appendix~\ref{app:propdowjons}.

       \begin{remark}
    A remark on the difference between \texttt{Nasdac} and \texttt{DowJons} is as follows. \texttt{Nasdac} is a two-stage approach. The two stages are (i) \texttt{SPA}-based SLF disaggregation and (ii) learned network-based SLF completion, respectively. 
    Both stages are lightweight to compute, but there may be error propagation from stage (i) to stage (ii).  Either stage alone could not guarantee recoverability of $\tX_\natural$, no matter how well the associated optimization problem (e.g., NMF) is solved.
    In contrast, for \texttt{DowJons}, if the optimization criterion in \eqref{eq:joint_opt} is solved, $\tX_\natural$ is ensured to be recovered with a reasonable bound on the estimation error. Hence, \texttt{DowJons} can be regarded as a one-stage method in the sense of ensuring recoverability guarantees. 
    Nonetheless, the recoverability advantages of \texttt{DowJons} come at the expense of a higher computational cost.
       \end{remark}

	\begin{algorithm}[t]
	\footnotesize
		\SetAlgoLined
		\KwData{$\tY, \tM_{\rm sens.}, R, \gamma, \tau, {\sf MaxIter}, \btheta$} 
		\KwResult{Estimated $\widehat{\tX}$, $\widehat{\S}$ and $\widehat{\C}$}
		\tcp{Initialization via \texttt{Nasdac}:}
		$\overline{\G} = {\rm unfold}(\tM_{\rm sens.} \circledast \tY)$ ;\\
		$\G =\overline{\G}(:, \bOmega_{\rm col})$;\\
		$\C^{(0)}, \S^{(0)} \leftarrow \texttt{Nasdac}(\tM_{\rm sens.} \circledast \tY, \bOmega, R, \f_{\theta}(\cdot) )$;\\
		\For{$r=1:R$}{
		     $\z_r^{(0)} \leftarrow \p_{\bm \theta_{\rm e}}(\S_r^{(0)})$;  \\
		}
		$k \leftarrow 0$; \\
		\While{$(f(\Z^{(k+1)}; \C^{(k+1)}) \leq f(\Z^{(k)}; \C^{(k)}) - \tau)$ \textbf{or} $(k \leq {\sf MaxIter})$}{
			
			$\C^{(k+1)} \leftarrow \texttt{NNLS}(\G, \S^{(k)}(:,\bOmega_{\rm col}))$;\\
			$\Z^{(k_0)} \leftarrow \Z^{(k)}$ ;\\
			\For{$t = 1:T$} {

		    $\Z^{(k_{t})} \leftarrow \Z^{(k_{t-1})} - \gamma^t \overline{\nabla}_{\Z} f(\Z^{(k_{t-1})};\C^{(k+1)}) ;$
			}
			$\Z^{(k+1)} \leftarrow \Z^{(k_{T})}$; \\
			\For{$r=1:R$}{
			$\S_r^{(k+1)} = \g_{\bm \theta_{d}}(\z_r^{(k+1)})$;
			}
			$k \leftarrow k+1$;
		}
		$\widehat{\tX} = \sum_{r=1}^R \widehat{\S}_r^{(k)} \circ \widehat{\c}_r^{(k)}$.
		\caption{\texttt{DowJons} } \label{algo:dow}
	\end{algorithm}
			
	\section{Experiments}\label{sec:exp}
		In this section, we use synthetic-data and real-data experiments to showcase the effectiveness of the proposed methods.
				
		\subsection{Synthetic-Data Experiments - Settings}\label{sec:exp_synthetic}
		\subsubsection{Baselines} We use a number of baselines to benchmark the proposed methods. To be specific, we compare the performance of our methods with that of the \texttt{TPS} method in \cite{ureten2012comparison}, the \texttt{LL1} method in \cite{zhang2020spectrum}, and the \texttt{DeepComp} method in \cite{teganya2020data}.  The \texttt{TPS} method uses 2D splines to interpolate between the observed entries, the \texttt{LL1} method is a block-term tensor based completion method, and the \texttt{DeepComp} method also uses deep generative models, but directly learns the aggregated radio map using a neural network; see our discussion in previous sections. 	 We hope to remark that the baseline \texttt{LL1} is implemented as a combination of emitter level disaggregation and classic interpolation (using \texttt{TPS}) based individual SLF completion. Therefore, it is similar to the proposed approach, but the deep network completion used in our framework is replaced by a classic SLF interpolation method. The comparison with \texttt{LL1} can therefore serve as an ablation study of the effectiveness of the learned deep priors in our approach. More comparisons with such ``SLF disaggregation and classic interpolation'' approaches using different interpolation methods can be found in the supplementary materials.

		\subsubsection{Data Generation}
		Our simulations and training set are generated following the joint path loss and log-normal shadowing model \cite{goldsmith2005wireless}.
		Under this model, the SLF of emitter $r$ is expressed as follows:
		\begin{equation}\label{eq:shadowmodel}
			\S_r({\bm y}) = \|{\bm y} - {\bm r}_r \|_2^{-\gamma_r} 10^{v_r({\bm y})/10},
		\end{equation}
		where ${\bm y}=(i,j)$ denotes the spatial coordinates, ${\bm r}_r$ is the location of $r$-th emitter, $\gamma_r$ is the path loss coefficient associated with emitter $r$, and $v_r({\bm y})$ is the correlated log-normal shadowing component. This component is sampled from zero-mean Gaussian distribution with variance $\eta_r$, and the auto-correlation between ${\bm y}$ and ${\bm y'}$ can be written as $\mathbb{E}(v_r({\bm y}), v_r({\bm y'})) = \eta_r \exp(-\|{\bm y}-{\bm y'}\|_2/{\rm d}_{\rm corr})$,	in which ${\rm d}_{\rm corr}$ is the so-called {\it decorrelation distance}.
		Smaller ${\rm d}_{\rm corr}$ and larger $\eta_r$ indicate more severe shadowing effects and thus harsher environments.
		For example, for an outdoor environment, the typical values of ${\rm d}_{\rm corr}$ and $\eta_r$ range from 50 to 100m and 4dB to 12dB, respectively \cite{goldsmith2005wireless}.

	    We consider a geographical region of $50 \times 50$m$^2$. We discretize the region into $50 \times 50$ grids, and consider $K$ frequency bins of interest, where $K=64$ unless specified otherwise.
		The PSDs of the emitters are assumed to be a combination of randomly scaled sinc functions. Specifically, the PSD of emitter $r$ is given by 
		 $   \c_r(k) = \sum_{i=1}^M a_i^{(r)} {\rm sinc}^2 ( \nicefrac{k-f_i^{(r)}}{w_i^{(r)}}),    $
		where $f_1^{(r)},\ldots,f_M^{(r)}$ are the central frequencies of $M$ subbands that are available to emitter $r$, where $M$ will be specified in different simulations;
		$w_i^{(r)}$ controls the width of the emitter $r$'s PSD's sidelobe at its $i$th subband, which is sampled from a uniform distribution between 2 and 4; $a_i^{(r)}$ is a random scaling of emitter $r$'s PSD intensity at its $i$th subband, which follows the uniform distribution between $0.5$ and $2.5$.
		After generating $\S_r$ and $\c_r$, $r=\{1, \dots, R\}$, we use \eqref{eq:factorization} to obtain the aggregate radio map $\tX$.  When noise is considered, we use $\tX= \sum_{r=1}^R \S_r \circ \c_r+\underline{\bm N}$, where $\underline{\bm N}$ represents the noise tensor.

		\subsubsection{Deep Network Configuration and Training Sample Generation}
		As mentioned, we employ a deep autoencoder for learning the generative model of the SLFs. The autoencoder has a convolutional encoder-decoder architecture with the latent dimension being $D=256$. More details can be seen in Appendix~\ref{app:DNN}.
		
	    To train our network, we generate 500,000 single emitter SLFs within the region of interest. The SLFs follow the model in \eqref{eq:shadowmodel}. For each sample, we randomly pick a position for the emitter from the $\bm r_r$ over the 50$\times $50 grids.
	    The parameters $\gamma_r$ for all $r$ follows uniform distribution from $2$ to $2.5$, which covers the range of path loss coefficients in scenarios of interest \cite{hindia2018outdoor}; $\eta_r = \eta$ for all $r$ and it is sampled uniformly from $3$ to $8$; and ${\rm d}_{\rm corr}$ is sampled uniformly from $30$ to $100$.
	    The selection of the latter two parameters is used to cover scenarios with moderate to severe shadowing effects.	

	    To generate training samples for \texttt{DeepComp},
	    we use $R$ such SLFs and randomly generated PSDs $\bm c_r$'s to construct $\tX$ following \eqref{eq:sigmodel}. 
	    The $\c_r$'s are generated using $M = 10$, so that most of the slabs contain at least one emitter, and other parameters set as described earlier.
	    We collect 500,000 $\tX(:,:,k)$'s as the training set with $R$ sampled randomly from $\{3,4,5,6\}$ for each $\tX$.

		\subsubsection{Performance Metric} 
		For an estimated $\widehat{\tX}$ of the ground-truth $\tX_\natural$, we measure \textit{squared reconstruction error} (SRE), i.e., $$
		    {\rm SRE} = \nicefrac{\| \widehat{\tX} - \tX_\natural \|_{\rm F}^2}{\| \tX_\natural \|_{\rm F}^2}.$$
		In order to evaluate the accuracy of $\widehat{\C}$, we use \textit{normalized absolute error} (NAE) to evaluate the estimates of $\C$ defined as follows:
		\begin{align}\label{eq:NAE}
		    {\rm NAE}_{\C} &= \frac{1}{R}\sum_{r=1}^{R} \left\| \frac{\c_r}{\|\c_r\|_1} - \frac{\widehat{\c}_{_r}}{\|\widehat{\c}_{_r}\|_1} \right\|_1,
		\end{align}
		where we have assumed that the column permutation between $\C$ and $\widehat{\C}$ has been removed, e.g., using the Hungarian algorithm (see Sec. VI in \cite{zhang2020spectrum}). For the SLFs, ${\rm NAE}_{\S}$ for $\S$ is defined in the same way; see \cite{zhang2020spectrum}.

		\subsubsection{Hyperparameters of Algorithms}
		Among the algorithms under test, \texttt{DowJons} and \texttt{LL1} have some hyperparameters to be determined in advance.
	    For \texttt{DowJons}, we stop the algorithm either when the relative change of the cost function is smaller than 0.003 or when the alternating optimization procedure reaches 10 outer iterations. For the inner iterations that tackle the $\Z$-subproblem, we set the maximum number of iterations $T$ to be 10. For tensor-based method \texttt{LL1}, we set the rank of SLFs to be up to 4 using trial-and-error type tuning. The regularization parameters for all the mode latent factors to be $10^{-3}$ . The algorithm is terminated when the relative change of its loss function less than $10^{-3}$ or when the number of iterations exceeds 50.
	    All the results are averaged from $30$ Monte Carlo trials.

	    \subsection{Synthetic-Data Experiments - Results}
		
			\begin{figure}[t!]
			    \centering
			    \includegraphics[width=0.7\linewidth]{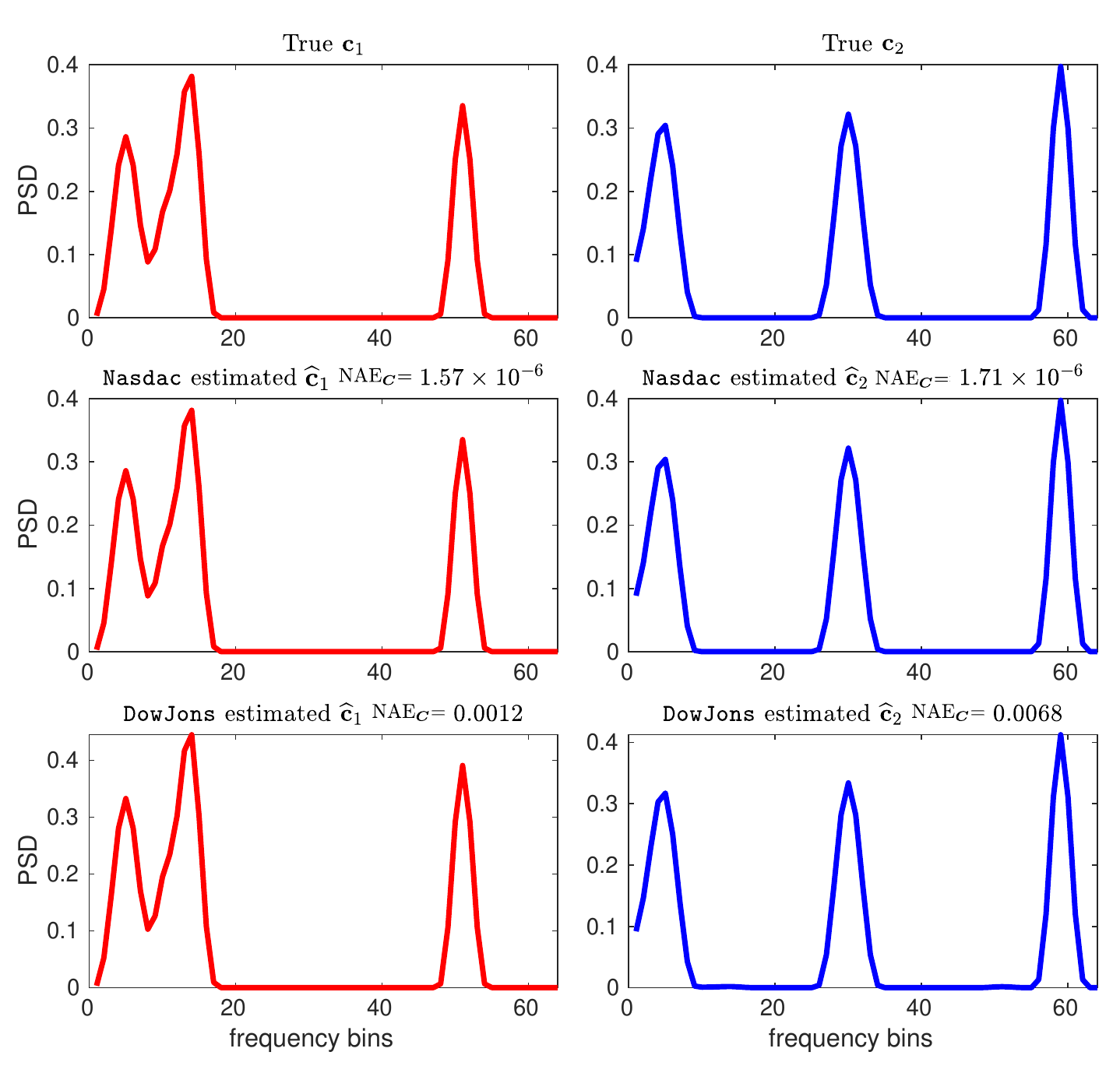}
			    \caption{Ground-truth and estimated PSDs by \texttt{Nasdac} and \texttt{DowJons}; $\rho=10\%$, $R=2$, $\eta= 5$, ${\rm d}_{\rm corr}=50$.}
			    \label{fig:sep_plots_psd}
			\end{figure}
			
			\begin{figure}[t!]
			    \centering
			    \includegraphics[width=0.8\linewidth]{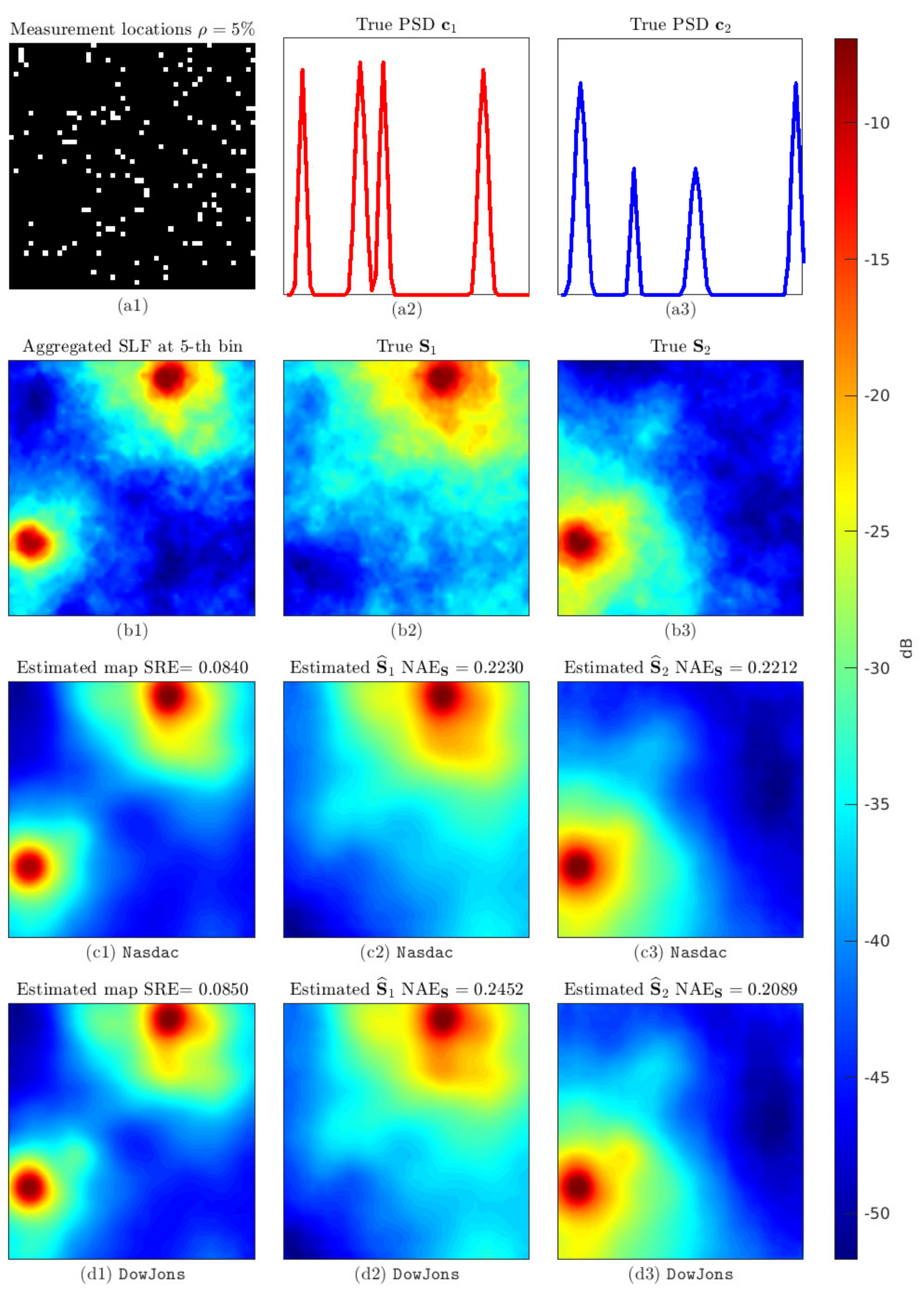}
			    \caption{Ground truth SLF ((b2), b(3)) and radio map (b1), and reconstructed SLF ((c2), (c3), (d2), (d3)) and radio map ((c1), (d1)) by \texttt{Nasdac} and \texttt{DowJons} under sparse spectral occupancy; $\rho=5\%$, $R=2$, $\eta=5$, and ${\rm d}_{\rm corr} = 50$;
			    the visualized maps are at the $5$th frequency bin.}
			    \label{fig:sep_plots_slf}
			\end{figure}
		    
			\subsubsection{Sparse Spectral Occupancy}
			Under the observation model in \eqref{eq:sensing_noiseless},
			we first test the algorithms under scenarios where the emitters use the spectral bands in a sparse manner such that Assumption \ref{ass:dominance} holds. 
			To ensure this, we first designate one frequency band for each emitter and allow to use up to $M = 16$ frequency bands out of the total remaining bands. We use this setting for all the tables and figures unless stated otherwise.

			Figs. \ref{fig:sep_plots_psd}-\ref{fig:sep_plots_slf} show an illustrative example, where $R=2$ emitters are in the region of interest. In this case, we randomly sample $\rho=|\bm \varOmega|/IJ\times 100\%=10\%$ of the 2,500 grids; i.e., $10\%$ of the $\tX(i,j,:)$'s are revealed to us by sensors that capturing the PSD over the $K=64$ frequency bands at $(i,j)$.
			We set $\eta_r=\eta=5$ for all $r$ and ${\rm d}_{\rm corr}=50$.  The SREs/ NAEs of the estimated radio maps/SLFs are also displayed in the figures, so that the reader can have a better understanding of the correspondence between the performance metrics and the visual quality of the recovered maps/SLFs. One can see that \texttt{Nasdac} and \texttt{DowJons} recover the PSDs and SLFs accurately in this case---and thus the aggregated radio map as well. 
		
			\begin{figure}[t!]
				\centering
				\includegraphics[width=0.8\linewidth]{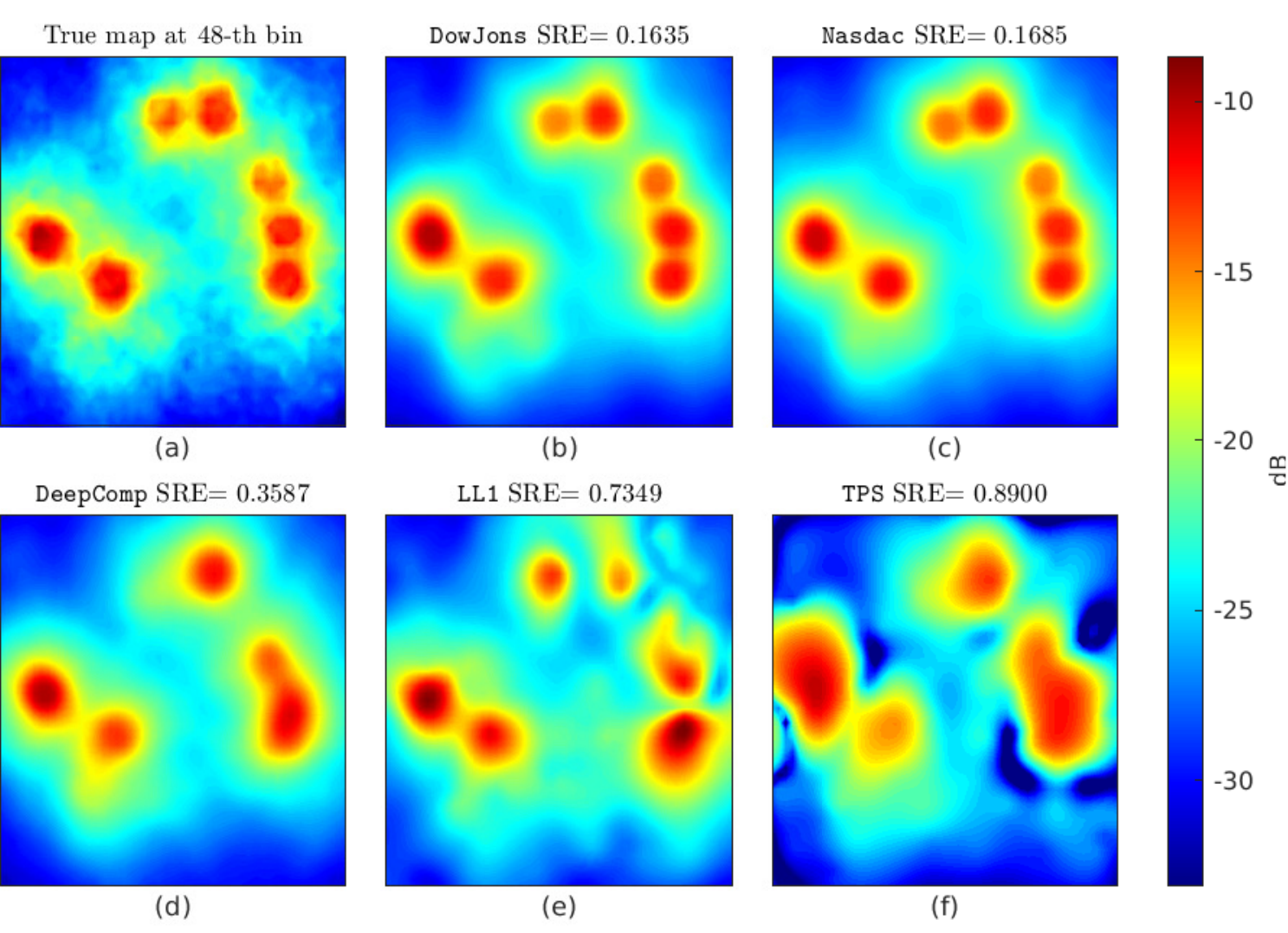}
				\caption{Ground-truth and reconstructed radio map by proposed algorithms and baselines under $R=10, \eta=5, \rho=5\%, {\rm d}_{\rm corr} = 50, M=4$ when Assumption \ref{ass:dominance} holds (see Fig.~\ref{fig:sep_plots_slf} (a2) and (a3) for the PSDs when Assumption \ref{ass:dominance} holds.). }
				\label{fig:sep_plots_map}
			\end{figure}

			Fig.~\ref{fig:sep_plots_map} shows the results of
			a much more challenging case, where $R=10$ emitters are present and only $\rho=5\%$ locations are sampled. The other settings are the same as before.
		    Fig.~\ref{fig:sep_plots_map} shows
			the ground truth and recovered radio maps by the algorithms at the 48th frequency band. One can see that both the proposed methods, i.e., \texttt{DowJons} and \texttt{Nasdac}, yield visually accurate maps as in the previous case. The \texttt{DeepComp} approach seems to miss some emitters, especially when the emitters are close to each other. This may be because $R=10$ emitters are present in this frequency, while \texttt{DeepComp} was trained using up to 6 emitters.
			Purely model-based methods (i.e., \texttt{LL1} and \texttt{TPS}) do not perform as well, possibly because the model assumptions (e.g., the SLFs being low-rank matrices in \texttt{LL1}) are grossly violated. 
			
			A remark is that although both \texttt{Nasdac} and \texttt{DowJons} work similarly in terms of recovery accuracy in Fig.~\ref{fig:sep_plots_map}, there is a notable difference in the runtime performance.
			Table \ref{tab:run_time} shows the runtime in the test stage of the algorithms. One can see that \texttt{Nasdac} takes only 0.069 seconds to accomplish the task. While \texttt{DowJons} takes 5.2 seconds. This is because there is no heavy optimization in the test stage of \texttt{Nasdac}, while \texttt{DowJons} involves solving a complex optimization problem in the test stage. As a tradeoff, \texttt{DowJons} can work with more challenging cases where Assumption~\ref{ass:dominance} does not hold---which will be seen in Sec.~\ref{sim:dense}.

		    \begin{table}[t!]
		        \scriptsize
		        \centering
		        \caption{Test time of algorithms for running the task in Fig.~\ref{fig:sep_plots_map}.}
                \begin{tabular}{|c|c|c|c|c|c|}
                    \hline
					 & \texttt{TPS}   & \texttt{LL1}   & \texttt{DeepComp}   & \texttt{Nasdac} & \texttt{DowJons}  \\ \hline
                     Runtime (sec.) & 0.203  &  6.8  &  0.112  &  \textbf{0.069}  &  5.2 \\ \hline
		        \end{tabular}
		        \label{tab:run_time}
		    \end{table}

            \begin{table}[t!]
				\scriptsize
				\centering
				\caption{SRE under various $\rho$'s; $R=7$, ${\rm d}_{\rm corr}=50$, $\eta=6$.}
				\begin{tabular}{|c|c|c|c|c|c|}
					
				    \hline
					$\rho$ & \texttt{TPS}   & \texttt{LL1}   & \texttt{DeepComp}   & \texttt{Nasdac} & \texttt{DowJons}  \\ \hline
           1\%  & 0.6882 &  0.8158  &  0.5257  &  0.4195  &   \textbf{0.4078} \\ \hline
           5\%  & 0.2974 &  0.3387  &  0.2047  &  \textbf{0.1319}  &   0.1332 \\ \hline
          10\%  & 0.1493 &  0.1347  &  0.1162  &  0.0918  &   \textbf{0.0896} \\ \hline
          15\%  & 0.1079 &  0.1085  &  0.0929  &  0.0798  &   \textbf{0.0713} \\ \hline
          20\%  & 0.0840 &  0.0710  &  0.0898  &  0.1013  &   \textbf{0.0599} \\ \hline
				\end{tabular}
				
				\label{tab:sep_omega}
			\end{table}
			
			Table \ref{tab:sep_omega} shows the performance of the algorithm under various $\rho$'s.  
			One can see that all algorithms favor more samples (sensing locations). 
		    The DL-based methods (\texttt{Nasdac}, \texttt{DowJons} and \texttt{DeepComp}) exhibit notable advantages over the model-based methods \texttt{LL1} and \texttt{TPS}, especially when $\rho\leq 5\%$. 
		    In addition, the proposed \texttt{Nasdac} and \texttt{DowJons} consistently outperform the baselines under most $\rho$'s.

        	\begin{figure}[t!]
        	    \centering
        	    \includegraphics[width=0.8\linewidth]{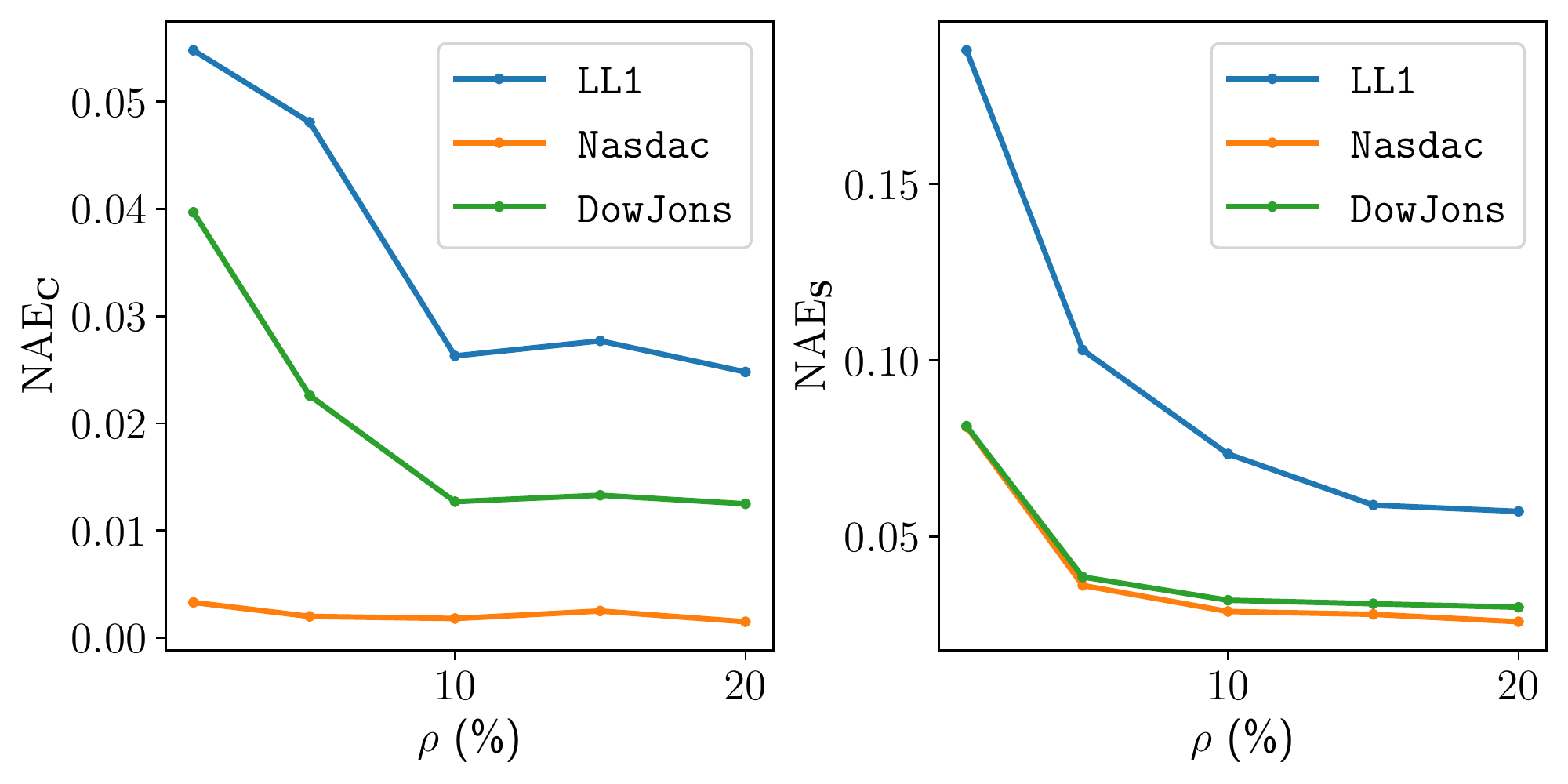}
            	\caption{ NAE under various $\rho$'s; $R=7$, ${\rm d}_{\rm corr}=50$, $\eta=6$.}
        	    \label{fig:sep-nae-omega}
        	\end{figure}

        	Fig.~\ref{fig:sep-nae-omega} shows the estimation accuracy of the individual PSDs and SLFs 
        	by pertinent algorithms under the same settings.
        	Note that only \texttt{LL1}, \texttt{Nasdac} and \texttt{DowJons} estimate such individual components, while other algorithms directly estimate the aggregated radio map.
        	Interestingly, the PSDs are estimated accurately by all the algorithms under test. However,
        	\texttt{LL1} performs poorly in terms of SLF estimation. This backs our motivation for this work---the low-rank matrix model that \texttt{LL1} leverages for estimating $\S_r$ does not always hold. On the other hand, \texttt{Nasdac} and \texttt{DowJons} are both able to estimate the SLFs more accurately. In particular, the proposed methods-output NAE$_{\S}$'s are at least improved by 50\% relative to the outputs of \texttt{LL1} when $\rho\leq 5\%$. 
        	
        	We also note that \texttt{Nasdac} exhibits higher estimation accuracy of the PSDs and SLFs relative to \texttt{DowJons}, even though the overall radio map reconstruction by the latter is more accurate. This may be because the optimization criterion of \texttt{DowJons} targets minimizing the error of the aggregated radio map, instead of the PSDs and SLFs. Also note the NAE definitions do not consider amplitudes of the quantities [$\bm c_r$ and $\bm S(r,:)$] under test [cf. Eq.~\eqref{eq:NAE}], but amplitudes are considered in the SRE measure when evaluating the reconstruction error. Hence, low NAEs do not directly imply good SREs, although they are related and are both meaningful metrics.

			\begin{table}[t!]
			\scriptsize
				\centering
				\caption{SRE under various $\eta$'s; $R=7$, ${\rm d}_{\rm corr}=50$, and $\rho=10\%$.}
				\begin{tabular}{|c|c|c|c|c|c|}
				    \hline
					$\eta$   & \texttt{TPS}      & \texttt{LL1}  & \texttt{DeepComp}          & \texttt{Nasdac} & \texttt{DowJons}  \\ \hline
            	    4  &  0.1415  &  0.1259  &  0.0852  &  0.0861  &  \textbf{0.0556} \\ \hline
            		5  &  0.1387  &  0.1245  &  0.0934  &  0.0979  &  \textbf{0.0684} \\ \hline
            		6  &  0.1507  &  0.1444  &  0.1184  &  0.1141  &  \textbf{0.0841} \\ \hline
            		7  &  0.1679  &  0.1581  &  0.1478  &  0.1536  &  \textbf{0.1225} \\ \hline
            		8  &  0.1850  &  0.1798  &  0.1644  &  0.1539  &  \textbf{0.1362} \\ \hline
				\end{tabular}
				\label{tab:sep_eta}
			\end{table}

			Tables \ref{tab:sep_eta}-\ref{tab:sep_dcorr} show the SREs under various levels of shadowing effects by changing $\eta$ and ${\rm d}_{\rm corr}$, respectively. Note that $\eta=4$ and ${\rm d}_{\rm corr}=90)$ correspond to relatively mild shadowing, while $\eta=8$ and ${\rm d}_{\rm corr}=30$ both correspond to severe shadowing. 
			From the two tables,
			one can see that \texttt{Nasdac} and \texttt{DowJons} output competitive SREs relative to the baselines. 
			We also note that \texttt{Nasdac} constantly produces reasonable results, but is less accurate compared \texttt{DowJons} in the most challenging cases, i.e., when $\eta=8$ or ${\rm d}_{\rm corr}=30$. This may be because \texttt{Nasdac} uses a two-stage approach. Challenging environments may make the first state (i.e., NMF-based disaggregation) less accurate---and the error propagates to the SLF completion stage. Nonetheless, using a one-shot learning criterion, \texttt{DowJons} does not have this issue. The price to pay is that \texttt{DowJons}'s optimization process is more costly relative to \texttt{Nasdac}.

            \begin{table}[t!]
                \scriptsize
				\centering
				\caption{SRE under various ${\rm d}_{\rm corr}$;  $R=7$, $\eta=6$ and $\rho=10\%$.}
				\begin{tabular}{|c|c|c|c|c|c|}
				\hline
					${\rm d}_{\rm corr}$   & \texttt{TPS}      & \texttt{LL1}  & \texttt{DeepComp}          & \texttt{Nasdac} & \texttt{DowJons}  \\ \hline
		30 & 0.1803  &  0.1948          &  0.1586  &  0.1317  &  \textbf{0.1279} \\ \hline
		50 & 0.1662  &  0.1741          &  0.1327  &  0.1058  &  \textbf{0.0912} \\ \hline
		70 & 0.1397  &  0.1437          &  0.1057  &  0.0867  &  \textbf{0.0735} \\ \hline
		90 & 0.1510  &  0.1463          &  0.1023  &  0.0706  &  \textbf{0.0611} \\ \hline
				\end{tabular}
				\label{tab:sep_dcorr}
			\end{table}

            \begin{table}[t!]
                \scriptsize
				\centering
				\caption{SRE under various $R$'s; $\eta=6$, ${\rm d}_{\rm corr}=50$, $\rho=5\%$.}
				\begin{tabular}{|c|c|c|c|c|c|}
				\hline
					$R$         & \texttt{TPS}     & \texttt{LL1}     & \texttt{DeepComp}   & \texttt{Nasdac}   & \texttt{DowJons}  \\ \hline
			    6   &  0.2355  &  0.2146  &  0.1426  &  0.0950  &  \textbf{0.0813} \\ \hline
				7   &  0.2353  &  0.2236  &  0.1616  &  0.1095  &  \textbf{0.0882} \\ \hline
				8   &  0.2237  &  0.2213  &  0.1715  &  \textbf{0.0845}  &  0.0872 \\ \hline
				9   &  0.2248  &  0.2459  &  0.1809  &  0.0879  &  \textbf{0.0830} \\ \hline
				10  &  0.2368  &  0.2500  &  0.1899  &  0.1029  &  \textbf{0.0986} \\ \hline
				11  &  0.2498  &  0.2528  &  0.1922  &  0.1081  &  \textbf{0.0901} \\ \hline   
				\end{tabular}
				\label{tab:sep_r}
			\end{table}

			Table \ref{tab:sep_r} shows the SREs of the algorithms when $R$ varies. 
			An important observation is that the performance of \texttt{DeepComp} deteriorates when $R$ increases. The reason is that \texttt{DeepComp} is trained using samples that contain up top $6$ emitters (which already makes the underlying state space overwhelmingly large, as we discussed; see, e.g., Remark~\ref{rmk:statespace}). Whenever a band contains more than $6$ emitters, the \texttt{DeepComp} approach tends to miss the extra emitters. Nevertheless, the proposed methods that model the emitters' SLFs individually do not have this problem---which supports our design goal of improving the generalization performance.

\begin{figure}[t!]
                \centering
                \includegraphics[width=0.6\linewidth]{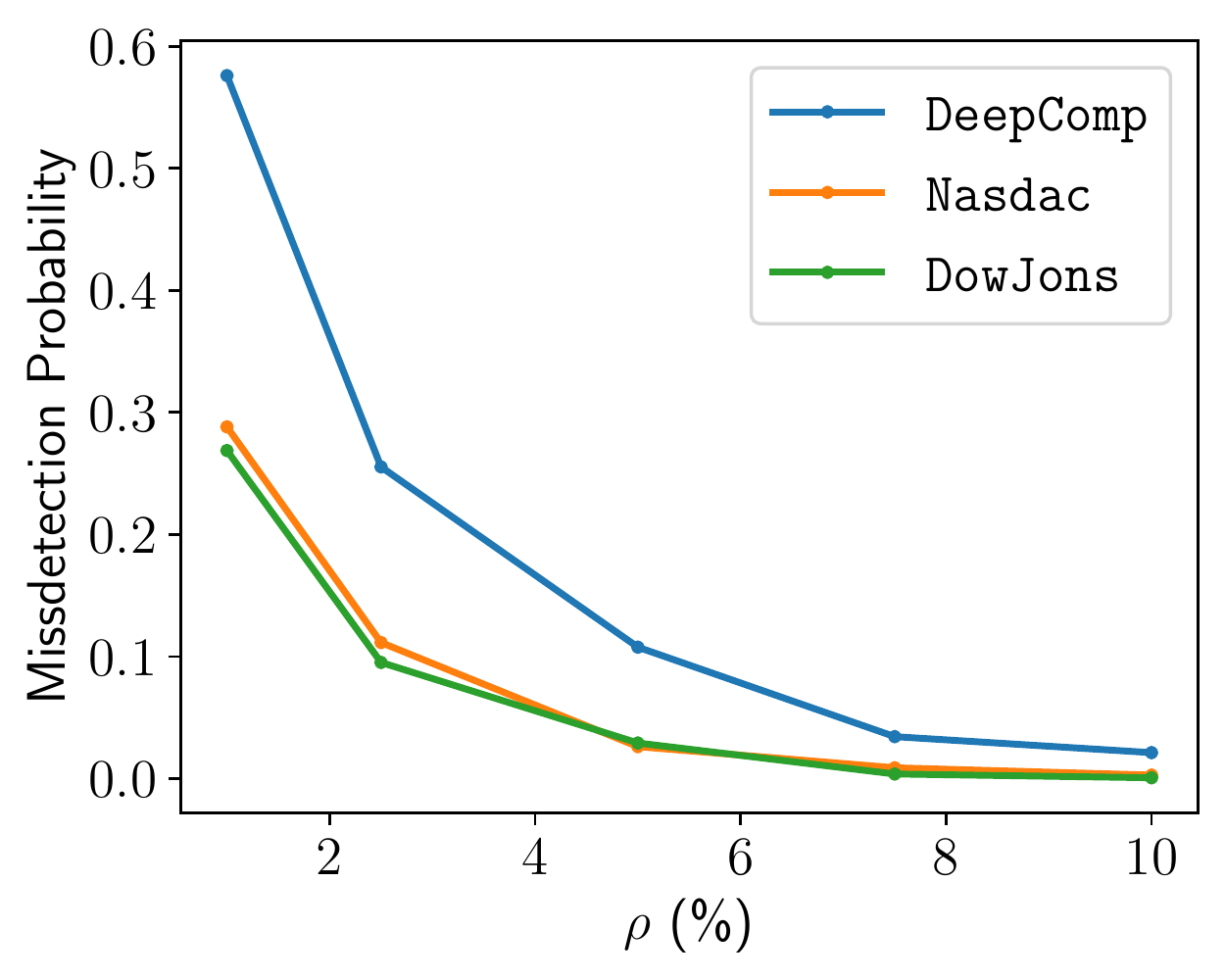}
                \caption{ Emitter misdetection probability under various $\rho$'s; $R=10$ $\eta=5$, ${\rm d}_{\rm corr} = 50$.} 
                \label{fig:miss_detect_omega}
            \end{figure}

			The results in Table~\ref{tab:sep_r} also articulate one noticeable challenge of \texttt{DeepComp}: It usually misses some emitters if more emitters than the number of emitters used in training samples appear in scene (also see Fig. \ref{fig:sep_plots_map}). 
			We define a misdetection case as follows: If the power of the reconstructed signal at the emitter location in a frequency drops below a certain threshold (25\% of the ground-truth signal strength in our experiments), we count this as a misdetection case. The misdetection probabilities of different DL-based methods under $R=10$, $\rho=5\%$, $\eta=5$ and $
			{\rm d}_{\rm corr}=50$ are shown in  Fig.~\ref{fig:miss_detect_omega}. One can see that \texttt{DeepComp} has a misdetection probability that is often much higher than that of the proposed methods.
			
            \begin{table}[t!]
                \scriptsize
				\centering
				\caption{SRE under various SNRs; $R = 7$, $\eta=6$, ${\rm d}_{\rm corr}=50$, and $\rho=10\%$.}
				\begin{tabular}{|c|c|c|c|c|c|}
				    \hline
					SNR(dB)         & \texttt{TPS}     & \texttt{LL1}     & \texttt{DeepComp}   & \texttt{Nasdac}   & \texttt{DowJons}  \\ \hline
    			    0   &  0.9019  &  0.8092 &  0.7745  &  0.9121  &  \textbf{0.6774} \\ \hline
    				10  &  0.1937  &  0.2101 &  0.1676  &  0.2521  &  \textbf{0.1413} \\ \hline
    				20  &  0.1540  &  0.1539 &  0.0947  &  0.1371  &  \textbf{0.0749} \\ \hline
    				30  &  0.1398  &  0.1238 &  0.0988  &  0.0853  &  \textbf{0.0747} \\ \hline
    				40  &  0.1181  &  0.0999 &  0.0807  &  0.0716  &  \textbf{0.0541} \\ \hline
				\end{tabular}
				\label{tab:sep_snr}
			\end{table}

			Table \ref{tab:sep_snr} shows the performance of all methods under different sensing noise levels under the noisy observation model in \eqref{eq:noisy}.
			In this simulation, all entries of the noise ther $\tN$ [cf. Eq.~\eqref{eq:noisy}] are sampled from a uniform distribution between $0$ and $1$, and then scaled to satisfy the pre-specified signal-to-noise ratios (SNRs). The SNR is defined as ${\rm SNR} = 10\log_{10}(\|\tX_\natural\|_{\rm F}^2/\|\tN\|_{\rm F}^2)$ (dB).
			One can see that \texttt{Nasdac} enjoys a high reconstruction accuracy when SNR$\geq 30$dB and outperforms all the baselines. However, it is less robust to heavily noise environments---which is, again, a consequence of using a two-stage approach. On the other hand, \texttt{DowJons} offers stable results across all SNRs under test, which also corroborates our noise robustness proof in Theorem~\ref{thm:rmse}.

		\subsubsection{Dense Spectral Occupancy}\label{sim:dense}
			\begin{figure}[t!]
			    \centering
			    \includegraphics[width=0.8\linewidth]{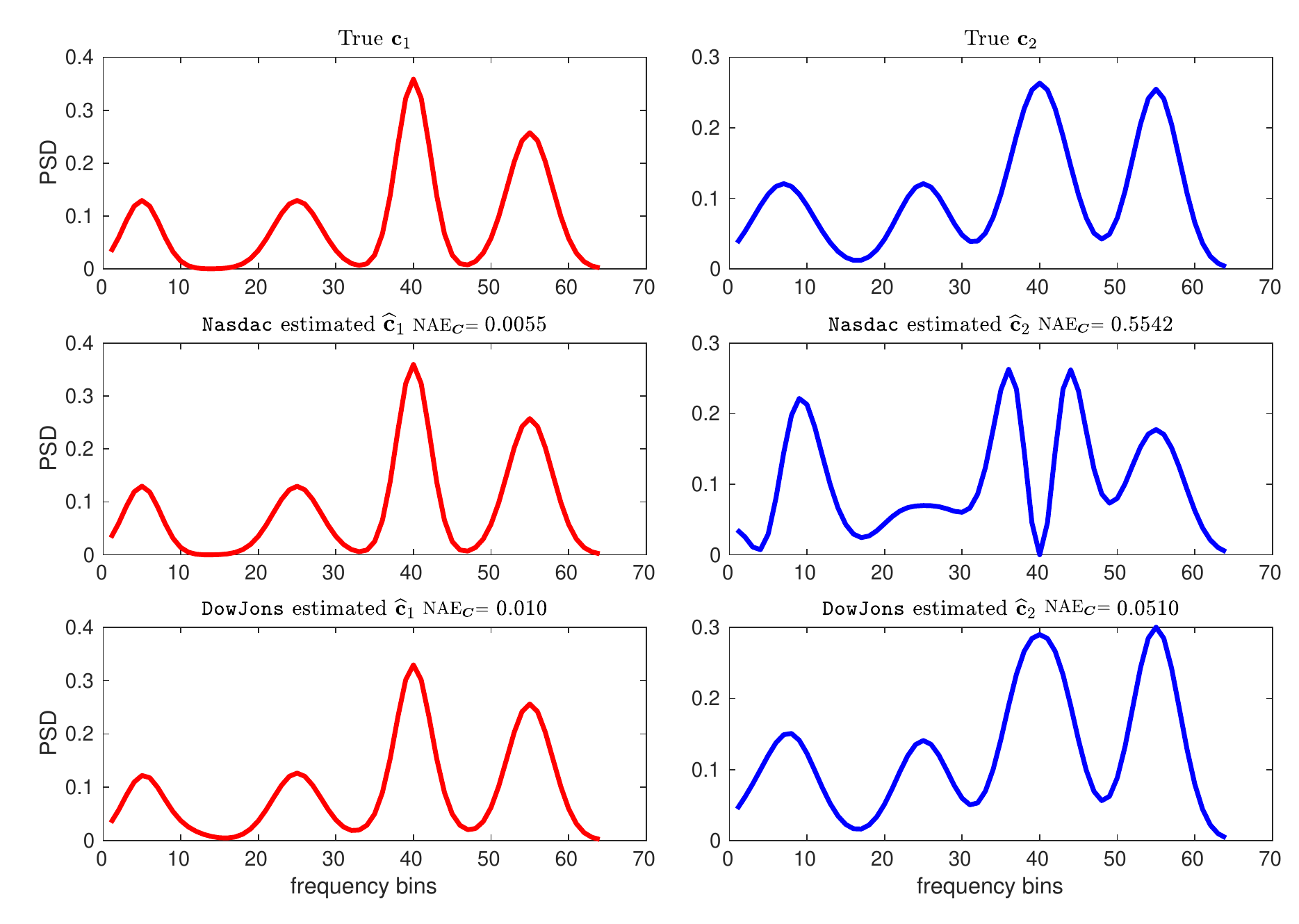}
			    \caption{Visualization of PSD reconstructed by \texttt{Nasdac} and \texttt{DowJons} from $10\%$ sampling when $\C$ is dense (does not satisfy Assumption \ref{ass:dominance}); $R=2$, $\eta=5$, ${\rm d}_{\rm corr}=50$.}
			    \label{fig:insep_plots_psd}
			\end{figure}
			\begin{figure}[t!]
				\centering
				\includegraphics[width=0.8\linewidth]{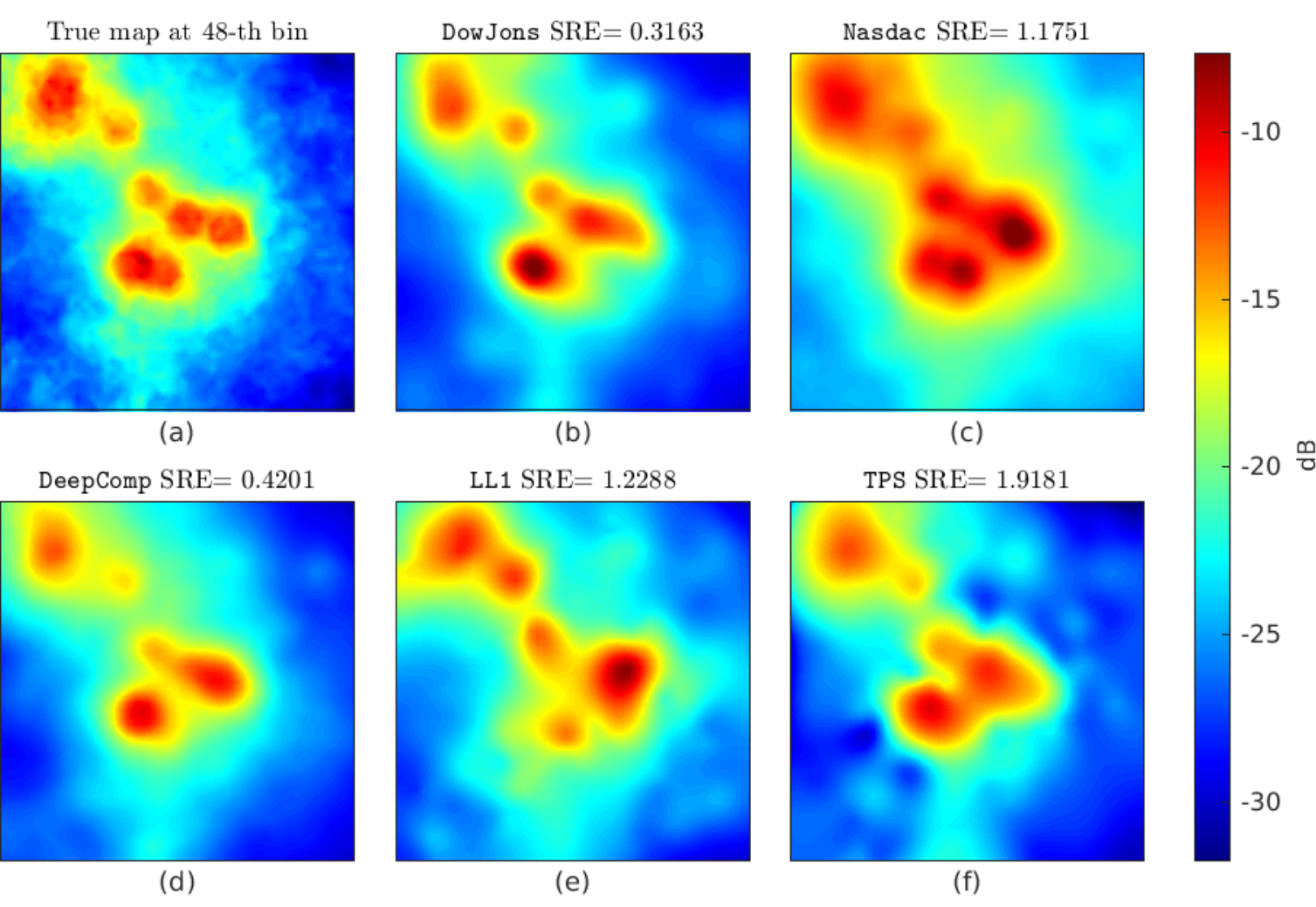}
				\caption{Ground-truth and reconstructed radio maps under $R=7$, $\eta=5$, $\rho=5\%$, $M=8$, and ${\rm d}_{\rm corr} = 50$, when Assumption \ref{ass:dominance} does not hold.}
				\label{fig:insep_plots_map}
			\end{figure}
            We also test the algorithms under the cases where the spectral bands are crowded---i.e., where
            Assumption \ref{ass:dominance} does not hold. Under such circumstances, \texttt{Nasdac} is expected to not work, while \texttt{DowJons} should not be affected.
            
            Fig. \ref{fig:insep_plots_psd} shows the estimated PSDs under $\rho = 10\%, R=2, \eta=5, {\rm d}_{\rm corr}=50$. In this case, the two emitters both use all the frequency bands and thus Assumption~\ref{ass:dominance} is clearly violated. As expected, \texttt{Nasdac} fails to estimate the PSDs. Nonetheless, \texttt{DowJons} works reasonably well in terms of reconstructing the PSDs.
            Fig \ref{fig:insep_plots_map} shows the reconstructed radio maps in a case where $R=7$, $\eta=5$, $\rho=5\%$, and ${\rm d}_{\rm corr} = 50$, where Assumption \ref{ass:dominance} is again violated. 
            Similar as the previous case, \texttt{DowJons} is essentially not affected by the violation of Assumption~\ref{ass:dominance}, but \texttt{Nasdac} could not output reasonable results.
            
            \begin{figure}[t!]
                \centering
                \includegraphics[width=0.6\linewidth]{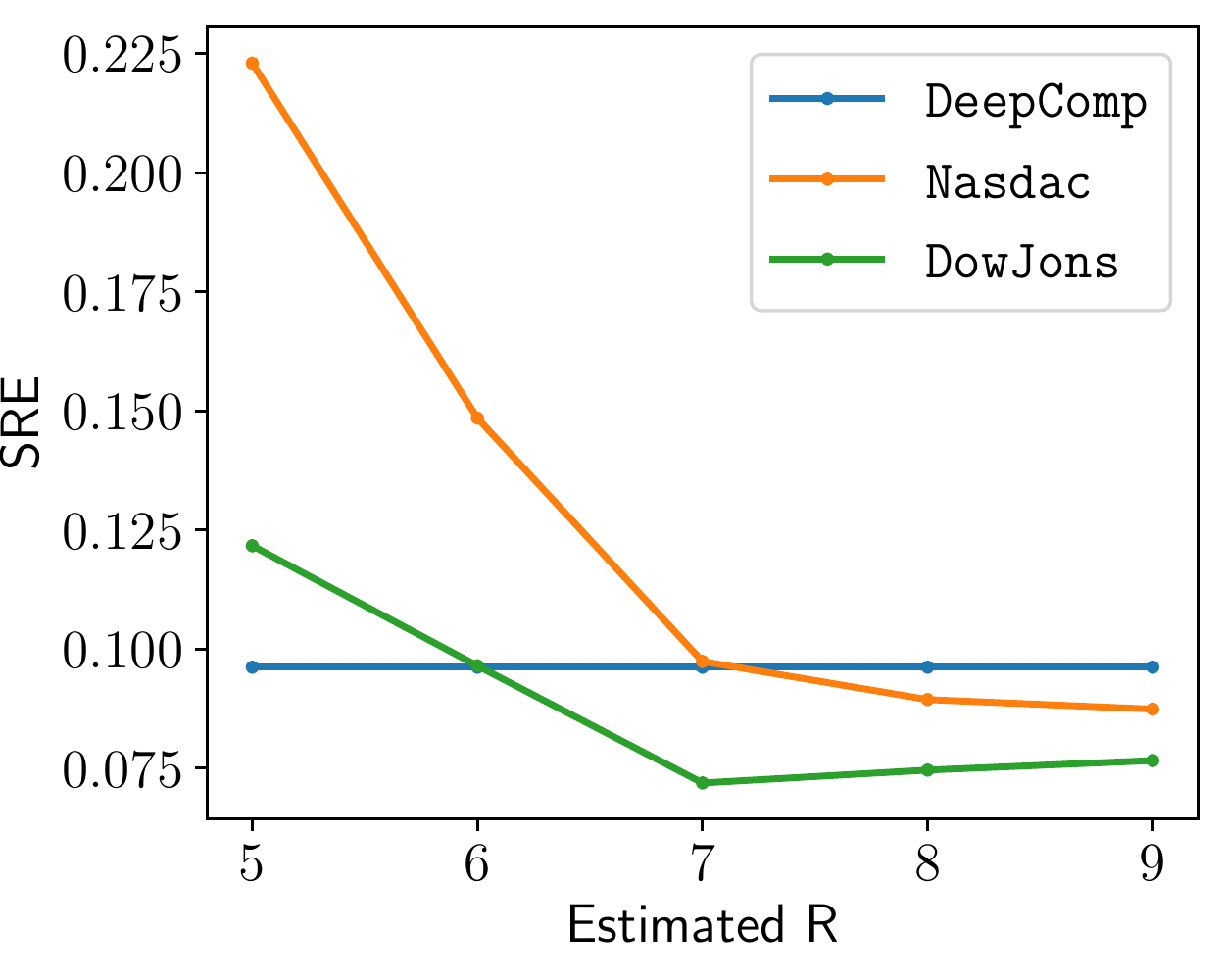}
                \caption{ SRE under various wrongly estimated $\widehat{R}$'s; true $R=7$, ${\rm d}_{\rm corr}=50$, $\eta=5$, and $\rho=10\%$.}
                \label{fig:emitter_misestimation}
            \end{figure}

            \subsubsection{Under Mis-specified $R$}
            Our algorithms use the knowledge of $R$ to separately model the individual SLFs of the emitters. Although model order selection is a mature technique in signal processing \cite{stoica2005spectral}, it is also of interest to observe how the algorithms perform under wrongly estimated $R$. 
            
            Fig.~\ref{fig:emitter_misestimation} shows the performance of DL-methods when $R$ is under or over-estimated. 
            Specifically, we test the algorithms under the case where $R=7$, but various wrongly estimated $\widehat{R}$'s ranging from 5 to 9 are used for the proposed algorithms. 
            The result of \texttt{DeepComp} is also presented as a baseline, which does not use $R$ and thus is a constant over all the cases.
            One can see that if $\widehat{R}<R$, then the SREs of the proposed algorithms tend to be high. Nonetheless, over-estimated $\widehat{R}$'s (i.e., $\widehat{R}\geq R$) seem to essentially not affect the SRE. 
            This phenomenon is understandable: under-estimated $\widehat{R}$'s lead to significant missing components, but over-estimated $\widehat{R}$'s may only introduce repeated or redundant components that may not affect the reconstruction.

            \subsubsection{Training-Testing Mismatch}
            It is natural to ask what happens if the shadowing environment in reality turned out to be very different from the ones that were using to train our neural models. To simulate this condition, we provide experiments where the testing scenarios are not covered in the training set of the neural models. Specifically, we train a set of neural models on simulated data generated from a limited range of shadowing parameters and test it on data generated from a wider range of shadowing parameters. 
            
            First, we use a narrow range of shadowing parameters, i.e., $\eta \sim {\rm Uniform}([3,5])$ and ${\rm d}_{\rm corr} \sim {\rm Uniform}([50, 60])$ to generate training data. The learned models are categorized as ``narrow range parameters-learned (NRP) models''. Second, we use a wider range of shadowing parameters, i.e., $\eta \sim {\rm Uniform}([3,8])$ and ${\rm d}_{\rm corr} \sim {\rm Uniform}([30, 100])$, to generate data to train another set of neural models. We categorize this set of models as ``wide range parameters-learned (WRP) models''. 
            We test the ``out-of-range'' cases (i.e., the parameters used for generating test data are out of the range of the parameters used in the training data) for both $\eta$ and $d_{\rm corr}$ under the NRP models. The results under the WRP models are used as benchmarks. The results are averaged over 30 trials. 
            
            Table~\ref{tab:outrange_eta} shows the SREs attained under various $\eta$ with $\rho=10\%, R=5, {\rm d}_{\rm corr} = 50$, and $M=20$. One can see that the NRP models admit a mismatch with the test data. One can see that \texttt{DeepComp} and \texttt{Nasdac} seem to have performance degradation when using NRP models. But \texttt{DowJons} is much more resilient to the training-testing model mismatch. This is perhaps because \texttt{DowJons} is an all-at-once optimization-based estimator, which is often more robust to modeling errors.
            
           Table~\ref{tab:outrange_dcorr} shows SREs attained under various ${\rm d}_{\rm corr}$ with $\rho=10\%, R=5, \eta=6, M=20$. Similar to the previous case, \texttt{DeepComp} and \texttt{Nasdac} suffer performance degradation but \texttt{DowJons} is robust to the model mismatch.

\begin{table}[t]
    \centering
    {
   
        \caption{ SRE under various $\eta$ for training-testing data mismatch case. $R=5, {\rm d}_{\rm corr}=50, \rho=10\%$.}  \label{tab:outrange_eta}
    \resizebox{\columnwidth}{!}{%
    \begin{tabular}{|c|c|c|c|c|c|c|c|}
        \hline
        & & \multicolumn{3}{|c|}{WRP models} & \multicolumn{3}{|c|}{NRP models} \\ \cline{3-8}
         $\eta$ & \texttt{LL1} & \texttt{DeepComp} & \texttt{Nasdac} & \texttt{DowJons} & \texttt{DeepComp} & \texttt{Nasdac} & \texttt{DowJons}   \\ \hline
   6 &  0.1521  &  0.1259  &  0.1139  &  \textbf{0.0954}  &  { 0.1610}  &  { 0.1416}  &  { 0.1005}  \\ \hline
    7 &  0.1533  &  0.1746  &  0.1247  &  0.1412  &  { 0.2007}  &  { 0.1604}  &  \textbf{{ 0.1337}}  \\ \hline
    8 &  0.2279  &  0.1861  &  0.1602  &  0.1549  &  { 0.2171}  &  { 0.2022}  &  \textbf{{ 0.1499}}  \\ \hline
    \end{tabular}
    }
    }
   
\end{table}

\begin{table}[t]
    \centering
    {
        \caption{SRE under various ${\rm d}_{\rm corr}$ for training-testing data mismatch case. $R=5, \eta=6, \rho=10\%$.}
    \label{tab:outrange_dcorr}
    \resizebox{\columnwidth}{!}{%
    \begin{tabular}{|c|c|c|c|c|c|c|c|}
        \hline
        & & \multicolumn{3}{|c|}{WRP models} & \multicolumn{3}{|c|}{NRP models} \\ \cline{3-8}
         ${\rm d}_{\rm corr}$ & \texttt{LL1} &\texttt{DeepComp} & \texttt{Nasdac} & \texttt{DowJons} & \texttt{DeepComp} & \texttt{Nasdac} & \texttt{DowJons}   \\ \hline
                    30 & 0.1471 & 0.1799 &  0.1408  &  \textbf{0.1316}  &  { 0.2101}  &  { 0.1847}  &  { 0.1382}  \\ \hline
                    70 & 0.1271 & 0.1227 &  0.1290  &  0.0828  &  { 0.1472}  &  { 0.1629}  &  \textbf{{ 0.0803}}  \\ \hline
                    90 & 0.1137 & 0.1060 &  0.1217  &  0.0830  &  { 0.1345}  &  { 0.1460}  & \textbf{{ 0.0746}}  \\ \hline
        \end{tabular}}}
\end{table}
        
        \subsection{Real Data Experiment}\label{sec:real_data_exp}
        In this subsection, we test the algorithms using real data.
            \subsubsection{Data Description} 
\begin{figure}[t!]
		        \centering
		        \includegraphics[width=0.7\linewidth]{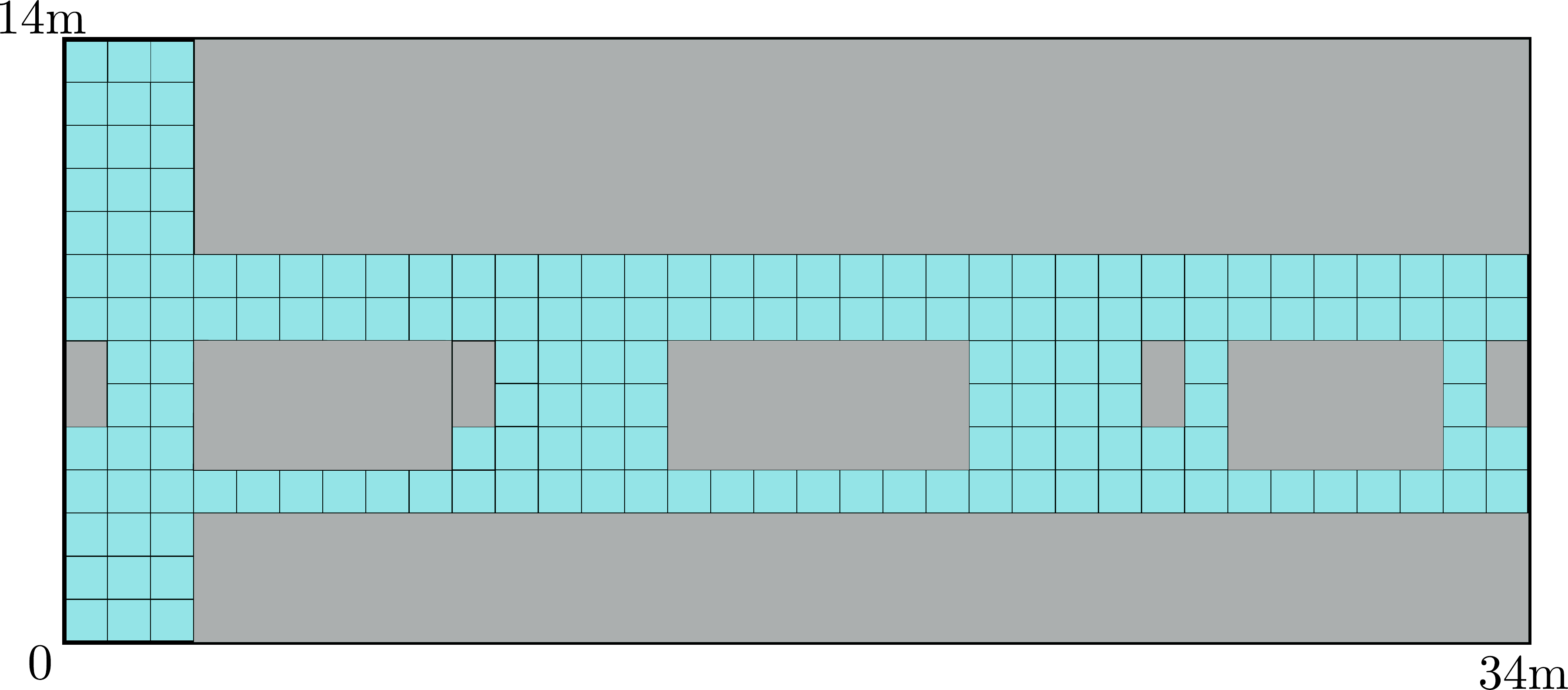}
		        \caption{Geometry of the $14 \times 34 {\rm m}^2$ indoor region according to \cite{king2008crawdad}. The region is divided into grids that have sizes of $1 \times 1 {\rm m}^2$. Measurements were taken at the center of each blue shaded cell. The dark shaded region are unobserved regions and contain rooms and pillars.}
		        \label{fig:floor_plan}
		    \end{figure}

\begin{figure}
		        \centering
		        \includegraphics[width=\linewidth]{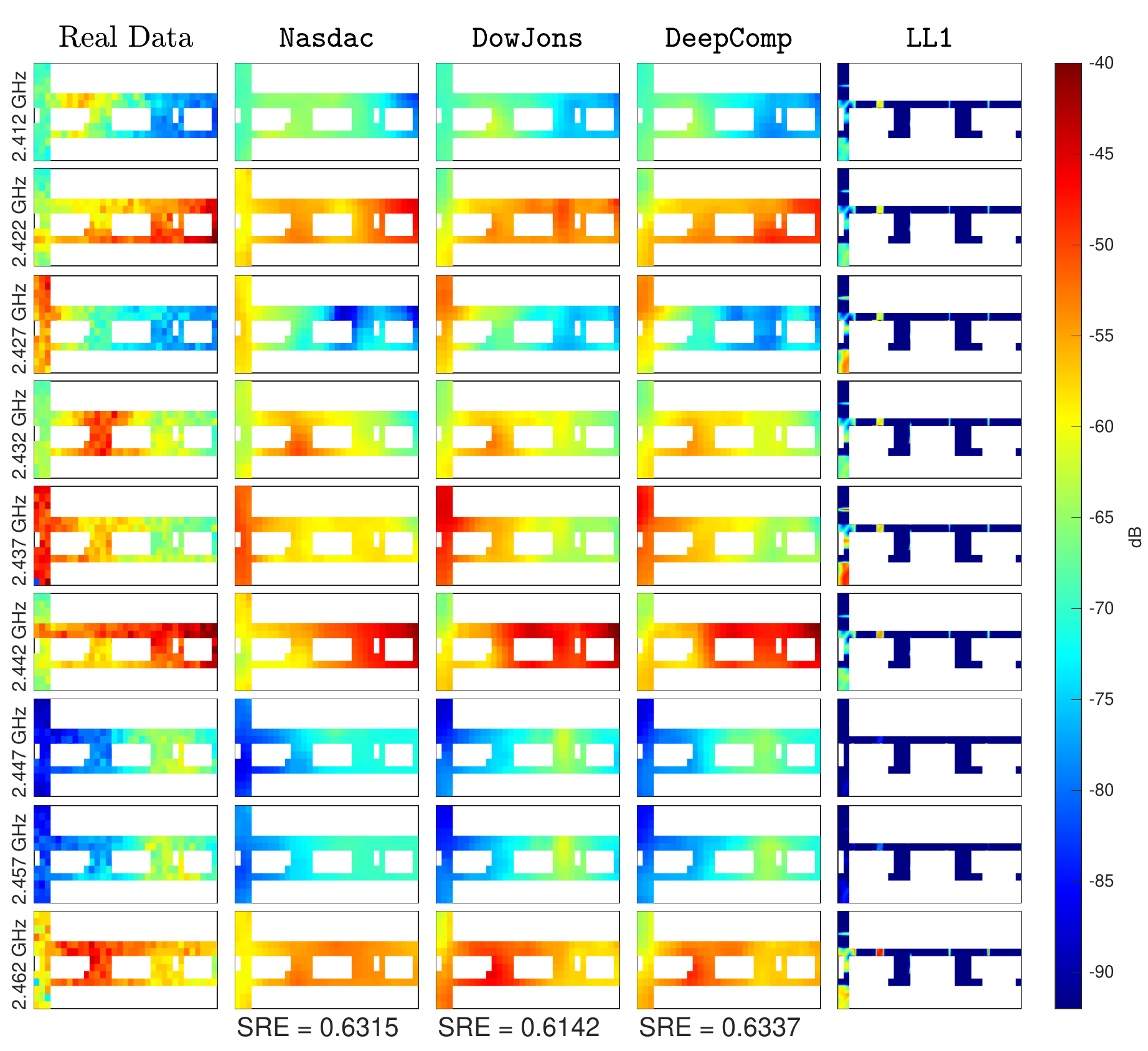}
		        \caption{ First Column: 166 ground-truth measurements covering the hallway for all frequency bands. Second, third, fourth, and fifth columns: reconstruction by \texttt{Nasdac},  \texttt{DowJons}, \texttt{DeepComp}, and \texttt{LL1}, respectively, across the hallway.}
		        \label{fig:real_data_severe_subsample}
		    \end{figure}

        The data was collected in Mannheim University in an office floor across 9 frequency bands \cite{king2008crawdad}. The frequency bins are centered at $2.412$GHz, $2.422$GHz, $2.427$GHz, $2.432$GHz, $2.437$GHz, $2.442$GHz, $2.447$GHz, $2.457$GHz, and $2.462$GHz, respectively. The indoor region is a $14 \times 34$m$^2$ area. The area is divided into $1 \times 1$m$^2$ grids. Sensors are placed at the center of 166 of those grid cells, denoted by blue shaded cells in Fig.~\ref{fig:floor_plan}, which roughly cover the hallway. The dataset constrains more than 400 observations for each (frequency, location) pair. We take the median as the observation for each (frequency, location) pair and obtain an incomplete radio map. All 9 slabs of the radio map are shown in left column of Fig \ref{fig:real_data_severe_subsample}, each corresponding to one frequency (with the top being 2.412GHz and the bottom being 2.462GHz).

            \subsubsection{Training Sample Generation}
            Since the region of interest is a narrow hallway in an indoor environment, we generate training samples with large $\eta$'s and small ${\rm d}_{\rm corr}$'s. Specifically, every training sample uses $\eta \in[8, 12]$ and ${\rm d}_{\rm corr} \in[5, 50]$, which are both sampled uniformly at random.
            The training samples cover a $14 \times 34$m$^2$ region that is the same as floor size. Note that emitters located outside the $14 \times 34$ region could also affect the measurements. In order to account for this, we let the emitters be located anywhere in $200 \times 200$m$^2$ region that contains the $14 \times 34$m$^2$ region at its center. We extract the SLF induced by the emitter in the $14 \times 34$m$^2$ region as our training example. The number of training examples and $\rho$ are set to be the same as in the synthetic data experiments (see section \ref{sec:exp_synthetic}). For \texttt{DeepComp}, each emitter use up to 4 frequency bands, which empirically ensures the existence of at least one emitter in each frequency slab. The number of emitters in the training examples are set to be the same as in the synthetic data experiments.
		    
		    \subsubsection{Result}
		    Since $R$ is unknown, we estimate $R$ using an existing model-order selection algorithm, namely, the \texttt{HySime} algorithm that was introduced for estimating the dimension of the signal subspace under noisy NMF models \cite{bioucas2008hyperspectral}. Using \texttt{HySime} on the measurements and the re-arranged NMF model \eqref{eq:factorization}, we obtain $\widehat{R}=7$. We use the estimated $\widehat{R}$ for \texttt{Nasdac}, \texttt{DowJons}, and \texttt{LL1}. For \texttt{LL1}, we set $L=4$ based on the visual reconstruction quality across the floor.
		    Fig. \ref{fig:real_data_severe_subsample} shows results when 8 random samples are used. One can see that the DL-based approaches, namely, \texttt{DowJons}, \texttt{Nasdac}, and \texttt{DeepComp} work better than the tensor model-based method \texttt{LL1}. This is understandable, since the low-rank model hinged on by \texttt{LL1} may be grossly violated in such an indoor environment. On the other hand, the DL-based methods may be better suited for such scenarios.
		    
	    Note that using 8 locations to recover the radio map presents an extremely challenging tensor completion problem, and the proposed methods output satisfactory performance, which is quite encouraging and shows the positive prospects of learning based methods in real world problems. The performance is even more promising when 16 locations are used; see Appendix \ref{app:real_data_supplement}.
	    Finally, using 166 ground-truth measurements in the first column, we evaluate the SRE for the DL-based methods---the SREs produced by \texttt{DowJons}, \texttt{Nasdac}, and \texttt{DeepComp} are $0.6142$,  $0.6315$, and $0.6337$, respectively.
	    
	    Table \ref{tab:real_data_rho} shows the SREs attained by DL-based methods under various $\rho$'s. The result is averaged over 30 random trials. We used the estimated $\widehat{R}=7$ again. One can see that that \texttt{DowJons} performs consistently better than other methods for all $\rho$'s.
	    
        \begin{table}[t]
            \centering
            \caption{SRE attained by DL-based methods on real data for various $\rho$.}
            \begin{tabular}{|c|c|c|c|}
                \hline
                 $\rho$ & \texttt{DeepComp} & \texttt{Nasdac} & \texttt{DowJons} \\ \hline
                 5\%  & 0.5459 & 0.5326 & \textbf{0.5080}  \\ \hline
                 10\% & 0.3985 & 0.4035 & \textbf{0.3876} \\ \hline
                 15\% & 0.3513 & 0.3605 & \textbf{0.3452} \\ \hline
                 20\% & 0.3046 & 0.3139 & \textbf{0.2936} \\ \hline
            \end{tabular}
            \label{tab:real_data_rho}
        \end{table}

	\section{Conclusion}
	In this work, a DL framework for blind spectrum cartography was proposed.
	Different than the existing DL-based SC methods that directly learn a data completion network for the observed incomplete radio map, our framework learns a deep network for completing the {\it individual} SLFs of the emitters.
	This way, both the offline training cost and the online prediction/completion error can be substantially reduced.
	A fast NMF-based data desegregation approach was proposed to extract the individual SLFs from the sensor measurements, which are used as inputs of the learned completion network. 
	This two-stage NMF-DL approach is fast and lightweight in the prediction/completion stage. However, it only works when the spectral occupancy of the emitters is relatively sparse, and it suffers from error propagation as other two-stage approaches.
	To enhance performance, a one-shot optimization-based data completion criterion was proposed, using the learned deep model for the SLFs as structural constraints. This approach was shown to guarantee the recoverability of the radio map tensor under more relaxed conditions relative to the NMF-DL approach---even if noise is present. Synthetic and real data experiments corroborate our design and recovery theory.
	
	\appendices
    \section{Neural Network Setting}\label{app:DNN}
    
\begin{figure}[h!]
    		\centering
    		\includegraphics[width=8.5cm]{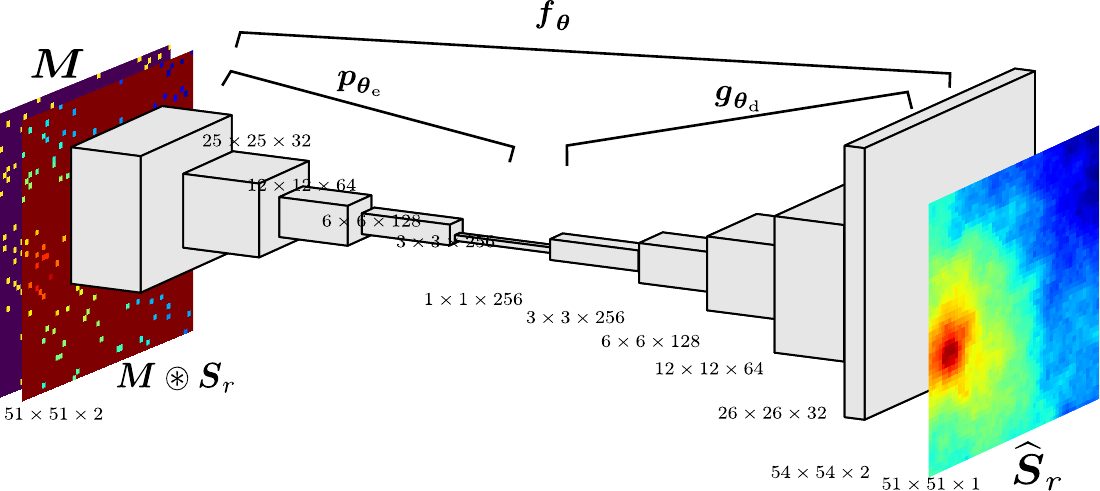}
    		\caption{The autoencoder network used in our proposed algorithms.}
    		\label{fig:autoencoder}
    	\end{figure}

    Table~\ref{tab:autoencoder} and Fig.~\ref{fig:autoencoder} offer details of the neural network architecture that is used in the experiments.
    
    In Table~\ref{tab:autoencoder}, ``Conv." and ``DeConv." stand for the convolutional and transposed convolutional operations, respectively. 
    In \cite{teganya2020data}, \texttt{DeepComp} essentially recovers the radio map frequency by frequency. Hence, the same network structure for recovering the SLFs in our method can be used in \texttt{DeepComp} for recovering the 2D radio maps, i.e., $\tX(:,:,k)$'s, one by one.

    	\begin{table}[h!]
    		\centering
    		\caption{Deep Completion Auto-encoder architecture. All activations are SELU except the last which is sigmoid. Appropriate padding is done to get desired output dimension.}
    		\begin{tabular}{c|c|c|c|c}
    			\textbf{SN} & \textbf{Layer}          & \textbf{Filter} & \textbf{\#Channels} & \textbf{Stride} \\ \hline \hline
    			1  & Conv.          & 4 $\times$ 4  & 32        & 2      \\ \hline
    			2  & Conv.          & 4 $\times$ 4  & 64        & 2      \\ \hline
    			3  & Conv.          & 4 $\times$ 4  & 128       & 2      \\ \hline
    			4  & Conv.          & 4 $\times$ 4  & 256       & 2      \\ \hline
    			5  & Conv.          & 3 $\times$ 3  & 256       & 1      \\ \hline
    			6  & DeConv.        & 3 $\times$ 3  & 256       & 1      \\ \hline
    			7  & DeConv.        & 4 $\times$ 4  & 128       & 2      \\ \hline
    			8  & DeConv.        & 4 $\times$ 4  & 64        & 2      \\ \hline
    			9  & DeConv.        & 4 $\times$ 4  & 32        & 2      \\ \hline
    			10 & DeConv.        & 4 $\times$ 4  & 2         & 2      \\ \hline
    			11 & Conv2d.        & 4 $\times$ 4  & 1         & 1      \\ \hline
    		\end{tabular}
    		\label{tab:autoencoder}
    	\end{table}
    		
    \section{Proof of Theorem~\ref{thm:nmf}}\label{app:thmnmf}
    Note that if $\S\in \mathbb{R}^{R\times IJ}$ is drawn from any joint absolutely continuous distribution, we have
    \[  {\sf Pr}\left({\rm rank}(\S(:,\bm \varOmega_{\rm col}))=R\right)=1,  \]
    under the assumption that $$|\bm \varOmega_{\rm col} |=|\bm \varOmega|\geq R.$$
    In addition, if Assumption~\ref{ass:dominance} is satisfied, ${\rm rank}(\C)=R$ since a scaled permutation matrix is a submatrix of $\bm C$.
    Therefore, $\bm G=\C\H$ is an NMF model with $\C$ and $\H$ being full column and full row rank, respectively. Under Assumption~\ref{ass:dominance}, $\bm C$ satisfies the separability condition. Hence, by standard identifiability analysis of separable NMF  \cite{fun2019nonnegative, fu2018identifiability, gillis2013fast, gillis2020nonnegative, fu2014self}  (particularly, \cite[Sec. III-IV]{fun2019nonnegative}), we reach the conclusion; see an example of separable NMF identification in Appendix~\ref{app:spa}.

    \bibliographystyle{IEEEtran}
    \bibliography{main}

    \begin{IEEEbiography}[{\includegraphics[width=1in,height=1.25in,clip,keepaspectratio]{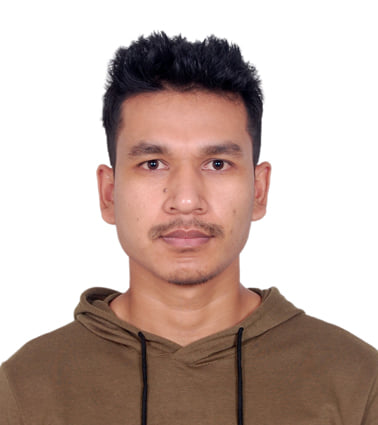}}]
	{Sagar Shrestha} received his B.Eng. in Electronics and Communication Engineering from Pulchowk Campus of Tribhuvan University, Kathmandu, Nepal, in 2016. He is currently working towards a Ph.D. degree at the School of Electrical Engineering and Computer Science, Oregon State University, Corvallis, OR, USA. His current research interests are in the
    broad area of statistical machine learning and signal
    processing.
    \end{IEEEbiography}
    \begin{IEEEbiography}[{\includegraphics[width=1in,height=1.25in,clip,keepaspectratio]{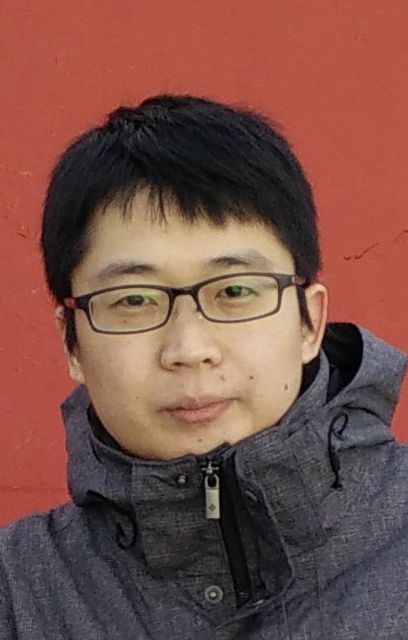}}]
        {Xiao Fu} (Senior Member, IEEE) received the B.Eng. and MSc. degrees from the University of Electronic Science and Technology of China (UESTC), Chengdu, China, in 2005 and 2010, respectively. He received the Ph.D. degree in
        Electronic Engineering from The Chinese University
        of Hong Kong (CUHK), Shatin, N.T., Hong Kong, in 2014. He was a Postdoctoral Associate with the Department of Electrical
        and Computer Engineering, University of Minnesota,
        Minneapolis, MN, USA, from 2014 to 2017. 
        Since 2017, he has been an Assistant Professor with the School of Electrical
        Engineering and Computer Science, Oregon State
        University, Corvallis, OR, USA. 
        His research interests include the broad area of signal
        processing and machine learning. 
        
        Dr. Fu received a Best Student Paper Award at ICASSP 2014, and was a recipient of the Outstanding Postdoctoral Scholar Award at University of Minnesota in 2016.
        His coauthored papers that received Best Student Paper Awards from IEEE CAMSAP 2015 and IEEE MLSP 2019, respectively. 
        He serves as a member of the Sensor Array and Multichannel Technical Committee  (SAM-TC) of the IEEE Signal Processing Society (SPS).
        He is also a member of the Signal Processing for Multisensor Systems Technical Area Committee (SPMuS-TAC) of EURASIP. He is the Treasurer of the IEEE SPS Oregon Chapter. He serves as an Editor of {\sc Signal Processing}. He was a tutorial speaker at ICASSP 2017 and SIAM Conference on Applied Linear Algebra 2021.
    \end{IEEEbiography}

    \begin{IEEEbiography}[{\includegraphics[width=1in,height=1.25in,clip,keepaspectratio]{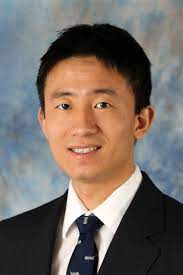}}]
       {Mingyi Hong}(Senior Member, IEEE) received the Ph.D. degree from the University of Virginia, Charlottesville, in 2011. He is an Associate Professor with the Department of Electrical and Computer Engineering, the University of Minnesota, Minneapolis. He is on the IEEE Signal Processing for Communications and Networking Technical Committee, and he is an Associate Editor for IEEE TRANSACTIONS ON SIGNAL PROCESSING. His research interests include optimization theory and applications in signal processing and machine learning.
    \end{IEEEbiography}
    


\clearpage

\begin{center}
{\bf Supplementary Material of ``Deep Spectrum Cartography: Completing Radio Map Tensors Using Learned Neural Models''}\\ S. Shrestha, X. Fu, and M. Hong    
\end{center}
		
    \section{Convergence Properties of \texttt{DowJons}}\label{app:propdowjons}
    We briefly discuss some convergence properties of \texttt{DowJons} in this appendix.
    For simplicity, we consider $\overline{\nabla}_{\bm Z} f(\bm Z,\bm C)={\nabla}_{\bm Z} f(\bm Z,\bm C)$, i.e., the plain vanilla gradient descent approach is used for updating $\bm Z$.  Our proof can be generalized to cover cases where stochastic gradient is used, e.g., by combining with proof ideas from fiber sampling-based stochastic tensor decomposition \cite{fu2020block,fu2020computing,battaglino2018practical}.
    Under this setting, we have the following result:
    
    \begin{proposition}\label{prop:dowjons}
    	Assume that the learned generative model $\bm g_{\bm \theta_{\rm d}}(\cdot)$ satisfies  $\|\bm g_{\bm \theta_{\rm d}}(\bm z_r)\|_2\leq \beta$ for all possible $\bm z_r$, that $\bm H^{(k)}$ is full rank for all $k$,
    	and $\|\bm z_r^{(k)}\|_2$ is uniformly upper bounded by $\alpha$ for all $k$.
    	Also assume that  at a given iteration $k$, the following holds:
    		    \begin{subequations}
    		     \begin{align}
    		  &\|\nabla_{\Z} f(\Z^{(k+1)},\C^{(k)}) \|\le \epsilon, \; \forall~k \label{eq:app:opt}\\
    		  & f(\Z^{(k+1)}, \C^{(k)}) \le f(\Z^{(k)}, \C^{(k)}),\; \forall~k \label{eq:descent}\\ 
    		  &\C^{(k+1)}  =\arg\min_{\C} f(\Z^{(k+1)}, \C),\; \forall~k.\label{eq:C}
    		    \end{align}
            \end{subequations}
    Then, the solution sequence produced by the \texttt{DowJons} algorithm converges to an $\epsilon$-stationary  point of \eqref{eq:joint_opt}.
    \end{proposition}
    
            First, if $\bm H^{(k)}$ is full rank, then the subproblem w.r.t. $\bm C$ is always strongly convex.
    		Hence, the following inequality holds with a certain $\gamma>0$:
    		\begin{align*}
    		    f(\Z^{(k+1)},\C^{(k+1)})\le f(\Z^{(k+1)},\C^{(k)}) - \frac{\gamma}{2}\|\C^{(k)}-\C^{(k+1)}\|^2.
    		\end{align*}
    		From \eqref{eq:descent} -- \eqref{eq:C}, it is clear that the objective is always decreasing. Since it is lower bounded, we know that the objective value converges, and that 
    		\begin{align}
    		    \|\C^{(k)}-\C^{(k+1)}\|^2\to 0.
    		\end{align}
    		Further, since both $\bm z_r^{(k)}$ and $\bm g_{\bm \theta_{\rm d}}(\bm z_r^{(k)})$ are bounded, it means that there exists a constant $\overline{L}<\infty$ such that $\nabla_{\Z}f(\Z, \C)$ is $\overline{L}$-Lipschitz continuous in $\C$ that uniformly holds for all iterations.
    		Following this argument, one can see the fillowing:
    		\begin{align*}
    		    &\|\nabla_{\Z}f(\Z^{(k+1)}, \C^{(k+1)})\|^2  \le  2\|\nabla_{\Z}f(\Z^{(k+1)}, \C^{(k)})\|^2\nonumber\\
    		    &\quad +2\|\nabla_{\Z}f(\Z^{(k+1)}, \C^{(k+1)})- \nabla_{\Z}f(\Z^{(k+1)}, \C^{(k)})\|^2\nonumber \\
    		    & \le 2 \epsilon^2 + \overline{L}^2 \|\C^{(k+1)}-\C^{(k)}\|^2
    		\end{align*}
    		Passing limit to both sides, we obtain
    		\begin{align*}
    		 \lim_{k\to\infty} \|\nabla_{\Z}f(\Z^{(k+1)}, \C^{(k+1)})\|^2  \le 2\epsilon^2.
    		\end{align*}
    		Further, from \eqref{eq:C}, we have
    		\begin{align*}
    		  \|\nabla_{\C}f(\Z^{(k+1)}, \C^{(k+1)})\|^2  =0, \; \forall~k.
    		\end{align*}
    		Combining the above two relations, we obtain:
    			\begin{align*}
    		 \lim_{k\to\infty} \|\nabla f(\Z^{(k+1)}, \C^{(k+1)})\|^2  \le 2\epsilon^2.
    		\end{align*}
    		This concludes the proof.

\section{Proof of Lemma \ref{lem:cover}}\label{app:prooflemma}
We first show the following lemma:
\begin{lemma}	\label{lemma:nn-cover}
	Let $\g_{\bm \theta_{d}}$, ${\cal X}_{R, \bm g_{\bm \theta_{\rm d}}}$, and ${\cal Z}\subseteq\mathbb{R}^D$ be defined as in Assumption~\ref{ass:existence}.  Also define ${\cal G}_{\bm \theta_{d}} = \{\g_{\bm \theta_{d}}(\z)~|~\forall \z \in {\cal Z}\subseteq\mathbb{R}^D \}$. Then, the following upper bound holds for the set ${\cal G}_{\bm \theta_{d}} $:  
	\begin{equation*}
		{\sf N}\left({\cal G}_{\bm \theta_{d}},  \frac{\varepsilon}{\sqrt{R}}\right) \leq \left(\frac{3Pq\sqrt{R}}{\varepsilon}\right)^D,
	\end{equation*}
	where $R$ is the number of emitters as before and $P$ is also defined in Assumption~\ref{ass:existence}.
\end{lemma}

\begin{IEEEproof}
	For any $\z, \z' \in {\cal Z}$, the following chain of inequalities holds:
	\begin{align}
		& \|\g_{\bm \theta_{d}}(\z) - \g_{\bm \theta_{d}} (\z') \|_{\rm F} \nonumber \\
		& =\left \| {\rm mat} ({\bm \zeta}_L (\A_L \ldots {\bm \zeta}(\A_1\z))) - {\rm mat}({\bm \zeta}_L(\A_L \ldots {\bm \zeta}(\A_1 \z'))) \right\|_{\rm F} \nonumber \\
		& = \| {\bm \zeta}_L (\A_L \dots {\bm \zeta}(\A_1 \z)) - {\bm \zeta}_L(\A_L \dots {\bm \zeta}(\A_1 \z')) \|_2 \nonumber \\
        & \leq \phi_L \| \A_L ({\bm \zeta}_{L-1} \dots {\bm \zeta}_1(\A_1 \z) - {\bm \zeta}_{L-1} \dots {\bm \zeta}_1(\A_1 \z'))\|_2 \nonumber \\
        & \leq \phi_L \| \A_L \|_2 \|{\bm \zeta}_{L-1} \dots {\bm \zeta}_1(\A_1 \z) - {\bm \zeta}_{L-1} \dots {\bm \zeta}_1(\A_1 \z')\|_2 \nonumber \\
        & \leq \prod_{i=1}^L \phi_i \|\A_i\|_2 \|\z - \z' \|_2 \leq P \|\z - \z '\|_2, \label{eq:Penet}
	\end{align}
	where we have used $ P =\prod_{i=1}^L \phi_i \|\A_i\|_2 $.
	
	Since ${\cal Z}$ is an Euclidean ball that has a radius of $q$ in the $D$-dimensional space, we know its covering number is upper bounded by the following \cite{pollard1990empirical}:
	\begin{equation*}
		{\sf N}\left({\cal Z},  \frac{\varepsilon}{P}\right) \leq \left(\frac{3Pq}{\varepsilon}\right)^D .
	\end{equation*}
	As shown in \eqref{eq:Penet}, an
	$\frac{\varepsilon}{P}$-net of ${\cal Z}$ is an $\varepsilon$-net of ${\cal G}_{\bm \theta_{d}}$.
	Hence, we have
	\begin{align*}
		{\sf N}({\cal G}_{\bm \theta_{d}},  \varepsilon) & \leq \left(\frac{3Pq}{\varepsilon}\right)^D,
	\end{align*}
	which also implies that the following holds:
	\begin{align*}
		{\sf N}\left({\cal G}_{\bm \theta_{d}},  \frac{\varepsilon}{\sqrt{R}}\right) & \leq \left(\frac{3Pq\sqrt{R}}{\varepsilon}\right)^D . 
	\end{align*}
	This completes the proof.
\end{IEEEproof}

Now we continue the proof of Lemma~\ref{lem:cover}. Consider the set ${\cal S}_{R} = \{\S \in \bbR^{R \times IJ}| \S=\begin{bmatrix}
	\s_1 \dots \s_R
\end{bmatrix}^{\T}, {\rm mat}(\s_r) = \g_{\bm \theta_{d}}(\z_r), \z_r \in {\cal Z}, r\in[R]\}$.
Let $\overline{\cal G}_{\bm \theta_{d}}$ denote an $\varepsilon/\sqrt{R}$-net of ${\cal G}_{\bm \theta_{d}}$.
 Then, for any $\S \in \cS_R$, we can construct a $\oS=[\os_1,\ldots,\os_R]^\T$ where ${\rm mat}(\os_r) \in \overline{\cal G}_{\bm \theta_{d}}$ for $r \in [R]$, such that the following holds:
\begin{align*}
	\|\S - \oS\|_{\rm F} = &\left(\sum_{i=1}^{R} \|\s_i - \os_i \|_2^2\right)^{1/2} \\
	\leq & \left( R \left(\frac{\varepsilon}{\sqrt{R}} \right)^2\right)^{1/2} \\
	 = & \varepsilon .
\end{align*}
 Hence, the set $\{ \oS=[\os_1,\ldots,\os_R]^\T : \os_r \in \overline{\cal G}_{\bm \theta_{d}}, r \in [R] \} $ is an $\varepsilon$-net of $\cS_R$. Note that the cardinality of such set is upper bounded by $({\sf N}({\cal G}_{\bm \theta_{d}},  \nicefrac{\varepsilon}{\sqrt{R}}))^R$, using Lemma~\ref{lemma:nn-cover}. Therefore, the following upper bound holds:
\begin{align*}
	{\sf N}({\cal S}_{R},  \varepsilon) & \leq \left(\frac{3Pq\sqrt{R}}{\varepsilon}\right)^{RD}.
\end{align*}

The above also means that
\begin{equation}
	{\sf N}\left({\cal S}_{R},  \frac{\varepsilon}{\sqrt{R}(\alpha+ \beta)}  \right) \leq \left(\frac{3PRq(\alpha+\beta)}{\varepsilon}\right)^{RD}.
	\label{eq:scover}
\end{equation}
 Next, we derive a bound for ${\sf N}(\cC_R,\varepsilon)$, where ${\cal C}_{R}= \{\C \in \bbR^{K \times R}~|~\C = \begin{bmatrix}
	\c_1 \dots \c_R
\end{bmatrix}, \|\c_i\|_2 \leq \alpha, \c_i \geq \zero,~i\in[R]\}$. First, the cardinality of  $\varepsilon/\sqrt{R}$-net used to ``cover'' the set $\{\c_r \in \bbR^K | \|\c_i\|_2 \leq \alpha, \c_i \geq \zero, i \in [R] \}$ can be upper bounded by $( \nicefrac{3 \sqrt{R}\alpha}{\varepsilon})^K$. 
 Then, following the same argument used to ``cover'' $\cS_R$, one can ``cover'' the set ${\cal C}_{R}$ using an $\varepsilon$-net with a cardinality of 
\begin{align*}
	{\sf N}({\cal C}_{R},  \varepsilon) & \leq \left(\frac{3\sqrt{R}\alpha}{\varepsilon}\right)^{RK},
\end{align*}
which leads to the following upper bound:
\begin{equation}
	{\sf N}\left({\cal C}_{R},  \frac{\varepsilon}{\sqrt{R}(\alpha+ \beta)}\right) \leq \left(\frac{3R\alpha(\alpha+\beta)}{\varepsilon}\right)^{RK} .
	\label{eq:ccover}
\end{equation}

Let $\overline{\cal C}_{R}$ and $\overline{\cal S}_{R}$  be an $\frac{\varepsilon}{\sqrt{R}(\alpha+ \beta)}$-net of ${\cal C}_{R}$ 
and an $\frac{\varepsilon}{\sqrt{R}(\alpha+ \beta)}$-net of ${\cal S}_{R}$, respectively.
We define a set 
$\overline{\cal X}_{R, \bm g_{\bm \theta_{\rm d}}} = \{ \otX = \sum_{r=1}^R {\rm mat}(\os_r) \circ \oc_r ~|~ [\os_1, \dots, \os_R]^\T \in \ocS_R, [\oc_1, \dots, \oc_R] \in \overline{\cC}_R, r \in [R]\}$.
Hence, the cardinality of $\overline{\cal X}_{R, \bm g_{\bm \theta_{\rm d}}}$ is upper bounded by
\begin{equation}\label{eq:final_cover_bound}
 \left| \overline{\cal X}_{R, \bm g_{\bm \theta_{\rm d}}}  \right| \leq  \left(\frac{3R(\alpha + \beta)}{\varepsilon}\right)^{R(K+D)} \alpha^{RK} (Pq)^{RD},
\end{equation}
where the right hand side is derived from the product of upper bounds in \eqref{eq:scover} and \eqref{eq:ccover}.
Let $\widetilde{\X}$ and $\oX$ be the matrix unfoldings of $\widetilde{\tX} \in {\cal X}_{R, \bm g_{\bm \theta_{\rm d}}}$ and $\otX \in \overline{\cal X}_{R, \bm g_{\bm \theta_{\rm d}}}$, respectively.
Note that the matrix unfoldings $\widetilde{\X}$ and $\oX$ can be represented as $\widetilde{\X} = \widetilde{\C} \widetilde{\S}, ~ \widetilde{\C} \in \cC_R, \widetilde{\S} \in \cS_R$ and $\oX = \oC \oS, ~ \oC \in \overline{\cC}_R, \oS \in \ocS_R$, respectively.
Hence, for any $\widetilde{\X}$, there exists $\oX\in\overline{\cal X}_{R, \bm g_{\bm \theta_{\rm d}}}$ such that the following holds:
\begin{align*}
	\|\widetilde{\X} - \oX\|_{\rm F} = & \|\widetilde{\C} \widetilde{ \S} - \oC \oS \|_{\rm F} \\
	= & \|\widetilde{\C} \widetilde{ \S} - \widetilde{\C} \oS + \widetilde{\C} \oS - \oC \oS \|_{\rm F} \\
	= & \|\widetilde{\C} ( \widetilde{ \S} - \oS) + (\widetilde{\C}  - \oC)\oS \|_{\rm F} \\
	\leq &  \|\widetilde{\C} \|_{\rm F} \|\widetilde{ \S}  - \oS\|_{\rm F} + \|\oS\|_{\rm F} \|\widetilde{\C} -\oC\|_{\rm F} \\
	\leq & \sqrt{R} \alpha \frac{\varepsilon}{\sqrt{R}(\alpha+ \beta)} + \sqrt{R} \beta \frac{\varepsilon}{\sqrt{R}(\alpha+ \beta)} \\
	= & \varepsilon .
\end{align*}
Therefore, $\overline{\cal X}_{R, \bm g_{\bm \theta_{\rm d}}}$ is an $\varepsilon$-net of ${\cal X}_{R, \bm g_{\bm \theta_{\rm d}}}$ with a covering number bounded by
\begin{equation*}
	{\sf N} \left( {\cal X}_{R, \bm g_{\bm \theta_{\rm d}}},  \varepsilon\right) \leq \left(\frac{3R(\alpha + \beta)}{\varepsilon}\right)^{R(K+D)} \alpha^{RK} (Pq)^{RD},
\end{equation*}
which follows from \eqref{eq:final_cover_bound}.

\section{Proof of Lemma \ref{lem:uc}}\label{app:uc}
To show the lemma, we adopt a proof technique that is often utilized in conventional matrix completion \cite{wang2012stability}, and adapt it to serve for tensor sampling-based completion with a deep generative model as the structural constraint.
The idea is to leverage the Sterfling's sampling-without-replacement extension of the Hoeffding's inequality
\cite{serfling1974probability}:
\begin{lemma}
	Let $X_1, X_2, \dots, X_w$ be a set of samples taken without replacement from $\{x_1, x_2, \dots, x_n\}$ of mean $\mu$. Denote $a= \min_i x_i$ and $b = \max_i x_i$. Then
	\begin{align*}
		& {\sf Pr} \left[ \left| \frac{1}{w} \sum_{i=1}^{w} X_i - \mu \right| \geq t \right] \\
		& \leq 2 \exp\left( - \frac{2wt^2}{(1-(w-1)/n)(b-a)^2}\right) .
	\end{align*}
	\label{lemma:hoeffding}
\end{lemma}
Consider a set of variables
$$\left\{ W_{ij} = \frac{1}{K}\left\|\tY(i,j,:) - \widetilde{\tX}(i,j,:)\right\|_2^2, \forall i \in [I], \forall j\in[J] \right\}.$$
Then, the sample mean of $W_{ij}$'s using $(i,j)\in\bm \varOmega$ is the empirical loss $\widehat{\sf Loss}(\widetilde{\tX})$ given by
\begin{align*}
	\widehat{\sf Loss}(\widetilde{\tX}) = & \frac{1}{|\bOmega|K} \sum_{i,j \in \bOmega} \left\|\tY(i,j,:) - \widetilde{\tX}(i,j,:) \right\|_2^2.
\end{align*} 
In addition, the actual mean of $W_{ij}$'s computed using all $i\in[I]$ and $j\in[J]$ is ${\sf Loss}(\widetilde{\tX})$ given by
\begin{equation*}
	{\sf Loss}(\widetilde{\tX}) = \frac{1}{IJK} \sum_{i,j \in [I] \times [J]} \left\|\tY(i,j,:) - \widetilde{\tX}(i,j,:) \right\|_2^2.
\end{equation*}

One can see that
\begin{align*}
    W_{ij} &= \frac{1}{K}\left\|\tY(i,j,:) - \widetilde{\tX}(i,j,:)\right\|_2^2 \\
    &\leq \frac{1}{K}\left(\left\|\tY(i,j,:)\right\|_2 + \left\| \widetilde{\tX}(i,j,:)\right\|_2\right)^2 \\
    & \stackrel{(a)}{\leq} \frac{1}{K}\left(\sqrt{K}(\upsilon + \nu) + \left\|\widetilde{\C} \widetilde{\S}(:,I(i-1)+j)\right\|_2\right)^2 \\
    &\leq \frac{1}{K}\left(\sqrt{K}(\upsilon + \nu) + \left\|\widetilde{\C}\right\|_F \left\|\widetilde{\S}(:,I(i-1)+j)\right\|_2\right)^2 \\
    &\leq \frac{1}{K}\left(\sqrt{K}(\upsilon + \nu) + \sqrt{R}\alpha \sqrt{R}\beta\right)^2 \\
    & = \xi,
\end{align*}
where (a) follows because $\max_{i,j,k} |\tY(i,j,k)| = \upsilon + \nu$.

To use Lemma~\ref{lemma:hoeffding}, one can see that  $a = \min_{(i,j) \in \bOmega} W_{ij} \geq 0$, and $b = \max_{(i,j) \in \bOmega} W_{ij} \leq \xi$. Hence, using Lemma~\ref{lemma:hoeffding}, we have 
\begin{align*}
	& {\sf Pr}[|\widehat{\sf Loss}(\widetilde{\tX}) - {\sf Loss}(\widetilde{\tX})| \geq t] \\
	& \leq 2 \exp \left( - \frac{2|\bOmega|t^2}{(1-(|\bOmega|-1)/IJ)\xi^2}\right) .
\end{align*}
Using the union bound on all $\otX \in \overline{\cal X}_{R, \bm g_{\bm \theta_{\rm d}}}$ yields
\begin{align*}
	& {\sf Pr} \left[\sup_{\otX \in \overline{\cal X}_{R, \bm g_{\bm \theta_{\rm d}}}}|\widehat{\sf Loss}(\otX) - {\sf Loss}(\otX)| \geq t \right] \\
	& \leq 2 |\overline{\cal X}_{R, \bm g_{\bm \theta_{\rm d}}}|\exp \left( - \frac{2|\bOmega|t^2}{(1-(|\bOmega|-1)/IJ)\xi^2}\right) . 
\end{align*}
Equivalently, with probability at least $1-\delta$, the following holds:
\begin{align*}
	& \sup_{\otX \in \overline{\cal X}_{R, \bm g_{\bm \theta_{\rm d}}}} \left|\widehat{\sf Loss}(\otX) - {\sf Loss}(\otX) \right| \\
	& \leq \sqrt{\frac{\xi^2 \log(2|\overline{\cal X}_{R, \bm g_{\bm \theta_{\rm d}}}|/\delta)}{2} \left(\frac{1}{|\bOmega|}- \frac{1}{IJ} + \frac{1}{IJ |\bOmega|}\right)} .
\end{align*}
Let $u(\bOmega) \triangleq \sup_{\otX \in \overline{\cal X}_{R, \bm g_{\bm \theta_{\rm d}}}} \left|\widehat{\sf Loss}(\otX) - {\sf Loss}(\otX) \right|$.  Since $\left|\sqrt{a} - \sqrt{b}\right| \leq \sqrt{|a-b|}$ holds for any nonnegative $a$ and $b$, one can see that
\begin{align}\label{eq:omega1}
	\sup_{\otX \in \overline{\cal X}_{R, \bm g_{\bm \theta_{\rm d}}}}\left|\sqrt{\widehat{\sf Loss}(\otX)} - \sqrt{{\sf Loss}(\otX)}\right| \leq \sqrt{u(\bOmega)} .
\end{align}

Also note that
\begin{align}
	& \left|\sqrt{{\sf Loss}(\widetilde{\tX})} - \sqrt{{\sf Loss}(\otX)}\right| \\
	= & \frac{1}{\sqrt{IJ}} \left|\|\tY-\widetilde{\tX}\|_{\rm F} - \|\tY - \otX\|_{\rm F}\right| \nonumber \\
	\leq & \frac{1}{\sqrt{IJ}} \|\tY-\widetilde{\tX} - \tY + \otX\|_{\rm F} \nonumber \\
	= & \frac{1}{\sqrt{IJ}} \|\widetilde{\tX} - \otX\|_{\rm F} \nonumber \\
	\leq & \frac{\varepsilon}{\sqrt{IJ}} . \label{eq:omega2}
\end{align}
Similarly, we have 
\begin{align}
	\left|\sqrt{\widehat{\sf Loss}(\widetilde{\tX})} - \sqrt{\widehat{\sf Loss}(\otX)}\right| \leq \frac{\varepsilon}{\sqrt{|\bOmega|}} . \label{eq:omega3}
\end{align}

Using \eqref{eq:omega1}, \eqref{eq:omega2}, and \eqref{eq:omega3},
we show that the following holds:
\begin{align*}
	&\sup_{\widetilde{\tX} \in {\cal X}_{R, \bm g_{\bm \theta_{\rm d}}}}\left|\sqrt{\widehat{\sf Loss}(\widetilde{\tX})} - \sqrt{{\sf Loss}(\widetilde{\tX})}\right| \\
	&\leq \sup_{\widetilde{\tX} \in {\cal X}_{R, \bm g_{\bm \theta_{\rm d}}}} \left|\sqrt{\widehat{\sf Loss}(\widetilde{\tX})} - \sqrt{\widehat{\sf Loss}(\otX)}\right|  \\
	& + \left|\sqrt{\widehat{\sf Loss}(\otX)} - \sqrt{{\sf Loss}(\otX)}\right| + \left|\sqrt{{\sf Loss}(\otX)} - \sqrt{{\sf Loss}(\widetilde{\tX})}\right| \\
	& \leq  \frac{\varepsilon}{\sqrt{|\bOmega|}} + \sqrt{u(\bOmega)} + \frac{\varepsilon}{\sqrt{IJ}} .
\end{align*}
Let $\varepsilon = cR$ for $c>0$. Then using the definition of $u(\bOmega)$, one can see that the following holds with probability of at least $1-\delta$:
\begin{align*}
	 \sup_{\widetilde{\tX} \in {\cal X}_{R, \bm g_{\bm \theta_{\rm d}}}} &\left|\sqrt{\widehat{\sf Loss}(\widetilde{\tX})} - \sqrt{{\sf Loss}(\widetilde{\tX})}\right| \\
	& \leq \frac{2\varepsilon}{\sqrt{|\bOmega|}} + \left(\frac{\xi^2 \log(2|\overline{\cal X}_{R, \bm g_{\bm \theta_{\rm d}}}| /\delta)}{2} \omega\right)^{1/4} \\
	& \leq \frac{2cR}{\sqrt{|\bOmega|}} + \left( \frac{\xi^2}{2} \log\left(\frac{2}{\delta}\left|\overline{\cal X}_{R, \bm g_{\bm \theta_{\rm d}}}\right|\right)\omega\right)^{1/4},
\end{align*}
where $\omega = \left(\frac{1}{|\bOmega|}- \frac{1}{IJ} + \frac{1}{IJ |\bOmega|}\right)$ and
$$ \left|\overline{\cal X}_{R, \bm g_{\bm \theta_{\rm d}}}\right| \leq \left(\frac{3(\alpha + \beta)}{c}\right)^{R(K+D)} \alpha^{RK}\left(q \prod_{i=1}^{L}\phi_i \| \A_i\|_2\right)^{RD},$$ which was shown by Lemma~\ref{lem:cover}.  This completes the proof.

\section{Proof of Theorem \ref{thm:rmse}}\label{app:rmse}
We first denote the empirical and actual losses associated with an optimal solution $\tX^{\star}$ as follows:
\begin{subequations}
	\begin{align}
		\sqrt{\widehat{\sf Loss}(\tX^{\star})} = & \frac{1}{\sqrt{|\bOmega|K}} \| \tM_{\rm sens.} \circledast (\tX^{\star} - \tY)\|_{\rm F} \\
		\sqrt{{\sf Loss}(\tX^{\star})} = & \frac{1}{\sqrt{IJK}} \|\tX^{\star} - \tY\|_{\rm F} .
	\end{align}
	\label{eq:matform}
\end{subequations}
We show that the following chain of inequalities holds:
\begin{align*}
	& \frac{1}{\sqrt{IJK}}  \|\tX^{\star} - \tX_{\natural}\|_{\rm F} = \frac{1}{\sqrt{IJK}}\|\tX^{\star} - \tY + \tN \|_{\rm F} \\
	\leq & \frac{1}{\sqrt{IJK}}\|\tX^{\star} - \tY\|_{\rm F} + \frac{1}{\sqrt{IJK}}\|\tN\|_{\rm F} \\
	\stackrel{\rm (a)}{\leq} & \frac{1}{\sqrt{|\bOmega|K}} \|\tM_{\rm sens.} \circledast (\tX^{\star} - \tY)\|_{\rm F}+ {\sf Gap}^\star({\bm \varOmega})\\
	& + \frac{1}{\sqrt{IJK}}\|\tN\|_{\rm F}  \\
	\stackrel{\rm (b)}{\leq} & \frac{1}{\sqrt{|\bOmega|K}} \|\tM_{\rm sens.} \circledast (\widetilde{\tX}^{\ast} - \tY)\|_{\rm F} + {\sf Gap}^\star({\bm \varOmega})\\
	&+ \frac{1}{\sqrt{IJK}}\|\tN\|_{\rm F}  \\
	\leq & \frac{\left( \|\tM_{\rm sens.} \circledast (\widetilde{\tX}^{\ast} -\tX_{\natural}) \|_{\rm F} + \|\tM_{\rm sens.} \circledast (\tX_{\natural} - \tY)\|_{\rm F} \right)}{\sqrt{|\bOmega|K}} \\
	& + {\sf Gap}^\star({\bm \varOmega}) + \frac{1}{\sqrt{IJK}}\|\tN\|_{\rm F} \\
	\leq &  \frac{\|\tN\|_{\rm F} }{\sqrt{IJK}}+ \frac{\left( {\sf Err}_{\rm rep} + \|\tM_{\rm sens.} \circledast \tN\|_{\rm F} \right)}{\sqrt{|\bOmega|K}}  + {\sf Gap}^\star({\bm \varOmega}),
\end{align*}
where $(a)$ follows from \eqref{eq:matform} and the fact that ${\sf Gap}^\star({\bm \varOmega}) \geq | \sqrt{\widehat{\sf Loss}(\tX^{\star})} - \sqrt{{\sf Loss}(\tX^{\star})} | $. 
To see the reason why $(b)$ holds, note that $\tX^{\star}$ is the global optimizer of the empirical loss, and thus also minimizes
$\|\tM_{\rm sens.} \circledast (\widetilde{\tX} - \tY)\|_{\rm F}$. Therefore, any other $\widetilde{\X}$ from ${\cal X}_{R, \bm g_{\bm \theta_{\rm d}}}$---including $\widetilde{\tX}^{\ast}$ ---leads to 
$$\|\tM_{\rm sens.} \circledast ({\tX}^\star - \tY)\|_{\rm F} \leq \|\tM_{\rm sens.} \circledast (\widetilde{\tX}- \tY)\|_{\rm F}.$$

\section{The SPA Algorithm for NMF}\label{app:spa}
	\begin{algorithm}[h]\label{algo:spa}
		\SetAlgoLined
		\KwData{$\F \in \bbR^{M \times N}, R$;} 
		\KwResult{$\widehat{\A}, \widehat{\B}$;}
		    \For{$\ell \leftarrow 1: N$}{
                $\widetilde{\bm F}(:,\ell) = \frac{\bm F(:,\ell)}{\|\bm F(:,\ell)\|_1}$; \label{spa:col_norm}\\
		    } 
		    $\widehat{\ell}_1 \leftarrow \arg \max_{\ell=1,\dots,K} \|\widetilde{\F}(:,\ell)\|_2$ \label{spa:1st index};\\
		    $\widehat{\a}_1 \leftarrow \widetilde{\F}(:,\widehat{\ell}_1)$; \label{spa:1st collect}\\
			\For{$r \leftarrow 1:R$}{
			    $\E \leftarrow \begin{bmatrix} \widehat{\a}_1, \dots \widehat{\a}_r \end{bmatrix}$  \label{spa:projection_collect};\\
			    $\bm P_{\E}^{\perp} \leftarrow {\bm I} - \E(\E^\T \E)^{-1} \E^{\T}$ \label{spa:projection};\\
			    $\widehat{\ell}_{r+1} \leftarrow \arg \max_{\ell=1,\dots,K} \|\bm P_{\E}^{\perp}\widetilde{\F}(:,\ell)\|_2$ \label{spa:next index};\\
		        $\widehat{\a}_{r+1} \leftarrow \widetilde{\F}(:,{\widehat{\ell}_{r+1}})$ \label{spa:next collect};\\
		    }
		    $\widehat{\B} \leftarrow (\widehat{\A}^\T \widehat{\A})^{-1} \widehat{\A}^\T \widetilde{\F}$; \label{spa:ls}\\
			return $\widehat{\A}, \widehat{\B}$.
		\caption{\texttt{SPA}}
	\end{algorithm}
   We briefly review the SPA algorithm that is widely used in the NMF literature; see, e.g., \cite{gillis2020nonnegative,gillis2013fast,fu2014self,fu2015blind}. Consider a low-rank matrix factorization model,
    \begin{equation}\label{eq:nmf_factor_model}
     {\bm F} = \A \B, \quad  \A \geq \zero, \B \geq \zero,
    \end{equation}
    where $\bm F \in \bbR^{M \times N}$, $\A \in \bbR^{M \times R}$ and $\B \in \bbR^{R \times N}$ with $R \leq {\rm min}(M,N)$. Generally, the factors $\A$ and $\B$ cannot be identified. To be specific, one can always use $\A\Q$ and $\Q^{-1} \B$ for any invertible $\Q$ to attain the same $\bm F$. Nonetheless, under some conditions one can recover the ground truth $\A$ and $\B$ up to column scaling and permutation ambiguities \cite{fun2019nonnegative}. 
	
	One of such conditions is the so-called \textit{separability} condition \cite{fun2019nonnegative,donoho2006compressed,gillis2020nonnegative}. Under this condition, there exists a column index $\ell_r$ for every $r = 1, \dots, R$ such that $\B(:,{\ell_r}) = \alpha_r \e_r$, where $\alpha_r>0$ is a scalar and $\e_r$ is the $r$th unit vector in $\bbR^R$. This implies that a column-scaled version of $\A$ appears in $\F$ as its submatrix. In \texttt{Nasdac}, the separability condition corresponds to Assumption \ref{ass:dominance}. To see this, consider $\G^\T = \S^\T \C^\T$ and let $\bm G^\T=\bm F$, $\bm S^\T=\bm A$ and $\bm C^\T=\bm B$. 
    Under Assumption \ref{ass:dominance}, apparently $\bm B$ satisfies the separability condition.
	
	Under separability, NMF boils down to identifying the column indices $\ell_r$'s, since
	\[ \widehat{\bm A} = \bm F(: ,\{\ell_1,\ldots,\ell_R\} ) =\bm A\bm \varSigma, \]
	where $\bm \varSigma$ is a diagonal matrix and $\bm \varSigma(r,r)=\alpha_r$.
	To identify $\{\ell_1,\ldots,\ell_R\}$, assume that $\B$ satisfies 
	\begin{equation}\label{eq:simplexconstraint}
	\B^\T \one = \one,~\bm B\geq \bm 0. 	    
	\end{equation}
    Note that for applications where $\bm B$ does not satisfy \eqref{eq:simplexconstraint}, a normalization step can be used to enforce it (as long as $\bm A\geq \bm 0$ holds). Specifically, using $\widetilde{\bm F}(:,\ell) = \frac{\bm F(:,\ell)}{\|\bm F(:,\ell)\|_1}$
	leads to a factorization model $\widetilde{\bm F}=\widetilde{\bm A}\widetilde{\bm B}$ whose right latent factor $\widetilde{\bm B}$'s columns reside in the probability simplex. Identifying $\widetilde{\bm A}$ and $\widetilde{\bm B}$ helps recovering $\bm A$ and $\bm B$ up to column permutation and scaling ambiguities;
	see \cite{fun2019nonnegative,gillis2013fast} for detailed explanation. 
	
    Under \eqref{eq:simplexconstraint}, assume that ${\rm rank}(\A)=R$ and the noise is absent. The following inequalities hold:
    \begin{align*}\label{eq:spa_derivation}
        \left\|\F(:,\ell) \right\|_2 = & \left\| \sum_{r=1}^R \A(:,r) \B(r,\ell) \right\|_2  \leq \sum_{r=1}^R \left\|\A(:,r) \B(r,\ell) \right\|_2 \nonumber\\
        = & \sum_{r=1}^R \B(r,\ell)\|\A(:,r) \|_2  \leq \max_{r=1,\dots, R} \|\A(:,r)\|_2.
    \end{align*}
    where the two inequalities become equalities simultaneously if and only if $\B(:,\ell) = \e_r$ for some $r$. Hence one can find an $\ell_r$ using the following:
    \begin{equation}\label{eq:identify_col}
     \widehat{\ell}_1 = \arg \max_{\ell=1,\dots,N} \|\F(:,\ell)\|_2.
    \end{equation}
    This leads to $\widehat{\ell}_1=\ell_r$ for some $r$. We let
    the first column in the estimated $\widehat{\A}$ be $\widehat{\A}(:,1)=\bm F(:,\widehat{\ell}_1)$.
    After $\widehat{\bm A}(:, 1:k )=\bm F(:,\{\widehat{\ell}_1,\ldots,\widehat{\ell}_k\})$ is formed, where $1\leq k\leq R$, all the data columns are projected to its orthogonal complement, so that the already identified columns in $\widehat{\bm A}(:, 1:k )$ will never come up again. Then, the same procedure in \eqref{eq:identify_col} is used to identify the next $\widehat{\ell}_{k+1}$.
    The algorithm is detailed in Algorithm~\ref{algo:spa}, where the normalization step is also included.

\section{Supplementary Experiment Results }
     
    \subsection{Synthetic Data: Additional Baselines}\label{app:new_baseline}
    \begin{table}[t!]
    \scriptsize
    \centering
    \caption{ SRE under various $\rho$'s; $R=5, {\rm d}_{\rm corr} = 50, \rho = 10\% $.}
    \resizebox{\columnwidth}{!}{
    \begin{tabular}{|c|c|c|c|c|c|c|c|}
        \hline
         $\eta$ &  \texttt{GLS}    & \texttt{NMF+TPS} & \texttt{TPS} & \texttt{LL1} & \texttt{DeepComp}  & \texttt{Nasdac} & \texttt{DowJons}    \\ \hline
            4    & 0.1363  &  0.1495    & 0.1366  & 0.1014    & 0.0948   & 0.1004   & \textbf{0.0626}  \\ \hline  
            5    & 0.1249  &  0.1968    & 0.1251  & 0.1166    & 0.0918   & 0.1160   & \textbf{0.0737}  \\ \hline  
            6    & 0.1713  &  0.1936    & 0.1712  & 0.1599    & 0.1339   & 0.1331   & \textbf{0.1101}  \\ \hline  
            7    & 0.1609  &  0.1706    & 0.1602  & 0.1469    & 0.1380   & 0.1285   & \textbf{0.1170}  \\ \hline  
            8    & 0.2029  &  0.3768    & 0.2018  & 0.1812    & 0.1673   & 0.1686   & \textbf{0.1454}  \\ \hline  
    \end{tabular}}
    \label{tab:new_baseline}
    \end{table}
    In this subsection, we present experiments with two more baselines, to showcase the performance gain that could be attained by deep learning-assisted methods relative to classic methods. The first baseline is \texttt{NMF + TPS}, where we replace the deep learning based completion stage of \texttt{Nasdac} with TPS based interpolation. The second baseline is group Lasso based interpolation (\texttt{GLS}) in \cite{bazerque2011group}. \texttt{GLS} represents the PSDs using an overcomplete dictionary. Therefore, it involves selecting a basis for the PSD. We provide \texttt{GLS} with the real basis of the simulated PSDs, i.e., ${\rm sinc}^2 ( \nicefrac{k-f_i^{(r)}}{w_i^{(r)}})$, where $f_i^{(r)} \in \cF, \forall i$ are the locations of the occupied frequency bins by emitter $r$.
    
    Table \ref{tab:new_baseline} shows the SREs attained under various $\eta$ with $R=5$, $(I,J,K)=(50, 50, 64)$, and ${\rm d}_{\rm corr} = 50$. The result is averaged over 30 trials. We let each emitter occupy $M=20$ frequency bins and randomly sample the width of the bins $w_i^{(r)}, \forall i,r$ uniformly from 2 to 4.5. One can see that the clear advantage of deep learning-based methods (i.e., \texttt{DeepComp}, \texttt{Nasdac}, and \texttt{DowJons}) over conventional interpolation (i.e., \texttt{TPS}) and the combination of disaggregation and classic interpolation (i.e., \texttt{LL1}, \texttt{GLS}, \texttt{NMF}+\texttt{TPS}).

    \subsection{Real Data: Additional Scenario}\label{app:real_data_supplement}
	\begin{figure}[t]
        \centering
        \includegraphics[width=\linewidth]{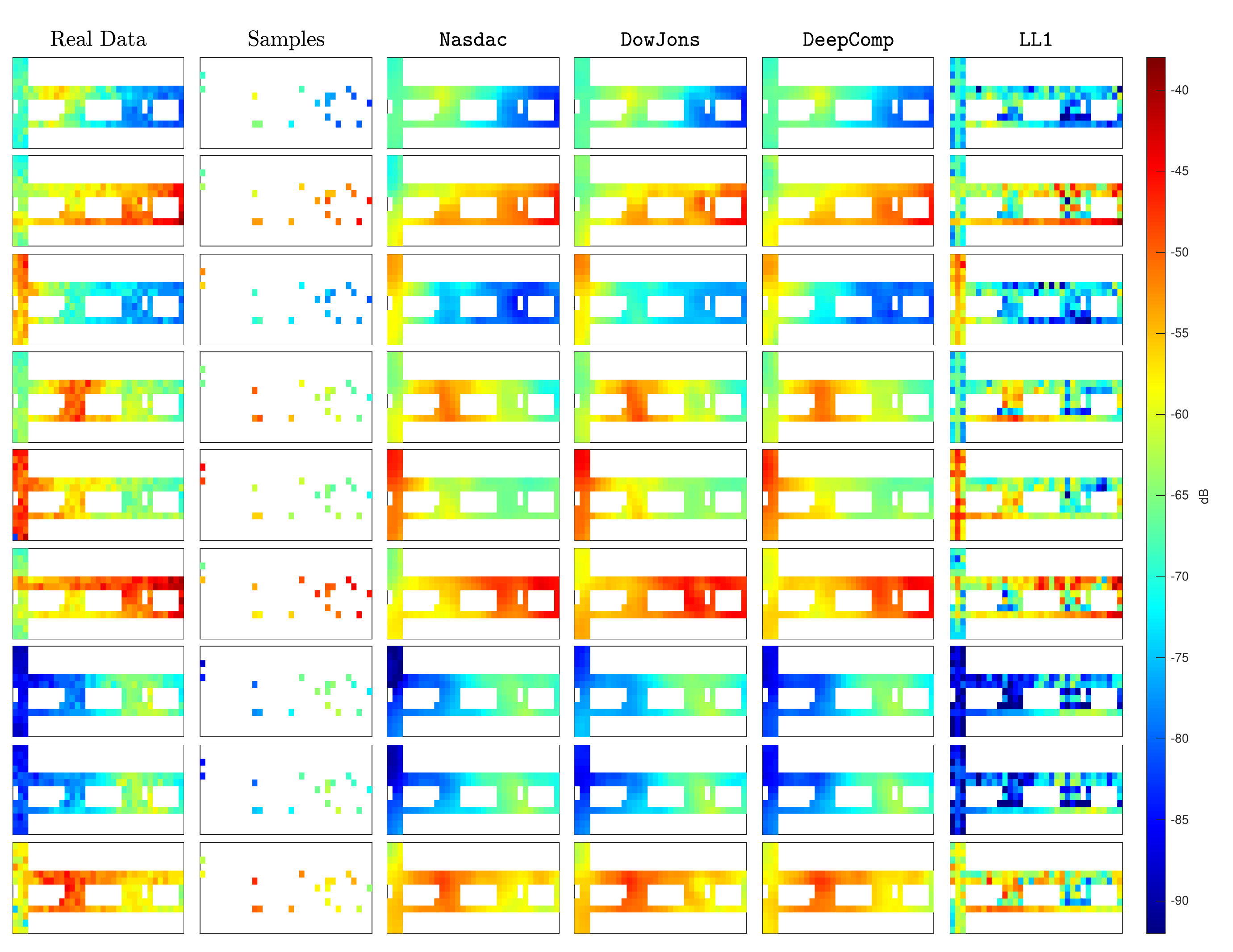}
        \caption{ Real data spectrum cartography task with 16 out of 166 observations as input to the algorithms.}
        \label{fig:10percent}
    \end{figure}
    For the real data experiment, we further observe the reconstruction performance of the algorithms using more samples (16 out of 166 measurements). The result is shown in Fig.~\ref{fig:10percent}. One can see that with 8 more measurements (compared to the experiment in Fig.~\ref{fig:real_data_severe_subsample}), the performance of the deep learning-based methods are much closer to the ground-truth data. The \texttt{LL1} method---which uses SLF disaggregation and classic interpolation---still struggles to output reasonable radio map estimations.


	\ifCLASSOPTIONcaptionsoff
	\newpage
	\fi

\end{document}

%% file: def.tex
\usepackage{steinmetz}
\renewcommand{\a}{\boldsymbol{a}}

\renewcommand{\c}{\boldsymbol{c}}

\newcommand{\e}{\boldsymbol{e}}
\newcommand{\f}{\boldsymbol{f}}
\newcommand{\g}{\boldsymbol{g}}

\newcommand{\p}{\boldsymbol{p}}

\newcommand{\s}{\boldsymbol{s}}

\newcommand{\y}{\boldsymbol{y}}
\newcommand{\x}{\boldsymbol{x}}
\newcommand{\z}{\boldsymbol{z}}

\newcommand{\btheta}{\boldsymbol{\theta}}

\newcommand{\zero}{\boldsymbol{0}}
\newcommand{\one}{\boldsymbol{1}}

\newcommand{\F}{\boldsymbol{F}}

\newcommand{\Y}{\boldsymbol{Y}}
\newcommand{\G}{\boldsymbol{G}}
\newcommand{\Q}{\boldsymbol{Q}}
\newcommand{\X}{\boldsymbol{X}}

\newcommand{\C}{\boldsymbol{C}}
\newcommand{\E}{\boldsymbol{E}}

\renewcommand{\H}{\boldsymbol{H}}
\newcommand{\M}{\boldsymbol{M}}
\newcommand{\A}{\boldsymbol{A}}
\newcommand{\B}{\boldsymbol{B}}

\renewcommand{\S}{\boldsymbol{S}}
\newcommand{\Z}{\boldsymbol{Z}}

\newcommand{\T}{{\!\top\!}}
\newcommand{\bOmega}{\bm{\varOmega}}

\newcommand{\tX}{\underline{\bm X}}
\newcommand{\tY}{\underline{\bm Y}}
\newcommand{\tZ}{\underline{\bm Z}}

\newcommand{\tM}{\underline{\bm M}}

\newcommand{\tN}{\underline{\bm N}}

\newcommand{\cC}{\mathcal{C}}

\newcommand{\cF}{\mathcal{F}}

\newcommand{\cS}{\mathcal{S}}

\newcommand{\ocS}{\overline{\cS}}
\newcommand{\oX}{\overline{\X}}
\newcommand{\otX}{\overline{\tX}}

\newcommand{\oC}{\overline{\C}}
\newcommand{\oS}{\overline{\S}}
\newcommand{\os}{\overline{\s}}
\newcommand{\oc}{\overline{\c}}

\newcommand{\bbR}{\mathbb{R}}

\newtheorem{theorem}{Theorem}
\newtheorem{assumption}{Assumption}
\newtheorem{lemma}{Lemma}
\newtheorem{proposition}{Proposition}

\newtheorem{definition}{Definition}
\newtheorem{remark}{Remark}

\usepackage{xcolor}
\usepackage{psfrag,framed}
\usepackage{lipsum}
\usepackage{chapterbib}
\PassOptionsToPackage{normalem}{ulem}
\usepackage{ulem}
\definecolor{shadecolor}{RGB}{220,220,220}